\documentclass[12pt]{article}
\usepackage{graphicx, setspace, lutypaper, simplewick, tcolorbox, enumitem}
\usepackage[export]{adjustbox} % for adjusting graphics in equations
\usepackage[normalem]{ulem}

\usepackage{caption}
\usepackage[titles]{tocloft}

\definecolor{hyperlinkbordercolor}{rgb}{.2,.7,.2}
\hypersetup{
    colorlinks = false,
    linkbordercolor = {hyperlinkbordercolor},
    citebordercolor = {hyperlinkbordercolor}
}

\makeatletter
\g@addto@macro\bfseries{\boldmath}
\makeatother
\numberwithin{equation}{section} % Equation numbering within sections

\definecolor{colorTC}{rgb}{.2,.7,.2}

\newcommand{\ii}{{\mathrm i}}
\newcommand{\no}[1]{{{} \!:\! {} #1 {} \!:\! {}}}

\renewcommand{\vec}[1]{\boldsymbol{#1}}

\begin{document}
% =============================================================================
% =============================================================================
% Title page
% =============================================================================
% =============================================================================
\begin{titlepage}
\title{Hamiltonian Truncation\\[-1pt]
Effective Theory}

\author{Timothy Cohen}
\address{\small Institute for Fundamental Science\\[-3pt] 
University of Oregon, Eugene, Oregon 97403, USA}

\author{Kara Farnsworth}

\address{\small CERCA, Department of Physics\\[-3pt]
 Case Western Reserve University, Cleveland, Ohio 44106, USA}

\author{Rachel Houtz}

\address{\small Institute for Particle Physics Phenomenology, Department of Physics\\[-3pt]
Durham University, Durham DH1 3LE, U.K.}

\author{Markus A. Luty}

\address{\small Center for Quantum Mathematics and Physics (QMAP)\\[-3pt]
University of California, Davis, California 95616, USA}

\begin{abstract}
\begin{spacing}{1.05}
Hamiltonian truncation is a non-perturbative numerical method for 
calculating observables of a quantum field theory. 
The starting point
for this method is to truncate
the interacting Hamiltonian 
to a finite-dimensional space of states spanned by the eigenvectors of the free 
Hamiltonian $H_0$
with eigenvalues below some energy cutoff $E_\text{max}$.
In this work,
we show how to treat Hamiltonian truncation systematically
using effective field theory methodology.
We define the finite-dimensional effective Hamiltonian by integrating out
the states above $E_\text{max}$.
The effective Hamiltonian can be computed by matching a transition amplitude
to the full theory,
and gives corrections order by order 
as an expansion in powers of $1/E_\text{max}$. 
The effective Hamiltonian is non-local, with the non-locality
controlled in an expansion in powers of $H_0/E_\text{max}$.
The effective Hamiltonian is also non-Hermitian, and we discuss whether
this is a necessary feature or an artifact of our definition.
We apply our formalism to 2D $\la\phi^4$ theory, and compute the
the leading $1/E_\text{max}^2$ corrections to the effective Hamiltonian.
We show that these corrections nontrivially satisfy the crucial 
property of separation of scales.
Numerical diagonalization of the effective Hamiltonian
gives residual errors of order $1/E_\text{max}^3$, as expected by 
our power counting.
We also present the power counting for 3D $\lambda \phi^4$ theory
and perform calculations that demonstrate the separation of scales 
in this theory.
%We also give some partial results on the application
%of our methods to 3D $\phi^4$ theory.
%\ML{Sounds weak to me...}
%\RH{Alternative:} \RHE{We include the first steps of applying our method to 3D $\phi^4$ theory}.
\end{spacing}
\end{abstract}

\end{titlepage}

\noindent
% =============================================================================
% =============================================================================
% =============================================================================
% =============================================================================
\setcounter{page}{2}
\setcounter{tocdepth}{2}
\begin{spacing}{.9}
\small
\tableofcontents
\end{spacing}

\section{Introduction}
\scl{Intro}
Numerical methods for studying strongly interacting quantum field theories and
quantum many-body systems are an important component of the modern
physics toolkit.
Two of the most commonly used methods are lattice Monte Carlo ({\it e.g\/}.~lattice
gauge theory) and the density matrix renormalization 
group (used mainly in condensed matter physics).
The focus of this paper is on a less-frequently used
approach known as Hamiltonian truncation, a numerical method
that diagonalizes the Hamiltonian projected onto a finite-dimensional
subspace of the full Hilbert space.
The method goes back to the earliest days of quantum mechanics,
where it is known as the Rayleigh-Ritz variational method.
Its first use in quantum field theory 
appears to be \Ref{Brooks:1983sb}.
The method was applied to renormalization group flows between
2D conformal field theories in \cite{Yurov:1989yu,Yurov:1991my},
where it was called the `truncated conformal space approach.'
This work demonstrated the effectiveness of the method
applied to 2D quantum field theories, 
and led to many applications both in elementary particle theory 
and condensed matter theory (see \cite{James:2017cpc} for a review).
More recently, there has been a
revival of interest in Hamiltonian truncation
in the quantum field theory literature following
the pioneering works
\Refs{Hogervorst:2014rta,Rychkov:2014eea,Katz:2014uoa}.
See \Refs{Rychkov:2015vap,Elias-Miro:2015bqk,Bajnok:2015bgw,Katz:2016hxp,Anand:2017yij, Elias-Miro:2017xxf, Elias-Miro:2017tup,Rutter:2018aog,Fitzpatrick:2018ttk, Hogervorst:2018otc,Delacretaz:2018xbn, 
Anand:2019lkt,Fitzpatrick:2019cif,EliasMiro:2020uvk,Anand:2020gnn,Anand:2020qnp,Hogervorst:2021spa}
for examples of subsequent developments and applications.

In this paper, we study a version of Hamiltonian truncation where the
finite-dimensional Hilbert space is defined using an energy cutoff,
since in this case we expect to be able to apply the 
ideas and techniques
of low-energy effective field theory \cite{Georgi:1993mps}.
Specifically, the full Hamiltonian is written
\[
H = H_0 + V,
\eql{FullTheoryH}
\]
where $H_0$ is a free Hamiltonian that can be diagonalized exactly.
The finite-dimensional Hilbert space  $\scr{H}_\text{eff}$ is defined to be linear combinations of
$H_0$ eigenstates 
\[
H_0 \ket{E_i} = E_i \ket{E_i},
\]
with $E_i \le E_\text{max}$.
We are interested in theories where the interactions in $V$ are weak
in the UV, for example theories with relevant couplings.
For such theories, we will show how to systematically construct
the effective Hamiltonian as an expansion in $1/E_\text{max}$
in perturbation theory.
We expect that physical quantities sensitive to energies
well below $E_\text{max}$ can be approximated by such an
effective Hamiltonian.

A fundamental limitation of Hamiltonian truncation is that the number of
states in $\scr{H}_\text{eff}$ grows exponentially with $E_\text{max}$, while the
accuracy is expected to decrease as a power of $1/E_\text{max}$.
Since the computational resources scale with the number of states,
the accuracy only improves logarithmically with computational
resources, at least for conventional computational methods.%\hspace{-3pt}
\footnote{Quantum computers can efficiently store an 
exponentially large Hilbert space with linear resources (qubits).
The development of a complete quantum algorithm for Hamiltonian
truncation is an interesting problem for future work.}
Despite this limitation, interesting levels of accuracy have been obtained using
Hamiltonian truncation in low-dimensional systems 
(reviewed in \cite{James:2017cpc}).
Furthermore, this method has the potential to perform
calculations in theories that are not easily treated with lattice 
methods, for example theories with chiral fermions 
\cite{Fitzpatrick:2019cif} or theories with
sign problems \cite{Bajnok:2015bgw, Rychkov:2015vap}.
This strongly motivates further study of Hamiltonian truncation
to determine its ultimate potential.
Our focus in the present paper is improving the convergence of 
the method as a function of the cutoff $E_\text{max}$,
a problem that has already been studied in
\cite{Lee:2000xna,Elias-Miro:2015bqk,Elias-Miro:2017tup,Elias-Miro:2017xxf,  
Rutter:2018aog}.
We will compare our method with these works in \sec{Previous}.

In this paper, we develop a systematic approach to Hamiltonian
truncation using the methodology of effective field theory.
We call the resulting formalism Hamiltonian Truncation Effective
Theory (HTET).
The essential idea is that the truncation parameter $E_\text{max}$ 
is treated as a UV cutoff, 
and the finite-dimensional effective Hamiltonian is defined by matching
to the full theory.
The resulting theory has many of the expected features of more conventional
effective field theories, but also has some unique features that result
from the nature of the cutoff:
\begin{itemize}
\item
We define the effective Hamiltonian by matching
a 
transition amplitude that is well-defined in finite volume
order by order in powers of $V$.
The matching can be carried out using a systematic diagrammatic expansion
similar to time-ordered perturbation theory.
\item 
The effective Hamiltonian is non-local.
This arises because the cutoff of the effective theory is non-local:
$E_\text{max}$ is the maximum value of the \emph{total} energy
(defined by $H_0$) of the system, which 
gets contributions from all excitations regardless of how far apart they are.
\item
The effective Hamiltonian is non-Hermitian.
Because time evolution with the full Hamiltonian mixes
states above and below the cutoff, there is no 
physical reason to expect the effective Hamiltonian 
to be Hermitian.
We discuss whether the non-Hermiticity is a necessary
feature, or an artifact of our definitions.
We give some arguments that non-Hermiticity is necessary to 
maintain desirable properties of the effective Hamiltonian,
but we leave a full discussion for future work.
\item 
We propose a power counting to all orders in the 
$1/E_\text{max}$ expansion in which the
non-Hermiticity and non-locality are controlled 
by an expansion in powers of $H_0/E_\text{max}$.
\item
Our
formalism can be applied without modification to theories with non-trivial UV divergences.
The renormalized fundamental theory gives predictions
that are finite and
independent of $E_\text{max}$, so matching is expected to
give an effective Hamiltonian whose predictions
are finite and independent of $E_\text{max}$.
\end{itemize}

The crucial property of any effective theory is that it factorizes
physical effects 
associated with different scales.
This means that the effective Hamiltonian depends on the properties of states above
the cutoff, and parameterizes the effects below the cutoff.
In perturbative matching calculations such as the one developed in this paper,
the separation of scales has a precise meaning:
corrections to the effective Hamiltonian are given by
sums over states that are dominated by states near the effective theory cutoff.
In particular, this implies that 
the effective theory matching corrections are insensitive to IR modifications of the theory,
such as masses and compact spatial dimensions, as long as the mass scale
of these modifications is small compared to the cutoff scale.
The calculations performed in this paper for $\la\phi^4$ theory in 2D and 3D
give a
nontrivial demonstration of this property, involving 2- and 3-
loop diagrams with overlapping UV/IR dominated regions
that cancel only with the correct operator definition 
and renormalization scheme.

To check that this formalism
actually improves the numerical convergence as expected, we
perform numerical calculations for 2D $\la\phi^4$
theory.
We find that the size of the numerical error is compatible with the
theoretically predicted scaling with powers of $1/E_\text{max}$, namely
$O(1/E_\text{max}^2)$ for the $O(\la)$ effective Hamiltonian and
$O(1/E_\text{max}^3)$ at $O(\la^2)$.
At this order, the effective Hamiltonian is local and Hermitian.
Going to higher orders is simply a matter of computing additional diagrams,
and we show by explicit computation
that there are non-local and non-Hermitian $1/E_\text{max}^3$ corrections to the effective 
Hamiltonian.
In future work, we plan to extend our calculations in
2D $\la\phi^4$ theory to higher order, and to perform numerical
calculations for 3D $\la\phi^4$ theory.
This will check that our formalism works when the effective
Hamiltonian is non-local, and in theories
with UV divergences.

The rest of this paper is organized as follows.
We derive a general formalism for matching onto the effective Hamiltonian in \sec{MatchingOntoHEff}.  
In \sec{diagrams} we derive a set of diagrammatic rules to compute the transition
amplitude that is used to perform the matching, using 2D $\la\phi^4$ theory as an example.
The renormalization and matching for 2D $\la \phi^4$ theory 
are presented in \sec{Renormalization} and \sec{Matching2DTheory} respectively.
We discuss the power counting of the effective theory in \sec{PowerCounting},
and give an explicit example of non-local and non-Hermitian terms that appear
at higher orders in the expansion.
Our numerical results for this theory are presented in  \sec{Numerics}.
In \sec{Previous}, we compare our formalism and results to previous 
results in the literature.
We conclude in \sec{Conc} with a discussion of future directions.
In Appendix A, we present some calculations for 
3D $ \la\phi^4$ theory
to illustrate the application of our methods
to theories with non-trivial UV divergences.
In Appendix B, we consider the possibility of defining
a Hermitian effective Hamiltonian by a similarity transformation.

\section{Effective Hamiltonian from Matching}
\scl{MatchingOntoHEff}
In this section, we 
define the effective Hamiltonian 
by matching
to the predictions of the full 
theory order by order in an expansion in powers of $V$.%
\footnote{Our definition differs from
the `exact effective Hamiltonian'
of \Refs{Hogervorst:2014rta,Rychkov:2014eea}, which 
is a function of the energy eigenvalue
that is being computed.
See \sec{ExactEffH}.}
This definition is the basis for the systematic computation
the effective Hamiltonian in an expansion in $1/E_\text{max}$,
which we present in \sec{PowerCounting}.
We make the separation of the full Hamiltonian into the free and 
interacting parts defined in \Eq{FullTheoryH}, and 
assume that the spectrum of $H_0$ (and $H$) is discrete, with
\[
\eql{eigenvals}
H_0 \ket{i} = E_i \ket{i},
\qquad
i = 0, 1, 2, \ldots
\]
We denote the full Hilbert space by $\scr{H}$, and the finite-dimensional subspace spanned by the states $\ket{i}$ with $E_i \le E_\text{max}$
by $\scr{H}_\text{eff}$. 
Our goal is to define an effective Hamiltonian $H_\text{eff}$ acting
on the effective Hilbert space
$\scr{H}_\text{eff}$, so that the low-lying eigenvalues of $H_\text{eff}$
approximate those of $H$.

The simplest approximation for the effective Hamiltonian is
\[
\eql{obviousapprox}
\bra{f} H_\text{eff} \ket{i}
\simeq \bra{f} H \ket{i}
\]
for $\ket{i},\ket{f} \in \scr{H}_\text{eff}$.
That is, $H_\text{eff}$ is the restriction of $H$ to the low-energy
subspace $\scr{H}_\text{eff}$.
The goal of our formalism is to systematically improve this 
approximation for theories in which the interaction $V$ can be treated
as a perturbation in the UV.
For 2D $\la \phi^4$ theory, \Eq{obviousapprox} is a good
starting approximation, and we will focus mainly on that case in this paper.
For theories with non-trivial UV divergences, one must go to higher orders in the matching
to obtain a good starting approximation.
We will discuss the example of 3D $\la\phi^4$ theory in
Appendix A.

The purpose of the effective theory
approach is to systematically improve the approximation
\Eq{obviousapprox} by including the effects of the states above the cutoff
$E_\text{max}$.
In theories that are weakly coupled in the UV, we expect that
the effects of the states above $E_\text{max}$ can be computed 
order by order in powers of $V$.
In the matching approach used here, this
is done by matching a physical quantity in the fundamental
and effective theory order by order in powers of $V$.
This means that we choose some observable that can be computed
in both the fundamental and effective theory, and 
define the  
effective Hamiltonian
by requiring that the physical quantities
agree. 
An obvious physical quantity to match in Hamiltonian truncation
is the spectrum of energy eigenvalues.
However, we will now show that this is not sufficient to completely
define the effective Hamiltonian.

\subsection{Matching the Spectrum}

\scl{matchspectrum}
For simplicity, we assume that the spectrum of $H_0$
is non-degenerate.
In perturbation theory, there is a one-to-one correspondence between
eigenvalues of $H$ and $H_0$: 
\[
H \ket{i} = \scr{E}_i \ket{i},
\eql{Heigenvalue}
\]
where 
\[
\scr{E}_i  = E_i + \scr{E}_{1i} + \scr{E}_{2i} + \cdots,
\qquad\text{with} \qquad
\scr{E}_{ni} = O(V^n). 
\]
Here $\ket{i} \in \scr{H}_\text{eff}$ and we have already set $\scr{E}_{0i} = E_i$ from \Eq{eigenvals} since we are working in the basis of the unperturbed Hamiltonian $H_0$. 
We write
\[
H_\text{eff} = H_0 + H_1 + H_2 + \cdots,
\qquad
H_n = O(V^n)
\]
and use standard Rayleigh-Schr\"odinger perturbation theory
compute the eigenvalues as an expansion in powers of $V$ in the effective
theory:
\begin{subequations}
\eql{Ematch}
\[
\scr{E}_{1i} &= \bra{i} H_1 \ket{i},
\eql{Ematch1}
\\[5pt]
\scr{E}_{2i} &= \sum_{\al \, \ne \, i}^<
\frac{\bigl| \bra{i} H_1 \ket{\al} \bigr|^2}{E_{i\al}}
+ \bra{i} H_2 \ket{i},
\\[5pt]
\!\!\!\!
\scr{E}_{3i} &= \sum_{\al,\be \, \ne \, i}^<
\frac{\bra{i} H_1 \ket{\al} \bra{\al} H_1 \ket{\be} \bra{\be} H_1 \ket{i}}
{E_{i\al} E_{i\be}
}
\nonumber\\
&\qquad{}
+ \sum_{\al \, \ne \, i}^< \frac{1}{E_{i\al}} \biggl[
\bra{i} H_1 \ket{\al} \bra{\al} H_2 \ket{i}
+ \bra{i} H_2 \ket{\al} \bra{\al} H_1 \ket{i}
\nonumber\\[-3pt]
&\qquad\qquad\qquad\qquad\quad{}
- \frac{\bra{i} H_1 \ket{\al} \bra{\al} H_1 \ket{i} \bra{i} H_1 \ket{i}}{E_{i\al}}
\biggr]
+ \bra{i} H_3 \ket{i},
\]
\end{subequations}
where we write
$E_{i\al} = E_i - E_\al$ and
\[
\sum_\al^< = \sum_{E_\al \, \le \, E_\text{max}}
\qquad \text{and}\qquad
\sum_\al^> = \sum_{E_\al \, > \, E_\text{max}}.
\]
These restricted sums appear because $H_\text{eff}$ is an operator on $\scr{H}_\text{eff}$,
so that the intermediate states in the effective theory
must be restricted to the low-energy subspace.

It is straightforward to see that \Eqs{Ematch} do not uniquely define
$H_\text{eff}$ order by order in the expansion in $V$.
We assume that $\scr{E}_{ni}$ have been computed
in the fundamental theory, and we are using \Eqs{Ematch} to determine 
the corrections $H_n$ in the effective theory.
At first order (see~\Eq{Ematch1}), the matching only determines the diagonal
elements of $H_1$.
At order $V^n$ with $n > 1$, the matching depends on the off-diagonal components
of $H_m$ with $m < n$, as well as the diagonal elements of $H_n$.
We therefore need additional relations to fully 
define the effective Hamiltonian.
In the following section, we show that matching a different quantity,
related to the $S$-matrix, 
completely defines
$H_\text{eff}$ order by order in perturbation theory.

\subsection{\label{sec:T}The Transition Matrix}
We showed 
above that matching the stationary states
does not uniquely define the effective Hamiltonian.
We need a complete definition for the effective Hamiltonian
as a starting point for systematic expansion.
In this subsection we will show that the effective Hamiltonian 
can be defined order by order in powers of $V$
by matching a transition amplitude.
It is natural to guess that the effective Hamiltonian
can be defined in this way.
In effective field theories in finite volume, the effective
Hamiltonian can be defined by matching the $S$-matrix,
the unitary time evolution operator between asymptotic 
scattering states.
The $S$-matrix is not well-defined in finite volume,
but we will show that we can define the effective Hamiltonian
by matching a similar observable.

We start with an initial state at $t_i = 0$.
We then turn off the interactions adiabatically for $t > 0$
by making the replacement
\[
V \to V e^{-\ep t},
\]
where $\ep > 0$ is an infinitesimal regulatory parameter with units of energy.
We then evolve the state to $t_f \to +\infty$, and 
compute its overlap with an eigenstate of $H_0$:
\[
\lim_{t_f \, \to \, \infty}
\bra{f} e^{-\ii H_\text{eff} t_f} \ket{i}
\stackrel{?}= \hspace{-5pt}
\lim_{\hspace{5pt}\ep \, \to \, 0^+}
\bra{f} \text{T}\ggap\text{exp}\left\{ -\ii \int_0^\infty \mathrm{d}t\ggap 
\bigl( H_0 + V e^{-\ep t} \bigr) \right\} \ket{i},
\]
where $\text{T}$ is the time ordering operator.
This is still not quite what we want, because it involves an ill-defined infinite
phase in the limit $t_f \to \infty$.
However, we can factor out this phase by working
in the interaction picture, 
defined in terms of Schr\"odinger picture by
\[
\ket{\Psi(t)}_\text{IP} = e^{\ii H_0 t} \ket{\Psi(t)}_\text{SP},
\qquad
\scr{O}_\text{IP}(t) = e^{\ii H_0 t}\ggap \scr{O}_\text{SP}\ggap e^{-\ii H_0 t}.
\]
The interaction picture time evolution operator is
\[
U_\text{IP}(t_f, t_i) 
&= \text{T}\ggap\text{exp}\left\{ -\ii \int_{t_i}^{t_f} \mathrm{d}t\ggap 
V_\text{IP}(t) \right\},
\eql{UIP}
\qquad\text{with}\qquad
V_\text{IP}(t) = e^{\ii H_0 t}\ggap V e^{-\ep t} e^{-\ii H_0 t}.
\]
We then define the operator
\[
\bra{f} \Si(\ep) \ket{i} \equiv \lim_{t_f \, \to \, \infty}
\bra{f} U_\text{IP}(t_f, 0) \ket{i},
\eql{defSigma}
\]
where we emphasize that $\ep \neq 0$ in the definition of $\Sigma$.

We now work out the perturbative expansion of $\Si$.
To do this, we note that the time evolution operator
written in \Eq{UIP} obeys
\[
\frac{\d}{\d t_i} U_\text{IP}(t_f, t_i)
= \ii\gap U_\text{IP}(t_f, t_i) V_\text{IP}(t_i),
\]
with boundary condition $U_\text{IP}(t_f, t_f) = \id$.
The solution is
\[
U_\text{IP}(t_f, t_i) = \id - \ii \int_{t_i}^{t_f} \mathrm{d}t\ggap
U_\text{IP}(t_f, t) V_\text{IP}(t) .
\]
This all-orders relation can be expanded iteratively in powers of $V$.
Applying this to \Eq{defSigma}, the leading terms are \[
\bra{f} \Si \ket{i} 
&= \de_{fi} + \frac{\bra{f} V \ket{i}}{E_{fi} + \ii\ep} 
+ \sum_\al \frac{\bra{f} V \ket{\al} \bra\al V \ket{i}}
{(E_{fi} + \ii\ep)(E_{f\al} + \ii \ep)}
+ O(V^3),
\eql{SigmaExpan}
\]
where $E_{f\al} \equiv E_f - E_\al$, as before.
This expression is not symmetric under the exchange $i \leftrightarrow f$ due to the fundamental time
asymmetry built into the definition of $\Sigma$ given in \Eq{defSigma}. 
Note that the \rhs\ of \Eq{SigmaExpan} is actually not well-defined because some of
the energy denominators vanish for $\ep = 0$.%
\footnote{We note that $\Si$ can be made well-defined in the $\ep \to 0$ limit if one chooses to
define $V$ so that it does not contain any diagonal terms when expressed in the $H_0$ eigenbasis.
That is, we make the replacements
\[
\eql{Vreplace}
H_0 \to H_0 + V_\text{diag},
\qquad % \qquad\text{and}\qquad
V \to V - V_\text{diag},
\]
where 
\[
V_\text{diag} = \sum_{i} \ket{i} \bra{i} V \ket{i} \bra{i},
\]
is the diagonal part of $V$.
As in perturbation theory for eigenvectors, this ensures that the
perturbation of the states are perpendicular to the unperturbed states.
The result for the effective Hamiltonian (\Eq{Heffmatch} below) is invariant
under this shift.
However, this approach is not convenient for the diagrammatic expansion
discussed in \S\ref{sec:diagrams} below, and so we will not pursue it further.
}
However, we will see below that these singular terms cancel in the matching.
We can therefore treat $\ep$ as an IR regulator,
taking the limit $\ep \to 0$ after performing the matching calculation.
The fact that the matching is insensitive to the IR details of the
theory is an important feature of the 
effective theory approach that will be discussed further
in \S\ref{sec:Matching2DTheory} below.

It is convenient to remove the $E_{fi}$
energy denominator that is common to all terms
by defining the $T$-matrix:
\[
\eql{Tdefn}
\bra{f} \Si \ket{i} = \de_{fi} 
+ \frac{\bra{f} \gap T \ket{i}}{E_{fi} + \ii\ep}.
\]
By analogy with the $S$-matrix, we will
refer to $T$ as the `transition matrix.'
The first few terms in the perturbative expansion of $T$ are then
\[
\bra{f} \gap T \ket{i} = \bra{f} V \ket{i}
+ \sum_\al \frac{\bra{f} V \ket{\al} \bra{\al} V \ket{i}}
{E_{f\al} + \ii\ep}
+ \sum_{\al,\be}
\frac{\bra{f} V \ket{\al} \bra{\al} V \ket{\be} \bra{\be} V \ket{i}}
{(E_{f\al} + \ii\ep)(E_{f\be} + \ii\ep)}
+ O(V^4),
\eql{Texpand}
\]
or in other words, the $O(V^n)$ contribution to $\bra{f} \gap T \ket{i}$ is given by
\[
\eql{Tn}
T_n = \!\! \sum_{\al_1,  \ldots,  \al_{n-1}}
\frac{\bra{f} V \ket{\al_1} \bra{\al_1} V \ket{\al_2}
\cdots \bra{\al_{n-1}} V \ket{i}}
{(E_{f\al_1} + \ii\ep) \cdots (E_{f\al_{n-1}} + \ii \ep)}.
\]
Note that the sum includes terms where the energy
denominators $E_{f\al}$ vanish,
and therefore the transition matrix diverges as $\ep \to 0$.
We can think of $\ep$ as an IR regulator, so these terms
are IR divergent.
However, we will show that these IR divergences cancel
in the matching, and the effective Hamiltonian is
completely defined by matching $T$ order by order in
powers of $V$.

\subsection{Matching the Transition Matrix}
\scl{match}
To match, we must compute the transition matrix $T$ in the effective theory.
In terms of
\[
V_\text{eff} = H_1 + H_2 + \cdots,
\]
the interaction picture time evolution operator is given by
\[
\hspace{-12pt} U_\text{eff,\gggap IP}(t_f, t_i) =
\text{T}\ggap\text{exp}\bigg(\! -\ii \int_{t_i}^{t_f} \!\mathrm{d}t\ggap 
V_\text{eff,\gggap IP}(t) \bigg),
\quad\text{  with  }\quad
V_\text{eff,\gggap IP}(t) = e^{\ii H_0 t}\ggap V_\text{eff} \ggap
e^{-\epsilon t} e^{-\ii H_0 t}.
\]
Sums over states in the effective theory are restricted to satisfy $E \le E_\text{max}$.
Matching the matrix elements of the transition matrix to the fundamental
theory, we then obtain our matching conditions:
\begin{subequations}
\eql{Heffmatch}
\begin{align}
\bra{f} H_1 \ket{i}_\text{eff} &= \bra{f} V \ket{i},
\eql{H1effmatch}
\\[5pt]
\bra{f} H_2 \ket{i}_\text{eff} &= 
 \sum_\al^> \frac{\bra f V \ket\al \bra\al V \ket i}{E_{f\al}},
 \eql{H2match}
\\[5pt]
\bra{f} H_3 \ket{i}_\text{eff} &= \sum_{\al,\, \be}^>
\frac{\bra f V \ket\al \bra\al V \ket\be \bra\be V \ket i}{E_{f\al} E_{f\be}}
- \sum_\al^< \sum_\be^>
\frac{\bra f V \ket\al \bra\al V \ket\be \bra\be V \ket i}{E_{\al\be} E_{f\be} }.
\end{align}%
\eql{Hmatch}%
\end{subequations}%
The matrix elements of the effective Hamiltonian are written with a subscript `eff'
to remind us that the matrix elements are evaluated in the finite-dimensional
Hilbert space $\scr{H}_\text{eff}$.
Note that for $E_i, E_f \ll E_\text{max}$ the energy denominators
in \Eq{Heffmatch} are of order $E_\text{max}$ or larger.
This reflects the fact that the matching calculation is insensitive
to the IR details of the theory, and is important for the
separation of scales, as we discuss below.
It also means that we can take the limit $\ep \to 0$ 
in the matching, so the IR divergences in the transition
matrix $T$ do not affect the matching.

$H_\text{eff}$ as given in \Eqs{Hmatch}
defines the effective
Hamiltonian order by order in powers of $V$.
We have verified to $O(V^4)$
that the eigenvalues computed using $H_\text{eff}$ agree
with those obtained from directly matching the eigenvalues 
\Eq{Ematch}.
A peculiar feature of the effective Hamiltonian defined here is that it is not Hermitian.
Note that time evolution with the full Hamiltonian $H$ mixes
states in the low-energy subspace $\scr{H}_\text{eff}$ with
states that are not in $\scr{H}_\text{eff}$.
Therefore, there is no reason {\it a priori} that the
effective Hamiltonian must be Hermitian.
However, it is 
interesting to explore if there exists
an alternative definition of the effective Hamiltonian
that is Hermitian.
Beyond the conceptual implications, this is 
motivated by the fact that numerical algorithms for
diagonalizing Hermitian matrices are more efficient than for
non-Hermitian matrices.

It is not obvious that $T$ is the `correct' observable
to match.
Is it possible to define a Hermitian effective Hamiltonian
that has the desired properties, such as separation of scales?
We do not have a complete answer, but we have
considered a few simple possibilities, and find that they
do not work.
For example, we can define a Hermitian effective Hamiltonian by
matching $T + T^\dagger$.
At $O(V^2)$ this gives an effective Hamiltonian that is related
to the Schrieffer-Wolf effective Hamiltonian \cite{PhysRev.149.491}),
but at $O(V^3)$ the effective Hamiltonian defined in this
way diverges in the limit $\ep \to 0$.%
\footnote{At $O(V^2)$, this gives the Hermitian average
of our effective Hamiltonian
\[
\bra f H_2 \ket i
= \frac 12 \sum_\al^> \bra f V \ket\al \bra\al V \ket i 
\left( \frac{1}{E_{f\al}} + \frac{1}{E_{i\al}} \right),
\]
but at $O(V^3)$ we obtain
\[
\bra{f} H_3 \ket{i}
&= \frac 12 \sum_{\al, \be}^>
\bra f V \ket\al \bra\al V \ket\be \bra\be V \ket i
\left( \frac{1}{E_{f\al} E_{f\be}} +
\frac{1}{E_{i\al} E_{i\be}} \right)
\nn
&\qquad{}
+ \frac 14 \sum_\al^> \sum_\be^<
\bra f V \ket\al \bra\al V \ket\be \bra\be V \ket i
\frac{E_{i\al}^2 + E_{i\be}E_{f\al} + E_{f\al}E_{\al\be}}
{E_{i\be} E_{i\al} E_{f\al} E_{\al\be}}
\nn
&\qquad{}
- \frac 14 \sum_\al^< \sum_\be^>
\bra f V \ket\al \bra\al V \ket\be \bra\be V \ket i
\frac{E_{f\be}^2 + E_{f\al}E_{i\be} - E_{i\be}E_{\al\be}}
{E_{f\al} E_{f\be} E_{i\be} E_{\al\be}}.
\]
We have omitted $\ii\ep$ terms for brevity.
This contains IR divergent terms where the energy denominators
vanish, for example the terms with $\al = f$ in the last line.}
This is an IR divergence, signaling a failure of separation
of scales.

Another possibility is to define a new effective Hamiltonian
from our $H_\text{eff}$ by
a similarity transformation
$H_\text{eff}' = G H_\text{eff}\gap G^{-1}$.
This does not change the spectrum, so $H_\text{eff}'$ is
a suitable effective Hamiltonian.
If $G$ is not unitary, it may be possible to obtain a
Hermitian effective Hamiltonian in this way.
In Appendix B, we present a choice of $G$ that removes
some but not all of the leading
non-Hermitian terms in 2D $\la\phi^4$ theory.

It would be be interesting to investigate alternative
definitions of the effective Hamiltonian,
but we leave this for future work.

\section{Diagrammatic Rules}
\scl{diagrams}
It is very useful to have a diagrammatic expansion
for the transition matrix $T$ to perform the matching.
In this section, we derive such a set of diagrammatic rules.
This requires specifying a model, and we use 
2D $\la\phi^4$ theory as an example.
This diagrammatic expansion is similar to `old-fashioned perturbation theory' for the $S$-matrix.

Not only will these rules serve as a useful 
calculational tool, but they also provide insight into the properties of the effective Hamiltonian.
For example, the apparent non-locality of the effective Hamiltonian that appears at subleading order in the $1/E_\text{max}$ expansion will have a simple diagrammatic interpretation.
Additionally, the diagrammatic expansion also illuminates the UV divergence
structure of the theory, making the interplay between matching and renormalization completely
transparent.

We consider
the Lagrangian density
\[
\scr{L} = \frac 12 (\d\phi)^2 - \frac 12 \ggap m^2 \phi^2 - \frac{\la}{4!}\gap \phi^4.
\]
The mass dimension of the fields and couplings in 2D
are
\[
[\phi] = 0,
\qquad 
[\la\gap] = [m^2] = 2.
\eql{dimensionalAnalysis}
\]
The $\phi^4$ coupling is relevant, meaning that this interaction is weak in the UV and
strong in the IR.

We quantize the theory on a spatial circle of radius $R$:
\[
\phi(x + 2\pi R, t) = \phi(x, t).
\]
We can therefore expand the fields in a discrete set of momentum modes:
\[
\phi(x, t) = \frac{1}{\sqrt{2\pi R}} \sum_{k \, \in \, \mathbb{Z}}
e^{\ii kx/R} \phi_k(t).
\]
The fact that $\phi$ is Hermitian implies
\[
\eql{phireal}
\phi_k^\dagger = \phi\sub{-k}.
\]

To simplify the calculations, it is useful to define the vertices
to be normal-ordered operators.
This means that we choose a `quantization mass' $m_\text{Q}$ and write
the Hamiltonian in terms of creation and annihilation operators
for Fock states with mass $m_\text{Q}$: 
\[
\eql{quantizationmassdefn}
H_0 = \sum_k \om_k a_k^\dagger a\sub{k},
\qquad\text{with} \qquad
\om_k = \sqrt{(k/R)^2 + m_\text{Q}^2},
\]
where creation and annihilation operators obey the standard
canonical commutation relations
\[
\bigl[a\sub{k}, a^\dagger_{k'} \bigr] = \de\sub{kk'}.
\]
The interaction term is then given by
\[
\eql{mVdefn}
V &= \myint \text{d} x\ggap \bigg[ \frac 12\gap m_V^2 \no{\phi^2}
+ \frac{\la}{4!} \no{\phi^4} \bigg],
\]
where $m_V$ is an `interaction mass' that contributes to $V$ 
and $\no{\cal{O}}$ denotes the normal ordering of the operator $\cal{O}$
with respect to the creation and annihilation operators defined above.
For the purpose of deriving the diagrammatic rules, we treat the
parameters $m_\text{Q}$ and $m_V$ as finite quantities;
the renormalization of the theory is discussed in \sec{Renormalization} below.
We write
\[
\eql{phimode}
\phi\sub{k} = \phi^{(+)}_k + \phi^{(-)}_{-k},
\]
where
\[
\phi^{(+)}_k = \frac{1}{\sqrt{2\om_k}}\gap a\sub{k},
\qquad 
\phi^{(-)}_k = \frac{1}{\sqrt{2\om_k}}\gap a^\dagger_{k}.
\eql{DefineCreateAndAnn}
\]
(Note that \Eq{phimode} satisfies \Eq{phireal}.)

We then obtain the diagrammatic rules using Wick's theorem.
The products of $V$ that appear in $T$ are not time-ordered (see \Eq{Tn}),
so the relevant contraction is defined by
\[
\phi_1 \phi_2 &= \no{\phi_1 \phi_2} + 
\bcontraction{}{\phi}{_1 }{\phi}
\phi_1 \phi_2.
\]
where
\[
\bcontraction{}{\phi}{_k }{\phi}
\phi_k \phi_{k'} = \frac{\de_{kk'}}{2\om_k}.
\]
Because we are writing $V$ in terms of normal-ordered operators,
the version of Wick's theorem we are using is
\[
\big(\no{ \scr{O}_1} \big) \cdots \big(\no{\scr{O}_n}\big)
= \no{ \big(\scr{O}_1 \cdots \scr{O}_n\big)}
+ \text{contractions}
\]
where contractions between fields in the same operator are omitted.
For example,
\[
\no{\phi_1^2} \no{\phi_2^2}
= \no{\phi_1^2 \phi_2^2}
+ 4 \bcontraction{}{\phi}{_1 }{\phi}
\phi_1 \phi_2
\no{\phi_1 \phi_2}
+ 2 \big(  \bcontraction{}{\phi}{_1 }{\phi}
\phi_1 \phi_2 \big)^2.
\]

The normal-ordered operators that appear in Wick's theorem consist of a
sum of terms with powers of $\phi^{(+)}$ acting on initial states to the right,
and powers of $\phi^{(-)}$ acting on final states to the left.
Each monomial in $\phi^{(\pm)}$ is represented by a sum of diagrams.
As usual, we denote the powers of $V$ by vertices, and the Wick contractions
by lines connecting to other vertices or external states.
The vertices in the diagram are ordered from left to right in the same order that
the insertions of $V$ appear in \Eq{Tn}.
The diagrammatic rules for $T$ at order $V^n$ can be summarized as follows:
\begin{itemize}
\item
Draw all possible diagrams with $n$ ordered vertices.
Each line is either connected to a different vertex,
to the initial state on the right,
or the final state on the left.  
Disconnected diagrams must be included. 
\item
Assign an independent mode number $k$ to each internal and external line.
Sum over the internal momenta.
\item
The rules for the vertices are
\[
\eql{vertexrules}
\includegraphics[valign=c,scale=0.75]{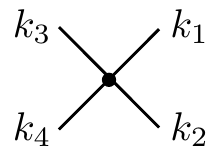} 
= \frac{\la}{2\pi R} \de_{k_1 + \cdots + k_4},
\qquad
\includegraphics[valign=c,scale=0.75]{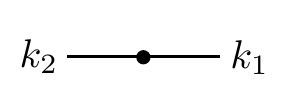} = m_V^2 \de_{k_1 k_2},
\]
where the momenta are taken to all flow into the vertex.
\item
Each internal line is associated with a factor of
\[
\includegraphics[valign=c,scale=0.75]{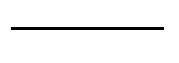}  = \frac{1}{2\om_k}.
\]
\item
A diagram with $n$ lines going to the initial state
and $m$ lines going to the final state contains the factor
\[
\eql{matrixelem}
\bra{f} \phi^{(-)}_{k_{n+m}} \cdots \phi^{(-)}_{k_{n+1}}
\phi^{(+)}_{k_n} \cdots \phi^{(+)}_{k_1} \ket{i}.
\]
\item
Each vertex is associated with an energy denominator, given by
\[
\frac{1}{E_{f\al} + \ii\ep} = \frac{1}{E_f - E_\al + \ii \ep},
\]
where $E_f$ is the energy of the final state,
and $E_\al$ is the energy of the state directly to the right
of the vertex.  
The energy denominator associated with the
rightmost vertex (which would give a factor of $1/E_{fi}$) is 
omitted (see \Eq{Tdefn}).
The initial and final states in general contain particles that do not
participate in the interaction, but these do not contribute to the
energy differences.
\item
Multiply by the symmetry factor
\[
S = \left( \frac{1}{4!} \right)^{\!\! n_4}
\left( \frac{1}{2} \right)^{\!\! n_2} \! C,
\]
where $n_4$ ($n_2$) is the number of $\phi^4$ ($\phi^2$) vertices, and
$C$ is the number of Wick contractions that give the same diagram.
To count the contractions, the initial and final state particles should be treated
as identical, but the initial
state particles can be distinguished from final state particles.
In terms of operators,
$C$ is the coefficient of the operator that appears
in front of the matrix element of the form in \Eq{matrixelem} when using Wick's theorem.%\hspace{-3pt}
\footnote{
This is the product of the coefficient of the normal-ordered operator
in Wick's theorem
and the coefficient of the operator in \Eq{matrixelem} in the normal
ordered operator.
For example,
\[
\no{\phi^4}
= \big( \phi^{(+)} \big)^4
+ 4 \phi^{(-)} \big( \phi^{(+)} \big)^3 
+ 6 \big( \phi^{(-)} \big)^2 \big( \phi^{(+)} \big)^2
+ 4 \big( \phi^{(-)} \big)^3 \phi^{(+)}
+ \big( \phi^{(-)} \big)^4.
\]}
\end{itemize}

We give some examples to illustrate these rules: 
\begin{subequations}
\eql{diagramexs}
\[
\eql{diag1}
\includegraphics[valign=c,scale=0.65]{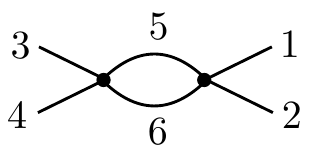}
&= \frac 18 \left( \frac{\la}{2\pi R} \right)^{\!\! 2}
\sum_{1, \ldots, 6} \de_{12,56} \de_{34,56}
\bra{f} \phi_4^{(-)} \phi_3^{(-)} \phi_2^{(+)} \phi_1^{(+)} \ket{i}
\nn
&\qquad\qquad\qquad\qquad{} \times
\frac{1}{2\om_5} \frac{1}{2\om_6}
\frac{1}{\om_3 + \om_4 - \om_5 - \om_6 + \ii\ep},
\\[10pt]
\eql{diag2}
\includegraphics[valign=c,scale=0.65]{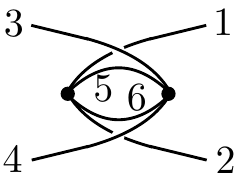}
&= \frac 18 \left( \frac{\la}{2\pi R} \right)^{\!\! 2}
\sum_{1, \ldots, 6} \de_{12,56} \de_{34,56}
\bra{f} \phi_4^{(-)} \phi_3^{(-)} \phi_2^{(+)} \phi_1^{(+)} \ket{i}
\nn
&\qquad\qquad\qquad\qquad{} \times
\frac{1}{2\om_5} \frac{1}{2\om_6}
\frac{1}{ - \om_1 - \om_2 - \om_5 - \om_6 + \ii\ep},
\\[10pt]
\eql{diag3}
\includegraphics[valign=c,scale=0.65]{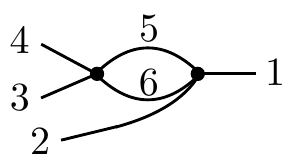}
&= \frac 14 \left( \frac{\la}{2\pi R} \right)^{\!\! 2}
\sum_{1, \ldots, 6}
\de_{1,256} \de_{34,56}
\bra{f} \phi_4^{(-)} \phi_3^{(-)} \phi_2^{(-)} \phi_1^{(+)} \ket{i}
\nn
&\qquad\qquad\qquad\qquad{} \times
\frac{1}{2\om_5} \frac{1}{2\om_6}
\frac{1}{ \om_3 + \om_4 - \om_5-\om_6 + \ii\ep}.
\]
Here the indices $5, 6$ label the internal lines, and we use the shorthand $\delta_{12,34} = \de_{k_1 + k_2, k_3 + k_4}$.
Note the difference between the energy denominators in \Eqs{diag1}
and \eq{diag2}, and the difference in the symmetry factor in \Eq{diag3}
compared to the previous two diagrams.
We also have disconnected diagrams such as
\[
\eql{diag4}
\includegraphics[valign=c,scale=0.5]{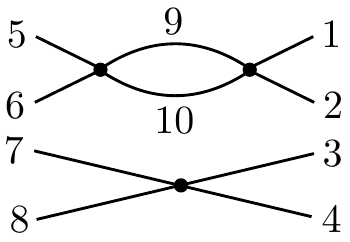}
&= \frac{1}{32} \left( \frac{\la}{2\pi R} \right)^3
\sum_{1, \ldots, 1\!\gap0} \de_{12,9 \gap1\!\gap0} \de_{9\gap 1\!\gap0,56} \de_{34,78}
\bra{f} \phi^{(-)}_8 \phi^{(-)}_7 \phi^{(-)}_6 \phi^{(-)}_5
\phi^{(+)}_4 \phi^{(+)}_3 \phi^{(+)}_2 \phi^{(+)}_1 \ket{i}
\nonumber\\[-5pt]
&\qquad\qquad\qquad\qquad{} \times
\frac{1}{2\om_{9}} \frac{1}{2\om_{1\!\gap0}}
\frac{1}{\om_5 + \om_6 - \om_9 - \om_{1\!\gap0} + \ii\ep}
\nn
&\qquad\qquad\qquad\qquad{} \times
\frac{1}{\om_5 + \om_6 + \om_7 + \om_8 - \om_3 - \om_4 - 
\om_9 - \om_{1\!\gap0} + \ii\ep}.  
\]
\eql{ExampleDiags}%
\end{subequations}

The diagrammatic rules given above are for the fundamental theory.
For the effective theory, the only difference is that the sums over
intermediate states are restricted to the low-energy subspace.
The intermediate states are associated with cuts between the vertices
of the diagram.
For each cut, we must include a step function that enforces the constraint
that the total energy is below $E_\text{max}$.
Note that unlike the energy differences in the denominators, this constraint
depends on the particles in the initial and final states that do not interact. 
For example, the step functions for the diagrams in \Eqs{diagramexs} 
evaluated for the effective theory are given by 
\begin{subequations}
\[
\includegraphics[valign=c,scale=0.65]{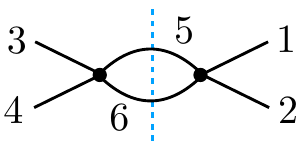} 
&:\ \Th(E_\text{max} - E_f + \om_3 + \om_4 - \om_5 - \om_6)
\\[5pt]
\includegraphics[valign=c,scale=0.65]{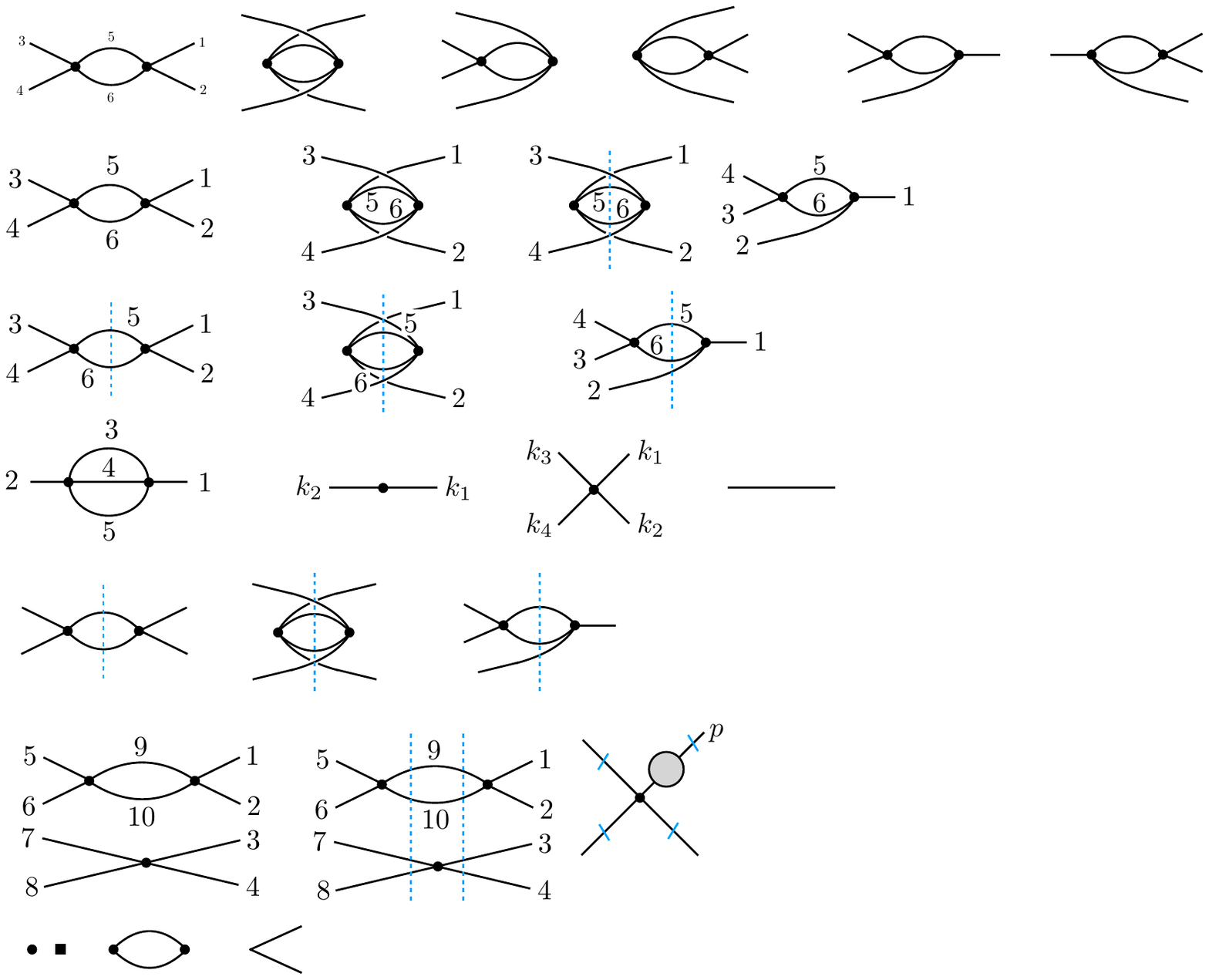} 
&:\ \Th(E_\text{max} - E_f  - \om_1 - \om_2 - \om_5 - \om_6)
\\
\includegraphics[valign=c,scale=0.65]{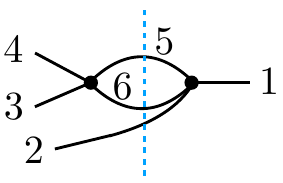} 
&:\ \Th(E_\text{max} - E_f + \om_3 + \om_4 - \om_5 - \om_6)
\\[5pt]
\includegraphics[valign=c,scale=0.65]{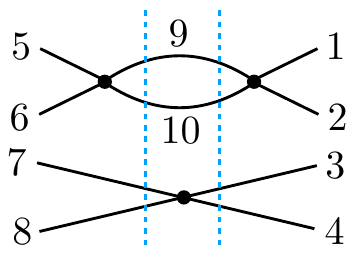} 
&:\ \Th(E_\text{max} - E_f + \om_5 + \om_6 - \om_9 - \om_{1\!\gap0})
\nonumber\\[-18pt]
&\qquad{}
\times \Th(E_\text{max} - E_f + \om_5 + \om_6 + \om_7 + \om_8 - \om_3 - \om_4
-\om_9 - \om_{1\!\gap0}).
\]
\eql{ThetaFnsEFT}%
\end{subequations}%
Recall that $E_f$ is the total energy of the final state, including 
the energy of particles that do not contract with
any vertex (which are not drawn in the diagrams).
The fact that diagrams depend on the energies of particles that do not
participate in the interaction 
is a manifestation of the non-locality of the effective Hamiltonian.

\section{Renormalization of 2D $\la\phi^4$ Theory}
\scl{Renormalization}
In this section, we discuss the renormalization of 2D $\la\phi^4$ theory.
In this theory, the coupling $\la$ has dimensions of mass-squared, see~\Eq{dimensionalAnalysis}.
The theory is therefore super-renormalizable, and in fact all UV
divergences can be eliminated by normal-ordering.
We will however also 
consider more general regulators and renormalization  schemes.
We do this for several reasons.
First, we will see below that separation of scales is manifest only in
a more general renormalization scheme for the fundamental theory.
Second, we wish to emphasize that renormalization of the fundamental theory
can be carried out independently of the Hamiltonian truncation.
This point will be important for theories with genuine UV divergences,
such as the 3D $\la\phi^4$ theory considered in Appendix A.

\subsection{Regularization}
Matching the results of the renormalized fundamental theory onto
the effective Hamiltonian requires considering the theory in finite volume.
We compactify the spatial direction on a circle, which breaks Lorentz invariance
and implies that particles carry discrete spatial momenta.
We  regularize the theory  using a hard momentum cutoff on the 
spatial momenta:
\[
\eql{fundcut}
\sum_k \quad\to\quad \sum_{k \, \le \, \La R},
\]
where $\La \gg E_\text{max}$;
we will eventually take $\La \to \infty$.
We choose this cutoff because makes it straightforward to compute
the quantity $T$ that we use to match to the effective theory.
The fact that this cutoff breaks Lorentz invariance does not cause
any significant complication for the simple 2D $\la \phi^4$ model studied here, as we will see below.%\hspace{-3pt}
\footnote{For more complicated models, it may be worthwhile to develop
the technology for using Lorentz invariant cutoffs (such as dimensional
regularization) to compute $T$ in finite volume.}

\subsection{UV Divergences}
Unlike the energy cutoff $E_\text{max}$ that we impose on the effective Hamiltonian,
the cutoff $\La$ on the fundamental theory is a local Wilsonian cutoff.
Therefore, the possible UV divergences can be classified by writing the possible local
counterterms, using dimensional analysis to determine the dependence
on the cutoff $\La$.
In 2D $\la\phi^4$ theory, the field $\phi$ is dimensionless.
However, at $O(\la^n)$ in perturbation theory, loops can generate
counterterms with at most $2n$ external $\phi$ lines.
Because the cutoff breaks Lorentz invariance, we have to allow for the
possibility of Lorentz violating counterterms.
The leading counterterms have the schematic form 
\[
\De\scr{L} \sim \myint \mathrm{d}^2 x\ggap \Bigl(
\la \ln\La
+ \la \ln\La \phi^2
+ \frac{\la^2}{\La^2} \big(\phi^2 + \phi^4\big)
+ \frac{\la^2}{\La^4} \bigl[ (\d_t \phi)^2 + \cdots \bigr]
+ \cdots \Bigr) .
\]
We see that only the vacuum energy and $\phi^2$ mass term receive UV divergent contributions.
The divergent vacuum energy in 
this theory means that only energy differences 
are physically meaningful.

We see that renormalizing the theory only requires introducing a bare mass parameter.  
Therefore, we write the bare Lagrangian as
\[
\scr{L} = \frac 12 (\d\phi)^2 - \frac 12 m_0^2\gap \phi^2
- \frac{\la}{4!} \phi^4.
\eql{bareL}
\]

\subsection{Renormalization}
\scl{mQ}
The fundamental and the effective theory must be defined 
on the same Fock space in order to carry out the matching
described in \sec{match}.
The unperturbed Hamiltonian for both theories is therefore given by
\Eq{quantizationmassdefn}, where $m_\text{Q}$ is a finite quantization mass 
that can be treated as a variational parameter to improve convergence
\cite{Rychkov:2015vap}.
The interaction term is then given by \Eq{mVdefn} with
the coefficient of $\no{\phi^2}$ given by
\[
\eql{mV2R}
m_V^2 = m_0^2 - m_\text{Q}^2 
+ \frac{\la}{8\pi R} \sum_{|k| \, \le \, \La R} \frac{1}{\om_k}.
\]
Here $m_V^2$ is a finite, renormalized quantity.%\hspace{-5pt}
\footnote{That is, the log UV divergence in the sum
\[
\sum_{|k| \, \le \, \La R} \frac{1}{\om_k}
\sim \int^\La \frac{\mathrm{d} k}{\om_k} \sim \ln\La
\]
is canceled by allowing the bare mass $m_0^2$ to depend on the cutoff.}
This definition of the renormalized mass is convenient for calculations,
since $m_V^2$ is the mass vertex that appears in our diagrammatic rules
(see \Eqs{mVdefn} and \eq{vertexrules}).
However, we will see that separation of scales is manifest only in a 
different renormalization scheme where the couplings are renormalized at a renormalization scale
$\mu \sim E_\text{max}$.

It is therefore also useful to define a renormalized mass $m_\text{R}^2$
that depends on an arbitrary renormalization scale $\mu$:
\[
\eql{renormmass}
m_\text{R}^2(\mu) =  m_0^2 
+ \frac{\la}{8\pi R} \sum_{\mu R \, < \, |k| \, \le \, \La R} \frac{1}{\om_k}.
\]
This is the standard definition of the renormalized mass parameter,
in which the contribution of modes with $k > \mu R$ have been absorbed
into the renormalized coupling.
We will see below that separation of scales is manifest in terms of
$m_\text{R}^2(\mu \sim E_\text{max})$.
The mass parameter $m_V^2$ is related to $m_\text{R}^2$ by 
\[
\eql{mv2renorm}
m_V^2 = m_\text{R}^2(\mu) - m_\text{Q}^2 +\frac{\la}{8\pi R} \sum_{|k| \leq \mu R} \frac{1}{\om_k} .
\]

Previous work on Hamiltonian truncation of this model works directly with the
normal-ordered Hamiltonian
in which case there are no UV divergences, and hence no renormalization is needed.
In our approach this corresponds to the choice $m_V^2 = 0$,
so that $V$ consists only of a normal-ordered $\no{\phi^4}$ term.
We then define the `normal ordered mass' by
\[
m_\text{NO} = m_\text{R}(\mu = 0).
\]
Using \Eq{mV2R} and \Eq{renormmass}, we see that this implies $m_\text{Q} = m_\text{NO}$
 since $m_V^2 = 0$. 
(To eliminate the $k = 0$ mode from the sum in \Eq{mv2renorm}, 
$\mu$ must actually be taken to be slightly negative.)
The normal ordered mass 
$m_\text{NO}$ is a convenient renormalization group
invariant parameter we can use to define the theory.

\section{Matching in 2D $\la\phi^4$ Theory}
\scl{Matching2DTheory}
Now that we have renormalized the UV theory, we can turn to applying the general
formalism given in~\Eq{Hmatch} to derive the effective theory matching corrections.  
In this section, we will  compute the leading terms in the low-energy expansion
of the effective Hamiltonian at
$O(V^2)$ in 2D $\la \phi^4$ theory. 
In this approximation, the effective Hamiltonian is local.
In \sec{PowerCounting} we will explain the power counting in powers of 
$1/E_\text{max}$
for this theory, and we will argue that these calculations give the leading 
$1/E_\text{max}^2$ corrections to the effective Hamiltonian.
The resulting effective Hamiltonian is therefore expected to have
errors of order $1/E_\text{max}^3$, which is confirmed by our
numerical results in \sec{Numerics}.

The calculations are a straightforward application of the diagrammatic rules
presented in \sec{diagrams} above,
but they illustrate some non-trivial features of the effective theory matching.
In particular, matching at $O(V^2)$ involves 2- and 3-loop diagrams with overlapping UV/IR sensitivity.
We will show that the IR sensitivity of the corrections 
is canceled by other contributions as required by separation 
of scales in the effective theory,
but this cancellation requires both the correct renormalization
prescription and definition of the operators in the effective Hamiltonian.

\subsection{Matching at $O(V)$: Operator Approach}
\scl{matchVop}
We begin at $O(V)$; the matching condition is simply given by (see \Eq{H1effmatch}) 
\[
\eql{H1matchagain}
\bra{f} H_1 \ket{i}_\text{eff} = \bra{f} V \ket{i}.
\]
We must be careful in interpreting this expression, since the \lhs\ is evaluated
in the effective theory with a truncated Hilbert space.
The correct way to match is to equate the coefficients of normal
ordered operators in the fundamental and effective theories.
This approach agrees with \Eq{H1matchagain}, 
because at this order the states above $E_\text{max}$ only impact 
the counterterm and normal ordering constants.
The normal-ordered full theory potential is given by \Eq{mVdefn},
with $m_V^2$ (the coefficient of $\no{\phi^2}$) given by \Eq{mV2R}.

The $O(V)$ correction to the effective theory has the form%\hspace{-3pt}
\footnote{Here we use the expressions 
\[
\eql{nodef}
\phi^2 = \no{\phi^2} + Z, \qquad{} \phi^4 = \no{\phi^4} + 6 Z \no{\phi^2} + 3Z^2,\]
 with $Z = \frac{1}{4\pi R} \sum_k \frac{1}{\om_k}$ for the 2D $\la \phi^4$ theory. }
\[
\eql{H1form}
H_1 &= \myint \mathrm{d}x\ggap \bigg[
\frac 12\gap m_1^2\gap \phi^2 + \frac{\la_1}{4!} \phi^4 \bigg]
\nn[5pt]
&= \myint \mathrm{d}x\ggap \bigg[
\frac 12\gap m_{V1}^2 \no{\phi^2} + \frac{\la_1}{4!} \no{\phi^4} \bigg]{}+ \text{constant},
\]
with
\[
m_{V1}^2 = m_1^2 + \frac{\la_1}{8\pi R} \sum_{|k| \, \le \, k_\text{max}}
\frac{1}{\om_k},
\]
where
\[
\eql{kmax}
k_\text{max} = R \sqrt{E_\text{max}^2 - m_\text{Q}^2} \simeq E_\text{max} R.
\]
Note that the normal-ordering in \Eq{H1form} is performed on the truncated Hilbert space.
We drop normal-ordering constants proportional to the identity operator,
since they do not contribute to differences of energy eigenvalues. 
(We will, however, need these contributions when we discuss the separation of scales
for the vacuum diagrams in \sec{MatchSep0Legs}.)

Matching the coefficients of $\no{\phi^4}$ and $\no{\phi^2}$ then gives 
simply
\[
\eql{H1match}
\la_1 = \la,
\qquad
m_{V1}^2 = m_V^2.
\]
Note that $m^2_{V1}$ is finite, as all couplings in
the effective theory must be, since we are matching onto a renormalized
fundamental theory.

We now discuss the separation of scales at this order.
Separation of scales should not apply to the coefficients of the
normal ordered operators, since normal ordering
depends on the split of the full Hamiltonian $H$ into `free' and
`interacting' terms.
Instead, we expect that separation of scales will hold for the
coefficients of non-normal-ordered operators.
Furthermore, effective field theory methodology tells us that 
separation of scales should be manifest in terms of couplings
renormalized at the matching scale, in this case $\mu \sim E_\text{max}$.
To see that these expectations are satisfied at $O(V)$, note that
the coefficient of $\phi^2$ in the effective theory at this order is given by
\[
m_\text{Q}^2 + m_1^2 = m_\text{R}^2(\mu = k_\text{max}/R).
\]
We will see a much more nontrivial check of separation of scales when we 
consider 2- and 3-loop diagrams in \sec{sepScales} and \sec{MatchSep0Legs}
below.

\subsection{Matching at $O(V)$: Diagrammatic Approach\label{sec:H1matchdiagram}}
Although the operator approach to the matching at $O(V)$ is very simple,
it becomes more cumbersome at higher orders.
This motivates us to re-derive it in terms of diagrams for illustration, since we 
will rely on the diagrammatic approach to compute the $O(V^2)$ corrections below.
Recall that we defined the diagrammatic expansion so that the
vertices are given by the coefficients of the normal-ordered operators; 
contractions (as defined using Wick's theorem) between the same vertex are omitted,
see \S\ref{sec:diagrams}.
The $O(V)$ contributions to $\bra{f} T \ket{i}$ in the fundamental theory
include the 4-point tree diagrams
\[
\includegraphics[valign=c,scale=0.65]{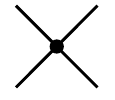}
+ \includegraphics[valign=c,scale=0.65]{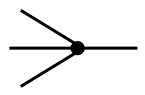}
+ \includegraphics[valign=c,scale=0.65]{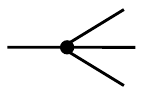}
+ \reflectbox{\includegraphics[valign=c,scale=0.65]{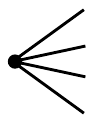}}
+ \includegraphics[valign=c,scale=0.65]{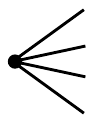}
= \frac{\la}{4!} \myint \mathrm{d}x \ggap
\bra{f} \no{\phi^4} \ket{i}.
\]
For contact interaction diagrams such as these,
the different ways of contracting the external lines to initial
and final states add together to build up the full contribution of the local operator.
In addition, there are diagrams given by the 2-point vertex 
(see \Eq{vertexrules}).
Recalling that the 2-point vertex is given by the coefficient of
$\no{\phi^2}$, we have for the fundamental theory 
\[
\includegraphics[valign=c,scale=0.65]{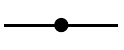}
+ \includegraphics[valign=c,scale=0.65]{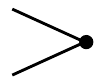}
+ \includegraphics[valign=c,scale=0.65]{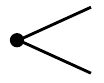}
= \frac 12\gap m_V^2 
\myint \mathrm{d}x \ggap \bra{f} \no{\phi^2} \ket{i}.
\]
In the effective theory, we have the same diagrams, 
with $\la \to \la_1$, and $m_V^2 \to m_{V1}^2$.
This reproduces \Eq{H1match}, as it must.

\subsection{Matching at $O(V^2)$: 4 Legs}
\scl{Match4legs}
We now consider the matching at $O(V^2)$ using the diagrammatic approach.
We begin with diagrams that have 4 external legs, because 
they are simpler to evaluate than the 2-point corrections.
The diagrams that contribute are given by
\[
\eql{T24diagrams}
\bra{f} T_{2,4} \ket{i}
&= \includegraphics[valign=c,scale=0.65]{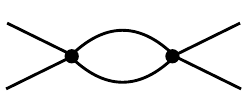}
+ \includegraphics[valign=c,scale=0.65]{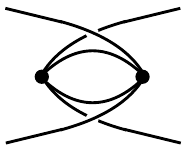}
+ \includegraphics[valign=c,scale=0.85]{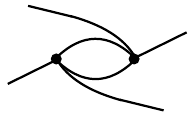}
\nn
&\qquad{}
+ \includegraphics[valign=c,scale=0.65]{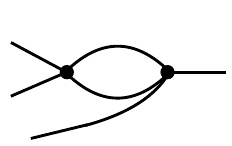}
+ \includegraphics[valign=c,scale=0.65]{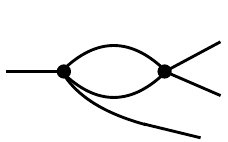}
+ \includegraphics[valign=c,scale=0.65]{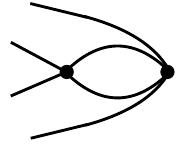}
+ \includegraphics[valign=c,scale=0.65]{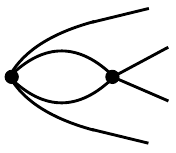} \, .
\]
Note that there are no diagrams with $\phi^2$ vertices at this order.

To match, we equate the diagrams in \Eq{T24diagrams} in the fundamental and
effective theories.
This means that the effective Hamiltonian depends on the difference
between the diagrams computed using the two descriptions of the theory.
For the first diagram on the \rhs\ of \Eq{T24diagrams},
the difference is given by (see \Eq{diag1})
\[
\!\!\!
\includegraphics[valign=c,scale=0.65]{Figs/V221_num_f}
- \biggl[ \includegraphics[valign=c,scale=0.65]{Figs/V221_num_f} \biggr]_\text{eff}
&= \frac{\la^2}{128\pi^2 R^2}  \sum_{1,\ldots, 4} \de_{12,34}
\bra{f} \phi_4^{(-)} \phi_3^{(-)} \phi_2^{(+)} \phi_1^{(+)} \ket{i}
\nn[4pt]
&\hspace{15pt} \times
\sum_{5,6} \de_{56,34} \frac{\Th(\om_5 + \om_6 - \om_3 - \om_4 + E_f - E_\text{max})}
{\om_5 \om_6 (\om_3 + \om_4 - \om_5 - \om_6)}.
\eql{diag1cont}
\]
We are omitting the $\ii\ep$ factors, which we showed do not affect the
matching in \sec{MatchingOntoHEff}.
Also, note that we are no longer explicitly including the UV
cutoff $\La$ in the sums over momenta.
Effective field theory methodology tells us that the effective Hamiltonian
should be determined by matching low-energy observables and expanding
in powers of IR scales \cite{Georgi:1993mps}.
In our case, the IR scales are given by
\[
\eql{localapprox}
m_\text{Q} \lsim \om_{1,2,3,4} \lsim E_{i,f} \ll E_\text{max}.
\]
The step function in \Eq{diag1cont} then implies that 
\[
\om_{5,6} \gsim E_\text{max}.
\]
That is, the matching is only sensitive to intermediate states whose energies are above the 
cutoff of the truncated theory.
This is a manifestation of the separation of scales in effective field theory.
The full theory diagram has a complicated dependence on the external
energies and momenta, but because we are matching matrix elements of 
low-lying states ($E_{i,f} \ll E_\text{max}$), 
this dependence can be expanded in a power series.
The first term that results from taking this expansion can be determined 
by setting all of the external momenta and energies to zero.
In this approximation, we have 
\[
\!\!\!
\includegraphics[valign=c,scale=0.65]{Figs/V221_num_f}
- \biggl[ \includegraphics[valign=c,scale=0.65]{Figs/V221_num_f} \biggr]_\text{eff}
&\simeq -\frac{\la^2}{128\pi R} 
\myint \mathrm{d}x\ggap \bra{f} \bigl[ \phi^{(-)} \bigr]^2 \bigl[ \phi^{(+)} \bigr]^2 \ket{i}
\nn[4pt]
&\hspace{23pt} \times
\sum_{k} \frac{\Th(2\om_k - E_\text{max})}
{\om_k^3}
\Bigl[ 1 + O(E_{i,f}/E_\text{max}) \Bigr],
\eql{diag1cont2}
\]
where we used
\[
\sum_{1,\ldots, 4} \de_{12,34} 
\bra{f} \phi_4^{(-)} \phi_3^{(-)} \phi_2^{(+)} \phi_1^{(+)} \ket{i}
= 2\pi R
\myint \mathrm{d}x\ggap \bra{f} \bigl[ \phi^{(-)} \bigr]^2 \bigl[ \phi^{(+)} \bigr]^2 \ket{i}.
\]
We see that in this approximation, the matrix element of the correction
is a local operator.
We therefore refer to this as the `local approximation.'

The other diagrams in \Eq{T24diagrams} differ from the one considered in \Eq{diag1cont2}
by its initial and final state contractions,
and its energy denominators, and (for the effective theory) the dependence of the 
$\Th$ functions that implement the cutoff.
But in the leading approximation discussed above, the latter two
effects disappear.
The diagrams therefore differ only by the matrix elements of fields
associated with the external lines.
Summing over all diagrams with all possible choices for the external lines
builds up the normal-ordered operator, as we found for the diagrammatic
matching at $O(V)$ in \S\ref{sec:H1matchdiagram}.
Putting it all together, we find
\[
\bra{f} T_{2,4} \ket{i}
- \bra{f} T_{2,4} \ket{i}_\text{eff}
&\simeq -\frac{\la^2}{128\pi R} \sum_{k} \frac{\Th(2\om_k - E_\text{max})}
{\om_k^3}
\myint \mathrm{d}x \ggap\bra{f} \no{\phi^4} \ket{i}.
\]
This gives a contribution to the effective Hamiltonian
\[
\eql{H24}
H_{2,4} \simeq \frac{\la_2}{4!}  \myint \mathrm{d}x
\ggap \no{\phi^4},
\]
where  
\[
\eql{lambda2}
\la_2 = -\frac{3 \la^2}{16\pi R}
\sum_{k} \frac{\Th(2\om_k - E_\text{max})}{\om_k^3}.
\]
Note that $\la_2$
depends only on the properties of states above the
cutoff, and therefore obeys the principle of separation of scales.
The sum is UV convergent, so it is dominated by the contribution
of states near the cutoff.

For illustration, we compute the sum in \Eq{lambda2}
in the large $E_\text{max}$ limit
by approximating it as an integral:
\[
\eql{sumEM}
\la_2 \simeq -\frac{3 \la^2}{16\pi R}
\times 2 \int_{\frac 12 E_\text{max}R}^{\infty} 
\frac{\mathrm{d}k}{\om_k^3} = -\frac{3\la^2}{4 \pi E_\text{max}^2} 
\Bigl[ 1 + O\big(m^2/E_\text{max}^2\big) \Bigr].
\]
Note that this has the form we expect from general power counting arguments
(see \Eq{effpwrcount}).
We can systematically correct the approximation \Eq{sumEM} 
using the Euler-MacLaurin summation formula \Eq{EM} to include
higher orders in the IR scales $R^{-1}/E_\text{max}$ and $m_\text{Q}/E_\text{max}$.
To obtain numerical results in \sec{Numerics}, we will directly evaluate 
sums such as \Eq{lambda2} numerically.

\subsection{Matching at $O(V^2)$: 2 Legs}
\scl{Match2Legs}
Now we consider the matching contribution at $O(V^2)$ from
diagrams with 2 external legs:
\[
\bra{f}T_{2,2}\ket{i} =\,\,
&\includegraphics[valign=c,scale=0.65]{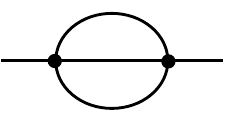} 
+ \includegraphics[valign=c,scale=0.65]{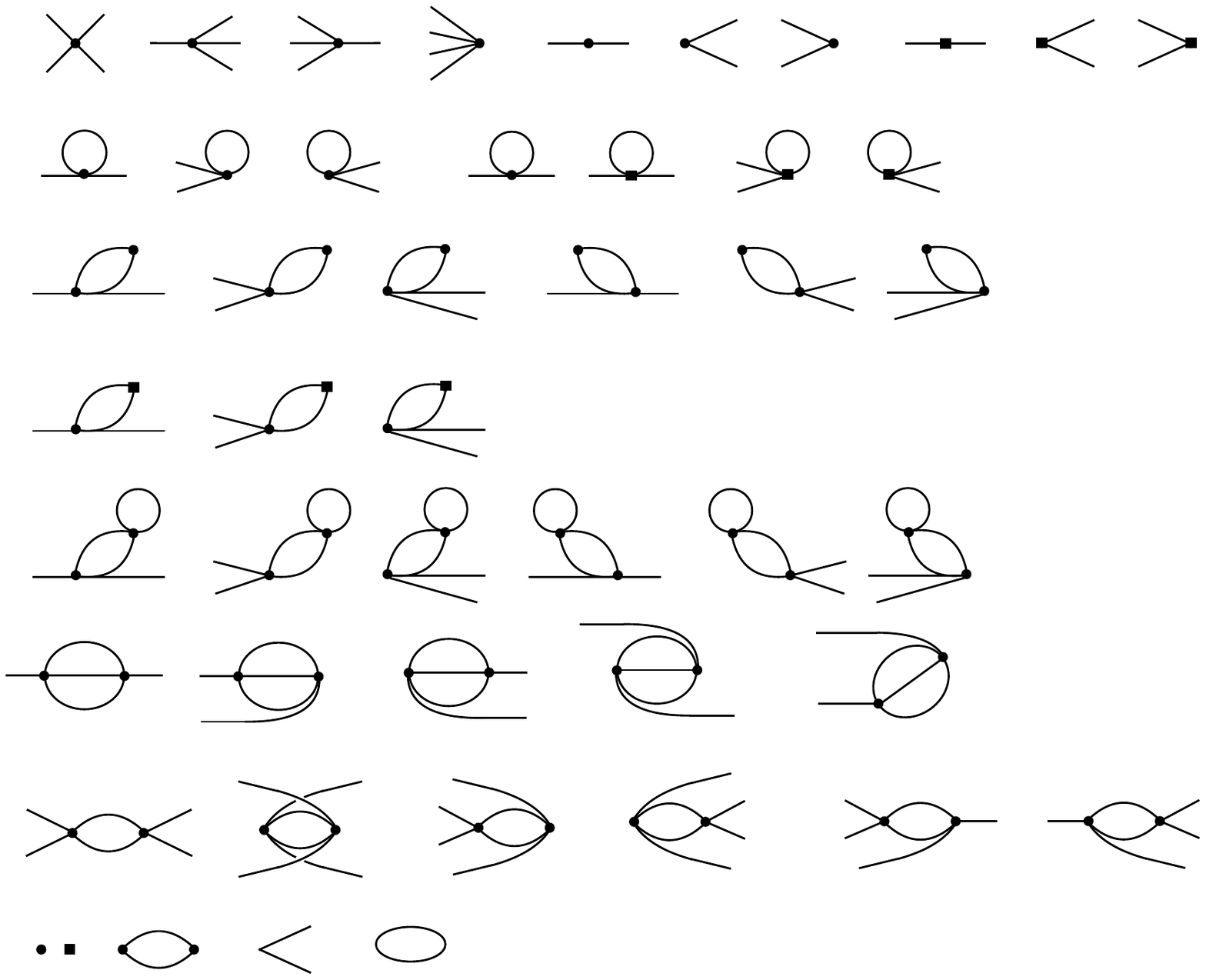}
+ \includegraphics[valign=c,scale=0.65]{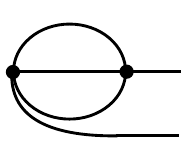}
+ \includegraphics[valign=c,scale=0.65]{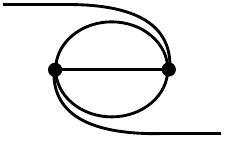}
\nonumber\\[3pt]
&\qquad{}
+ \includegraphics[valign=c,scale=0.65]{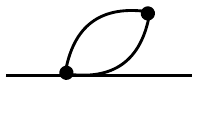}
+ \includegraphics[valign=c,scale=0.65]{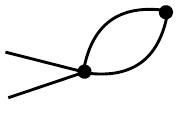}
+ \includegraphics[valign=c,scale=0.65]{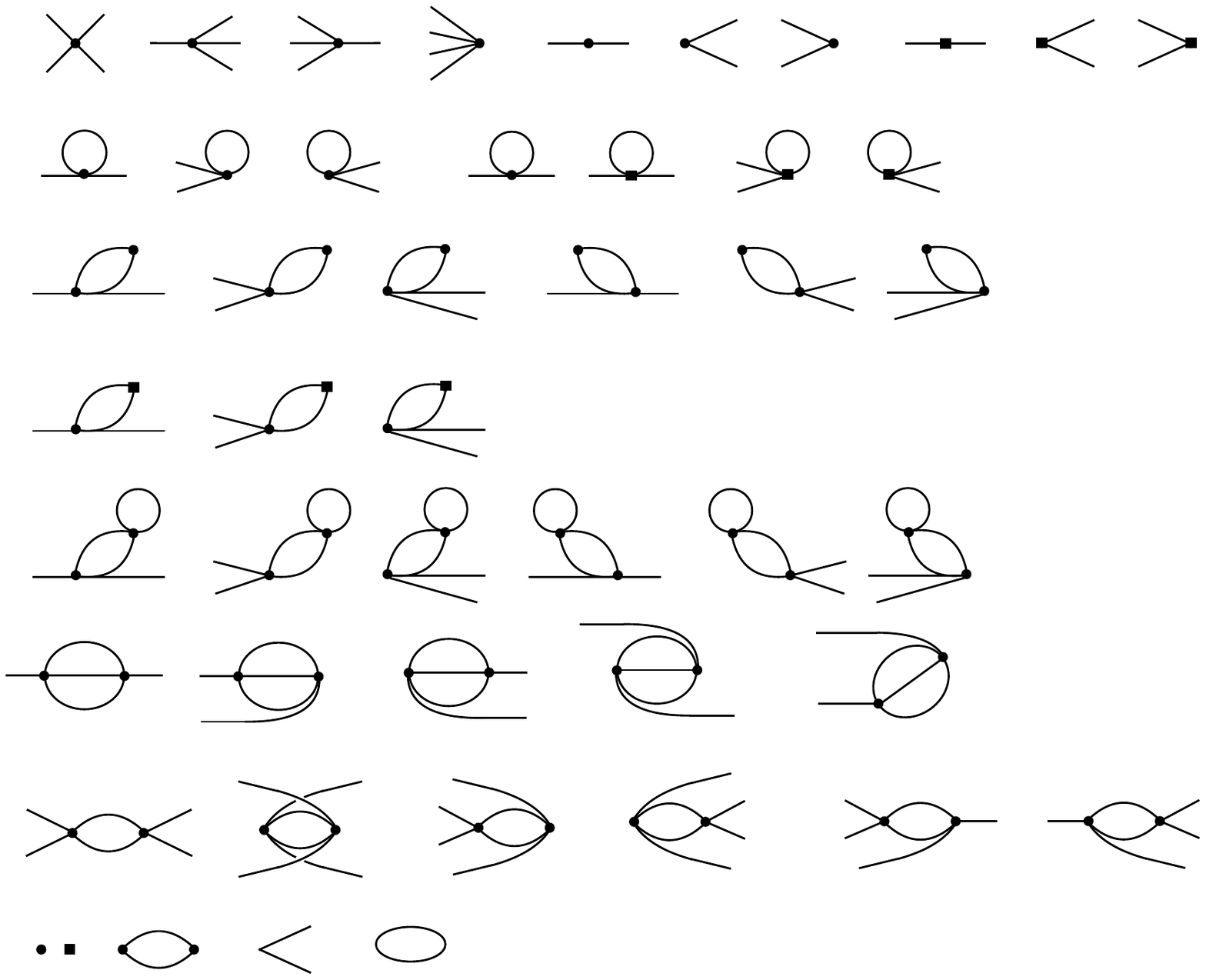}
+ \reflectbox{\includegraphics[valign=c,scale=0.65]{Figs/snail_f}}
+ \reflectbox{\includegraphics[valign=c,scale=0.65]{Figs/snail2_f}}
+ \reflectbox{\includegraphics[valign=c,scale=0.65]{Figs/snail3_f}}
 \,\, .
\eql{V2diagrams}
\]
Let us begin with the first 2-loop diagram shown in \Eq{V2diagrams}.
As we discussed in the previous section, 
matching is determined by the difference between the diagrams
in the fundamental and the effective theory:
\[
\hspace{-5pt}
\includegraphics[valign=c,scale=0.65]{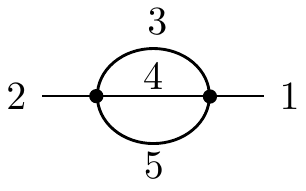}
- \biggl[ \includegraphics[valign=c,scale=0.65]{Figs/sunset_num_f}
\biggr]_\text{eff}
&= \frac 16 \left( \frac{\la}{2\pi R} \right)^2
\sum_{1,\ldots, 5} \de_{1,2} \gap \de_{1,345}
\bra{f} \phi_2^{(-)} \phi_1^{(+)} \ket{i}
\nn[4pt]
&\hspace{72pt} \times
\frac{1}{2\om_3} \frac{1}{2\om_4} \frac{1}{2\om_5}
\frac{\Th(\om_3 + \om_4 + \om_5 - E_\text{max})}
{\om_2 - \om_3 - \om_4 - \om_5}.
\]
Note that, unlike the 1-loop diagrams analyzed in the previous subsection, 
the sum over the intermediate momenta $k_{3,4,5}$ includes mixed UV/IR 
regions where 
some of the momenta are large, while others are small.
Therefore, the separation of UV and IR scales is not manifest
diagram by diagram at this order.%\hspace{-3pt}
\footnote{In the context of renormalization theory, this is the
problem of overlapping divergences.}
However, we will show in \sec{sepScales} that this important 
property manifests when we add all the diagrams together
and use the correct renormalization prescription and
definition of operators.
This justifies our use of the approximation 
neglecting the external energies and momenta since they are much smaller than
the intermediate momenta we are integrating out.
As above, we refer to this simplifying choice as the `local approximation.'

For the diagram above, the local approximation gives
\[
\includegraphics[valign=c,scale=0.65]{Figs/sunset_num_f}
- \biggl[ \includegraphics[valign=c,scale=0.65]{Figs/sunset_num_f} \biggr]_\text{eff}
&\simeq \frac{\la^2}{192\pi^2 R^2}
\myint \mathrm{d}x\ggap \bra{f} \phi_2^{(-)} \phi_1^{(+)} \ket{i}
\nn
&\hspace{16pt} \times
\sum_{3,4,5} \de_{345,0} 
\frac{\Th(\om_3 + \om_4 + \om_5 - E_\text{max})}
{\om_3 \om_4 \om_5 (- \om_3 - \om_4 - \om_5)} .
\]
Note that the dependence on the external momenta factors out,
implying that the matrix element is local.
The first 4 diagrams in \Eq{V2diagrams} are the same up
to the factors that depend on the initial and final states, and
combining them gives
\[
\includegraphics[valign=c,scale=0.65]{Figs/sunset_f} 
&+ \includegraphics[valign=c,scale=0.65]{Figs/sunset2_f}
+ \includegraphics[valign=c,scale=0.65]{Figs/sunset3_f}
+ \includegraphics[valign=c,scale=0.65]{Figs/sunset4_f}
- \biggl[\includegraphics[valign=c,scale=0.65]{Figs/sunset_f} 
+ \includegraphics[valign=c,scale=0.65]{Figs/sunset2_f}
+ \includegraphics[valign=c,scale=0.65]{Figs/sunset3_f}
+ \includegraphics[valign=c,scale=0.65]{Figs/sunset4_f}\, \biggr]_\text{eff}
\nn[8pt]
&\simeq \frac{\la^2}{192\pi^2 R^2} 
\myint \mathrm{d}x \ggap \bra{f} \no{\phi^2} \ket{i} \times
\sum_{3,4,5} \de_{345,0} 
\frac{\Th(\om_3 + \om_4 + \om_5 - E_\text{max})}
{\om_3 \om_4 \om_5 (- \om_3 - \om_4 - \om_5)}.
\eql{sunsetsum}
\]

Finally, we compute the remaining 6 diagrams in \Eq{V2diagrams}
in the local approximation: 
\[
\includegraphics[valign=c,scale=0.65]{Figs/snail_f}
&+ \includegraphics[valign=c,scale=0.65]{Figs/snail2_f}
+ \cdots 
- \biggl[ \includegraphics[valign=c,scale=0.65]{Figs/snail_f}
+ \includegraphics[valign=c,scale=0.65]{Figs/snail2_f}
+ \cdots 
\biggr]_{\text{eff}}
\nn[5pt]
&
\simeq -\frac{\la}{32\pi R} \myint \mathrm{d}x\ggap
\bra{f} \no{\phi^2} \ket{i}
\Biggl[
m_V^2 \sum_k \frac{1}{\om_k^3}
- m_{V1}^2
\sum_{|k| \, \le \, k_\text{max}}
\frac{\Th(E_\text{max} - 2\om_k)}{\om_k^3}
\Biggr]
\nn[8pt]
&= -\frac{\la m_V^2}{32\pi R} \myint \mathrm{d}x\ggap \bra{f} \no{\phi^2} \ket{i}
\sum_k \frac{\Th(2\om_k - E_\text{max})}{\om_k^3}.
\]
As above,
combining the diagrams together yields the matrix element of the $\no{\phi^2}$ 
operator.
We have used the $O(V)$ matching condition
given in~\Eq{H1match} to derive the last line.
Combining this with \Eq{sunsetsum}, we obtain
(in the local approximation)
\[
\eql{H22NO}
H_{2,2} \simeq \frac 12 m^2_{V2} \myint \mathrm{d}x\ggap  \no{\phi^2},
\]
where
\[
\eql{Deltam22}
m^2_{V2} &= -\frac{\la}{16\pi R} \biggl[
\frac{\la}{6\pi R} 
\sum_{3,4,5} \de_{345,0} 
\frac{\Th(\om_3 + \om_4 + \om_5 - E_\text{max})}
{\om_3 \om_4 \om_5 (- \om_3 - \om_4 - \om_5)}
+ m_V^2
\sum_k \frac{\Th(2\om_k - E_\text{max})}{\om_k^3}
\biggr].
\]
These expressions are quite complicated, so for numerical studies
we numerically evaluate the sums in
the matching corrections, making sure that the sums have adequately converged.
The matching correction is dominated by the contribution of states near the
$E_\text{max}$ cutoff, so we believe it is important to treat 
these states accurately in the matching, as we have done here.

\subsection{Separation of Scales at $O(V^2)$: 2 Legs}
\scl{sepScales}
The result \Eq{Deltam22} does not appear to satisfy the separation
of scales principle in the effective theory, since both terms contain sums 
with arbitrarily small momenta (since $m_V^2$ contains an IR sum, 
see \Eq{mv2renorm}). 
The expectation based on effective field theory methodology is that
the cancelation of such overlapping UV/IR regions takes place only
if all diagrams are included, and if the renormalization scale is chosen
to be near the cutoff of the effective theory \cite{Georgi:1993mps}.
As already discussed in  \S\ref{sec:matchVop}, the normal ordered operators
have a renormalization scale near the IR, and so separation of scales
applies to the coefficients of non-normal ordered operators defined with
a renormalization scale $\mu \sim E_\text{max}$.
We therefore write
\[
H_2 = \myint \mathrm{d}x\ggap \biggl[
\frac 12\gap m^2_2\gap 
\phi^2
+ \frac{\la_2}{4!} \phi^4 \biggr],
\]
where (see \Eqs{H24}, \eq{H22NO} and \eq{nodef})
\[
\eql{Deltambar2}
m_2^2 = m_{V2}^2
- \frac{\la_2}{8\pi R} \sum_{|k| \, \le \, k_\text{max}}
\frac{1}{\om_k}.
\]
Our goal is to show that all the contributions to $m^2_2$ that
involve products of UV and IR dominated sums cancel when all the 
contributions are included.
For example, the sum in the first line of \Eq{Deltam22} includes a region
where one of $k_{1,2,3}$ is much smaller than the other two:
\[
\sum_{3,4,5} \de_{345,0} 
\frac{\Th(\om_3 + \om_4 + \om_5 - E_\text{max})}
{\om_3 \om_4 \om_5 (- \om_3 - \om_4 - \om_5 + \ii \ep)}
\simeq \, 3 \!\!\! \sum_{|k'| \, \ll \, k_\text{max}} \frac{1}{\om_{k'}}
\times \sum_k \frac{\Th(2\om_k - E_\text{max})}{\om_k^2 (-2\om_k)}
+ \cdots
\]
The factor of 3 comes from the 3 distinct regions where one of the $k_{1,2,3}$ is small.
We see that all of the mixed UV/IR contributions to $m_2^2$ 
have the same factorized form:
\[
\eql{2dsep}
m_2^2
\simeq -\frac{\la^2}{32\pi^2 R^2}
\biggl(\, \underbrace{\frac 12 + \frac 14 - \frac 34}_{{}  = \, 0}\, \biggr)
\sum_{|k'| \, \ll \, k_\text{max}} \frac{1}{\om_{k'}}
\times \sum_k \frac{\Th(2\om_k - E_\text{max})}{\om_k^3} + \cdots
\]
Here, the first two terms in the parentheses are 
contributions to $m_2^2$,
and the third comes from
un-normal-ordering $\phi^4$ (see \Eq{nodef}).
The fact that the mixed UV/IR terms cancel shows that the 2-loop correction to the 
2-point function does in fact manifest the separation of scales. 
Not only is this a critical check that our matching procedure is well defined
in the limit that the IR scales are taken to zero, it also justifies our
use of the local approximation 
in the matching calculation above.%
\footnote{%
It would be very interesting to check the separation
of scales beyond the local approximation.}

Note that the contribution from states below the cutoff does not
cancel exactly in the sums above.
Separation of scales requires only that the dominant contribution
to effective couplings
comes from states \emph{near} the cutoff, and this is what we have 
demonstrated above.

\subsection{Matching and Separation of Scales at $O(V^2)$: 0 Legs}
\scl{MatchSep0Legs}
In this section, we consider the vacuum diagrams at $O(V^2)$.
Recall that our diagrammatic rules are defined so that there are no contractions
between lines at the same vertex, so there are only 2 diagrams:
\[
\bra{f}T_{2,0}\ket{i} = \ggap
\includegraphics[valign=c,scale=0.65]{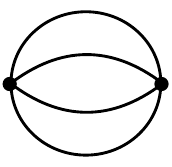}
+ \includegraphics[valign=c,scale=0.65]{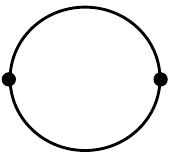}
\eql{0LegsDiags}
\]
In the local approximation, these yield a contribution proportional to
the identity operator in the effective Hamiltonian.
We will compute these diagrams to give an additional 
test of the separation of scales.

The 3-loop vacuum diagram in the local approximation gives a matching
contribution
\[
\!\!\!\!
\includegraphics[valign=c,scale=0.65]{Figs/bubbled_f}
- \biggl[ 
\includegraphics[valign=c,scale=0.65]{Figs/bubbled_f} \biggr]_{\text{eff}}
&= \braket{f}{i}
\frac{1}{24} \left(\frac{\lambda}{2\pi R}\right)^{\! 2}\sum_{1,\ldots,4}\de_{1234,0} \frac{\Theta(\om_1+\om_2+\om_3+\om_4 - E_\text{max})}{16\om_1\om_2\om_3\om_4(-\om_1-\om_2-\om_3-\om_4)}.
\eql{3LoopMatching}
\]
This diagram contains two different kinds of regions with overlapping
UV/IR dominated sums:
one where 2 momenta are large and 2 are small (with multiplicity 6)
and one where 3 momenta are large and 1 is small (with multiplicity 4).
These are given by
\[
\!\!\!\!
\includegraphics[valign=c,scale=0.65]{Figs/bubbled_f}
- \biggl[ 
\includegraphics[valign=c,scale=0.65]{Figs/bubbled_f} \biggr]_{\text{eff}}
&\simeq \braket{f}{i} \frac{6}{24} \left(\frac{\lambda}{2\pi R}\right)^{\! 2}\sum_{k} \frac{\Theta(2\om_k- E_\text{max})}{16\om_k^2(-2\om_k)}
\sum_{|k'|,|k''| \, \ll \, k_\text{max}}\frac{1}{\om_{k'} \om_{k''}}
\nn[5pt]
&\qquad{}
+ \braket{f}{i} \frac{4}{24} \left(\frac{\lambda}{2\pi R}\right)^{\! 2}\sum_{k}
\frac{\Theta(2\om_k+\om_{2k}- E_\text{max})}{16\om_k^2\om_{2k}(-2\om_{k}-\om_{2k})}
\sum_{|k'| \, \ll \, k_\text{max}}\frac{1}{\om_{k'}}
\nn
&\qquad{}
+ \cdots
\eql{UVIRbubb}
\]
The first term is $\sim (\ln m_\text{Q})^2/E_\text{max}^2$, while the second
is $\sim (\ln m_\text{Q})/E_\text{max}^3$.

The 1-loop vacuum diagram gives a matching
contribution
\[
\includegraphics[valign=c,scale=0.65]{Figs/bubble_f}
- \bigg[ \includegraphics[valign=c,scale=0.65]{Figs/bubble_f} \bigg]_\text{eff}
&= \braket{f}{i} \frac{1}{2} \left(m_V^2\right)^{\! 2}\sum_k \frac{\Theta(2\om_k - E_\text{max})}{4\om_k^2(-2\om_k)}.
\eql{vacloop}
\]
We expect separation of scales to be manifest only if
we use a renormalization scheme in the fundamental theory where
$\mu \sim E_\text{max}$.
In this case, \Eq{vacloop} has an overlapping UV/IR dominated region
because $m_V^2$ is IR dominated (see \Eq{mv2renorm}).
The overlapping region gives a contribution $\sim (\ln m_\text{Q})^2/E_\text{max}^2$.

As we have previously discussed, we also expect
separation of scales to be manifest for the effective Hamiltonian written 
in terms of operators that are not normal ordered.
This effective Hamiltonian has additional contributions to the identity operator
from un-normal-ordering the matching contributions computed above using \Eq{nodef}.
Writing
\[
H_{2,0} = C_2 \myint \text{d}x \ggap \id,
\]
the additional contributions to $C_2$ from un-normal-ordering are
\[
\De C_2 &= \left(\frac{\la}{2\pi R}\right)^{\! 2}\frac{1}{16}\frac{1}{4\pi R}  
\bigg[\frac{1}{3}\sum_k \frac{\Theta(2\om_k + \om_{2k} - E_\text{max})}
{\om_k^2 \om_{2k}(2\om_k+\om_{2k})} 
\!\! \sum_{|k'| \, \ll \, k_\text{max}} \frac{1}{\om_{k'}}
\nn[5pt]
&\qquad\qquad\qquad\qquad\quad{} 
+ \frac{3}{8} \sum_k \frac{\Theta(2\om_k - E_\text{max})}{\om_k^3}
\!\! \sum_{|k'|,|k''| \, \ll \, k_\text{max}} \frac{1}{\om_{k'} \om_{k''}}
\bigg] + \cdots
\eql{C2nnorm}
\]
Combining these contributions, 
the overlapping UV/IR contributions to $C_2$ are given by
\[
C_2 &\simeq \frac{\la^2}{128\pi^2 R^2}
\biggl(\, \underbrace{\frac 13 - \frac 13}_{{}  = \, 0}\, \biggr)
\sum_k \frac{\Theta(2\om_k + \om_{2k} - E_\text{max})}{\om_k^2 \om_{2k} (2\om_k+\om_{2k})}
\!\!\sum_{|k''| \,  \ll \, k_\text{max}} \frac{1}{\om_{k''}} 
\nn[5pt]
&\qquad{}
+ \frac{\la^2}{128\pi^2 R^2}
\biggl(\, \underbrace{\frac 38 - \frac 18 - \frac 14}_{{}  = \, 0}\, \biggr)\sum_k \frac{\Theta(2\om_k - E_\text{max})}{\om_k^3}  
\!\! \sum_{|k|,|k''| \, \ll \, k_\text{max}} \frac{1}{\om_{k'}\om_{k''}} 
+ \cdots
\]
The positive contributions are from \Eq{C2nnorm},
and the negative contributions are from the diagrams in \Eqs{UVIRbubb} and \eq{vacloop}.
Once again, we see that overlapping UV/IR regions cancel,
leaving a effective coefficient that is dominated by the contributions of states near the cutoff.

% ==================================================================================
\section{Power Counting and Locality}
\scl{PowerCounting}
%\ML{I felt that there was too much emphasis on the local approximation
%in the introduction to this section.} \TC{Okay with these cuts} \RH{Agree}
%\MLE{\sout{We have seen that the leading $O(V^2)$ corrections to the effective
%Hamiltonian in 2D $\la\phi^4$ theory are local and Hermitian.
%This arises because the corrections to the effective Hamiltonian
%come from states with energies ($H_0$ eigenvalues) above $E_\text{max}$,
%and for such states we can neglect the energies and momenta of the
%external states, which are assumed to have energies
%$E_{i,f} \ll E_\text{max}$. 
%However, we \KFE{generically} expect that the effective Hamiltonian will have 
%non-Hermitian and non-local
%contributions when we go to higher 
%orders in the expansion in $1/E_\text{max}$. 
%%\KF{numerical results don't probe non-local stuff}
%Intuitively, the cutoff is non-local because it is a cutoff on the
%\emph{total} energy, which counts contributions from  non-interacting particles.
%The effective theory approximates a local fundamental theory with a
%non-local cutoff, and therefore we expect that the corrections to
%the effective Hamiltonian will be non-local at some level.}}

The idea of Hamiltonian truncation is that the low-lying states 
of the system can be well-approximated by a truncated Hilbert space 
with maximum energy (measured by $H_0$) $E_\text{max}$.
Effective field theory ideas suggest that we should therefore
be able to compute low-energy
observables in a systematic expansion in powers of $1/E_\text{max}$.
In this section, we present a power-counting scheme that we
believe gives the general form of the effective Hamiltonian
defined above.
An important feature of this power counting is that the non-locality
of the effective Hamiltonian is controlled in the $1/E_\text{max}$
expansion.

Our discussion will be for 2D $\la\phi^4$ theory, but it is
straightforward to generalize to other theories.
When we present numerical results in \sec{Numerics}, we will
see that the errors are consistent with the predictions
of the power counting presented here.

\subsection{Non-locality and Non-Hermiticity at $O(V^2)$}
Before presenting the general power counting, we present calculations
of non-local (and non-Hermitian) 
contributions to the effective Hamiltonian.
This will help motivate the general power counting we present in the
following subsection.

We will compute the $O(E_f)$ correction to 
the matching contribution in \Eq{diag1cont}.
To simplify the calculation, we assume that
\[
\eql{simpapprox}
\om_{1,2,3,4} \ll E_{i,f} \ll E_\text{max},
\]
so the sum over internal momenta in \Eq{diag1cont} can be approximated by
\[
\eql{sumexEf}
\sum_{5,6} \de_{56,34} \frac{\Th(\om_5 + \om_6 - \om_3 - \om_4 + E_f - E_\text{max})}
{\om_5 \om_6 (\om_3 + \om_4 - \om_5 - \om_6)}
&\simeq \sum_k \frac{\Th(2\om_k + E_f - E_\text{max})}{\om_k^2 (-2\om_k)}.
\]
The approximation \Eq{simpapprox} allows us to neglect the masses, so
the sum over $k$ can be taken over the range
\[
|k| > \frac{(E_\text{max} - E_f)R}{2}.
\]
This can be evaluated as a series in $1/E_\text{max}$
using the Euler-Maclauren summation formula
\[
\eql{EM}
\sum_{k \, = \, k_\text{min}}^\infty f(k)
= \int_{k_\text{min}}^\infty \text{d}k\ggap f(k)
+ \sfrac 12 f(k_\text{min})
- \sum_{r \, = \, 1}^\infty \frac{B_{2r}}{(2r)!} f^{(2r)}(k_\text{min}),
\]
where $B_{2r}$ are the Bernoulli numbers.
\Eq{EM} is  an asymptotic expansion valid for any sufficiently smooth
function $f$ such that $f(k)$ and its derivatives vanish 
as $k \to \infty$.
Applying this to the sum in \Eq{sumexEf} gives
\[
\sum_k &\frac{\Th(2\om_k + E_f - E_\text{max})}{\om_k^3} 
\nn[4pt]
&\qquad {}= 
\frac{4R}{E_\text{max}^2}
\bigg[ 1 - \frac{2 E_f}{E_\text{max}}
+ O(1/E_\text{max} R)
+ O(E_f / E_\text{max}^2 R)
+ O(E_f^2/E_\text{max}^2) \bigg].
\]
There are similar contributions from the other diagrams.
Including the $O(E_f)$ corrections, the effective Hamiltonian
has a contribution (compare to \Eq{H24})
\[
H_{2,4} = \frac{\la_2}{4!} \myint \mathrm{d}x\ggap 
\bigg[ \no{\phi^4} - \frac{2 H_0}{E_\text{max}} \no{\phi^4} + \cdots
\bigg],
\eql{H24NonLocal}
\]
where $\la_2$ is given by \Eq{lambda2}.
Note that the $E_f$ dependence has been written in terms of $H_0$.
The term in $H_{2,4}$ that depends on $H_0$ is non-local, because
$H_0$ is itself given by an integral over $x$. 
From the diagrammatic expansion, we expect that all of the
non-locality arises from the energy dependence in the step functions
that define the cutoff of the effective Hamiltonian.
Expanding these in powers of the energies of the external states
gives powers of $H_0$, as in this example.
Therefore, we expect that \emph{all} of the non-locality of
the effective Hamiltonian can be parameterized by powers of
$H_0 / E_\text{max}$.
Note that the non-local term in \Eq{H24NonLocal} 
is also not Hermitian, which is consistent with the fact that
$H_\text{eff}$ is non-Hermitian at $O(V^2)$ (see \Eq{H2match}).

\subsection{General Form of the Effective Hamiltonian}
We now discuss what we expect for the general form of the effective
Hamiltonian.
The cutoff on the effective theory is defined by $H_0$, and this is the
only source of non-locality in the theory.
We therefore conjecture that all of the non-locality
in the effective theory can be parameterized by the free Hamiltonian $H_0$.
This leads naturally to the following general form for the $O(V^n)$
term in the effective Hamiltonian:
\[
\eql{effpwrcount}
H_n \sim \frac{\la^n}{E_\text{max}^{2n-2}} \sum_a
\myint \gap \mathrm{d}x\, C_{na} \!\ggap
\biggl( \frac{H_0}{E_\text{max}},\ggap \frac{R^{-1}}{E_\text{max}},\ggap
\frac{m_\text{Q}}{E_\text{max}} \biggr)
\scr{O}_{na}\!\ggap\biggl(\frac{\d_x}{E_\text{max}},\ggap
\frac{\d_t}{E_\text{max}},\ggap \phi \biggr),
\]
where we have factored out powers of $E_\text{max}$ so that $C_{na}$ and $\scr{O}_{na}$
are dimensionless.
Here $\scr{O}_{na}$ is a local Hermitian
operator made of the field $\phi$ and its
derivatives, and the real coefficient $C_{na}$ depends on the free Hamiltonian $H_0$, as
well as the IR mass scales $m_\text{Q}$ and $R^{-1}$.
Expanding $\scr{O}_n$ in derivatives, and $C_n$ in powers of $m_\text{Q}$,
$R^{-1}$, and $H_0$ gives the effective Hamiltonian as an expansion in $1/E_\text{max}$.
Note that the powers of $H_0$ in \Eq{effpwrcount} are to the left of the
operator $\scr{O}_n$. 
There is no loss of generality in writing it this way, since
$[H_0, \phi] = -\ii \dot\phi$.

In fact, the form of the effective Hamiltonian is a bit more
constrained than \Eq{effpwrcount}.
The $\phi \mapsto -\phi$ symmetry implies that only
even powers of $\phi$ can appear in $\scr{O}_n$.
Second, parity invariance implies 
that only even powers of $\d_x$ can appear in $\scr{O}_n$.
Time reversal allows linear terms in time derivatives, since
time reversal includes complex conjugation, so that $\ii\d_t$
is invariant. 
(Factors of $\ii\d_t$ in the effective Hamiltonian
correspond to factors of single-particle energies $\om_k$ in the 
transition amplitude.)
The diagrammatic rules depend on $m_\text{Q}$ only 
quadratically (through $\om_k$), so $C_{na}$ involves only
even powers of $m_\text{Q}$.
Finally, the number of powers of $\phi$ is limited by the
fact that the interaction contains only 4 powers of the fields;
for example, at $O(V^2)$ we have only $\phi^2$ and $\phi^4$ terms,
as we have already seen above. 
Taking all this into account, the power counting rule predicts the parametric form
\[
H_2 \sim \frac{\la^2}{E_\text{max}^2} \myint \mathrm{d} x
\ggap\bigg[ & \phi^2 + \phi^4
+ \frac{R^{-1} + H_0}{E_\text{max}} \big( 1 + \phi^2 + \phi^4 \big)
+ \frac{1}{E_\text{max}} \big( \phi + \phi^3 \big) \ii\dot\phi
\nn[5pt]
&\quad{}
+ O(1/E_\text{max}^2) \bigg].
\eql{H2E3}
\]
\Eq{effpwrcount} gives $H_3 \sim \la^3/E_\text{max}^4$, 
so the leading corrections to the effective theory beyond
the order computed in  \sec{Match4legs} and \sec{Match2Legs} above
comes from  $1/E_\text{max}^3$ corrections to $H_2$.
These can be computed by going beyond the
local approximation.
This calculation is straightforward, but not trivial.
For example, note that \Eq{H2E3} contains a 3-loop contribution
(from the diagram \Eq{3LoopMatching}) that is $\sim \la^2 H_0 / E_\text{max}^3$.
Computing the full $1/E_\text{max}^3$ corrections and checking that they
further reduce the truncation error as indicated by the
power counting is well worth doing.
We plan to address this in future work.

\section{Numerical Results for 2D $\la\phi^4$ Theory}
\scl{Numerics}
In this section, we present numerical results for 2D $\la\phi^4$ theory
using the improved effective Hamiltonian described above.
Our main focus is on the convergence of the results as a function 
of the cutoff $E_\text{max}$, rather than the determination of
physically interesting quantities.
We therefore choose the dimensionful parameters $\la$, $m^2$, and $R$
so that the theory has a single physical IR scale $M_\text{IR}$.
The only large parameter in the Hamiltonian truncation is then
$E_\text{max}/M_\text{IR}$, where $E_\text{max}$ is the cutoff
on the Hamiltonian truncation.
The dimensionful parameters $\la$, $m^2$, and $R$ are related to the
physical IR scale $M_\text{IR}$ by dimensional analysis.
However, some of the dimensionless constants of proportionality
are expected to differ significantly from 1, as we now explain.

\subsection{Explanation of Dimensionless Factors}
First, we consider the coupling $\la$, which has mass dimension 2.
This means that it gets strong at a scale $M_\text{IR} \propto \sqrt{\la}$.
To estimate the constant of proportionality, we use the idea that
the scale $M_\text{IR}$ is the IR cutoff scale where perturbative loop
corrections are the same size as tree-level effects 
(`\naive dimensional analysis')~\cite{Manohar:1983md}.
In 2D, the perturbative loop corrections are proportional to
\[
\la \myint \frac{\mathrm{d}^2 p}{(2\pi)^2} f(p^2)
= \frac{\la}{4\pi} \int_{0}^\infty \mathrm{d} p^2 \ggap f(p^2).
\]
(Note that $\la$ is defined to be the coefficient of $\phi^4/4!$
in the Lagrangian, so the symmetry factors in loop corrections are order 1.)
Therefore we assume that the physical IR scale defined by the coupling $\la$ 
is given by
\[
M_\text{IR}^2 \sim \frac{\la}{4\pi}.
\]

Next, we consider the compactification radius $R$.
In perturbation theory, the effect of this is to give a series
of Kaluza-Klein excitations with energy differences $1/R$.
We therefore choose
\[
R^{-1} \sim M_\text{IR} .
\]
Note that the finite-size corrections for particles of mass $m$ in 1
spatial dimension are proportional to $e^{-2\pi R m}$ {\cite{Luscher:1985dn}},
so this choice may be sufficient to neglect finite
size effects.

Finally we consider the mass parameter $m^2$.
At tree-level, it is clear that $M_\text{IR} \sim m$, but we must consider
the possibility of large loop effects.
However, as shown in \sec{Renormalization}, the renormalized
coefficient of $\phi^2$ in the effective
Hamiltonian is independent of the renormalization scheme used for the
fundamental theory.
To present our numerical results,
we therefore mostly use the `normal-ordered scheme' defined in \sec{mQ},
and we assume
\[
m_\text{NO} \sim M_\text{IR},
\]
where $m_\text{NO}$ is the coefficient of $\no{\phi^2}$ in the fundamental
theory.

\subsection{Numerical Results}

We now present the numerical results.
Collecting the results above, the effective Hamiltonian is given by
\[
H_\text{eff}
&= H_0 + \int \text{d} x \left[ \frac12 (m_V^2 + m_{V2}^2)  \no{\phi^2} 
		+ \frac{\la + \la_2} {4!}  \no{\phi^4} \right], 
\]
where the corrections are calculated in
\S\ref{sec:matchVop} and \S\ref{sec:Match4legs}:
\begin{subequations}
\[
m_{V2}^2 &= \frac{\la}{16\pi R} \biggl[
\frac{\la}{6\pi R} 
\! \sum_{3,4,5} \! \de_{345,0} 
\frac{\Th(\om_3 + \om_4 + \om_5 - E_\text{max})}
{\om_3 \om_4 \om_5 ( \om_3 + \om_4 + \om_5)}
\! - \! m_V^2
\! \sum_k \! \frac{\Th(2\om_k - E_\text{max})}{\om_k^3}
\biggr],
\\[5pt]
\la_2 &= -\frac{3 \la^2}{16\pi R}
\sum_{k} \frac{\Th(2\om_k - E_\text{max})}{\om_k^3}
\]
\end{subequations}
The frequencies $\om_k$ are computed using the `quantization mass'
$m_Q$, while $m_V$ is defined in \Eq{mVdefn} and evaluated in \Eq{mv2renorm}.
The sums in $m_{V2}^2$ and $\la_2$ are computed numerically
by summing over $k$ up to $k_\text{UV} = 1000$,
which guarantees the inclusion of single particle states
with $E \gg E_\text{max}$.
According to the power counting discussed in \sec{PowerCounting},
the error without the terms $\la_2$ and $m_{V2}^2$
should be $O(1/E_\text{max}^2)$, and including the
terms $\la_2$ and $m_{V2}^2$ should reduce the error 
to $O(1/E_\text{max}^3)$.

Most of the calculations we perform use
the reference values
\begin{align}
1 = \sqrt{\frac{\la}{4\pi}} = m_\text{NO} = \frac{2\pi R}{10},
\eql{benchmark}
\end{align}
in arbitrary units.
The dimensions can always be restored by inserting factors of $R$
(for example) using dimensional analysis.

For the numerical truncation, we work in a Fock basis of states defined by creation operators \Eq{DefineCreateAndAnn}:
\[
\ket{\{n\}} &= \left(\prod_k \frac{1}{\sqrt{n_k!}} \big(a^\dagger_k \big)^{n_k}\right) |0\rangle,
\]
where $n_k$ is the number of particles with momentum $k$. 
The states are labeled by their eigenvalues with respect to the free Hamiltonian $H_0$ in
\Eq{quantizationmassdefn} and the spatial momentum $P$:
\begin{subequations}
\[
H_0 |\{n\} \rangle &= \sum_k n_k \om_k |\{n\} \rangle\\[5pt]
P   |\{n\} \rangle &= \sum_k n_k k |\{n\} \rangle .
\]
\end{subequations}
In addition, the theory has a parity symmetry
\[
\phi(x, t) \mapsto \phi(-x, t)
\]
and a $\mathbb{Z}_2$ symmetry
\[
\phi(x, t) \mapsto -\phi(x, t).
\]
The $\mathbb{Z}_2$ symmetry acts on our basis states as
\[
\eql{Zchargedefn}
Z \ket{\{n\}} = (-1)^{\sum_k \! n_k} \ket{\{n\}}.
\]
We will present the results for states with $P = 0$, and quote values of 
the $\mathbb{Z}_2$ charge for excited states.
For simplicity, we do not keep track of parity in this work.

\begin{figure}[t!]
	\centering
	\begin{minipage}{.9\textwidth}
    \centering
    \includegraphics[width=.8\textwidth]{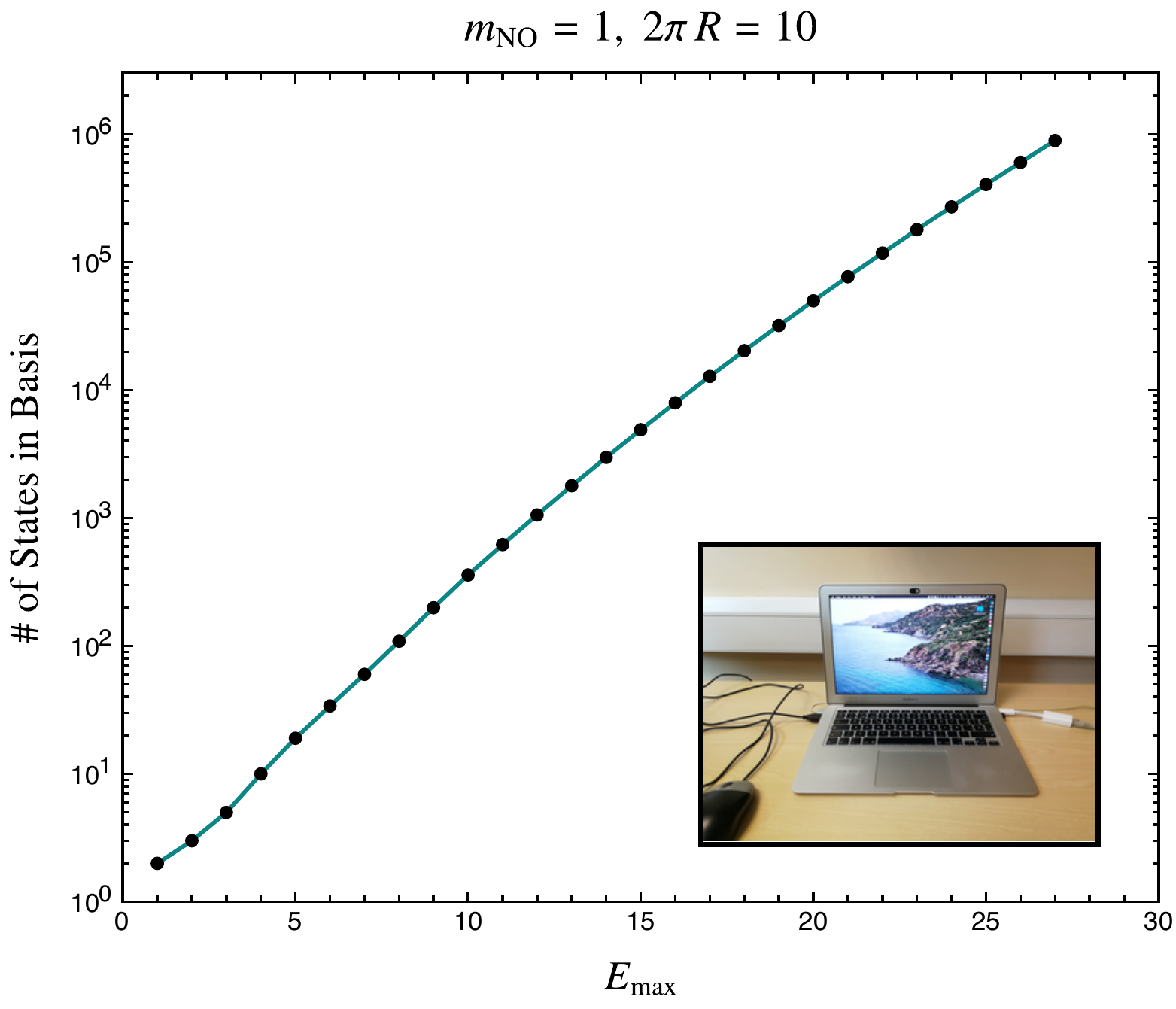}
    \hfill
    \caption{Number of states in the basis for the $H$ matrix computed on a laptop [inset] in this work. The highest point in the plot corresponds to $E_\text{max} = 27$ with $\sim9\times10^5$ states.     }\label{fig:computation-power}
    \end{minipage}
\end{figure}

The calculations were
%\ML{I prefer original} \RH{I was just smoothing out the transition from the new text added. If we're moving the text earlier, we can keep the old wording}
performed using very modest computing resources,
namely a laptop computer running {\tt Python}
with matrix diagonalization performed using the package \texttt{scipy.sparse.linalg}.%
 %\hspace{-3pt}
%
\footnote{The code is available at \href{https://github.com/rahoutz/hamiltonian-truncation}{\texttt{github.com/rahoutz/hamiltonian-truncation}}.}
The largest cutoff explored was $E_\text{max} = 27$, corresponding to
$\sim9\times10^5$ states in the truncated Hilbert space for $m_\text{NO} = 1$
(see \sec{mQ}). The most computationally intensive step is the construction of the 
Hamiltonian matrix, which grows in size exponentially as $E_\text{max}$ is increased, see~\Fig{computation-power}.

\begin{figure}[t!]
    \centering
    \begin{minipage}{.9\textwidth}
    \centering
    \includegraphics[trim= 0cm	0cm	0cm	0cm, clip=true,  width=.9\textwidth]{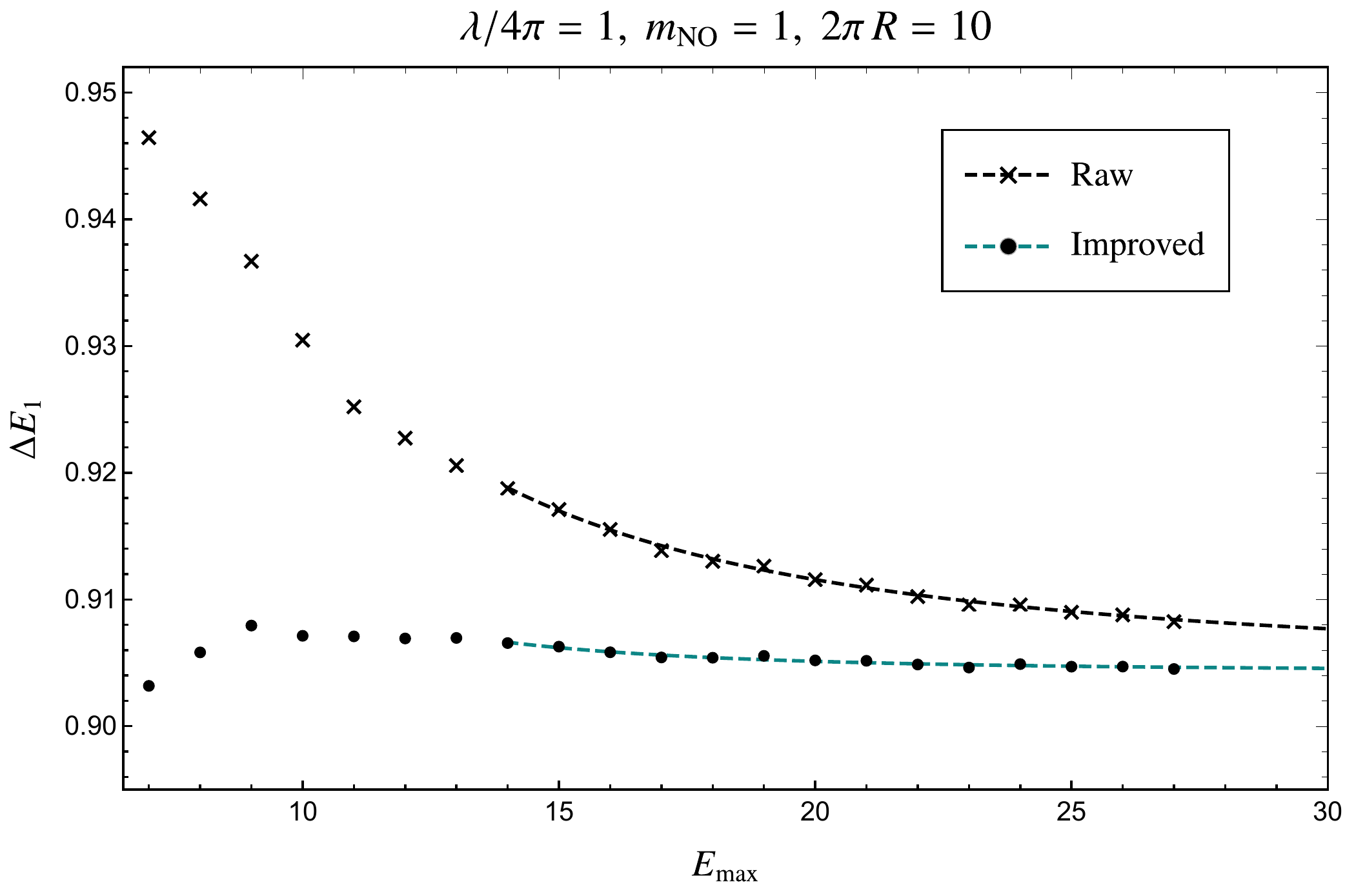}
    \caption{The ground state excitation energy $\Delta E_1$ as a function of the (dimensionless) energy cutoff $E_\text{max}$ for both the raw truncated [crosses] and improved [dots] results. For $E_\text{max} \geq14$, the ground state excitation energy of the raw and improved theories are fit to $1/E_\text{max}^2$ and $1/E_\text{max}^3$, respectively.}
    \label{fig:E10_fit}
    \end{minipage}

	\centering
	\begin{minipage}{.9\textwidth}
\vspace{30pt}
    \centering
    \raisebox{.02cm}{\includegraphics[trim= 0cm	0cm	0cm	0cm, clip=true,  width=0.4665\textwidth]{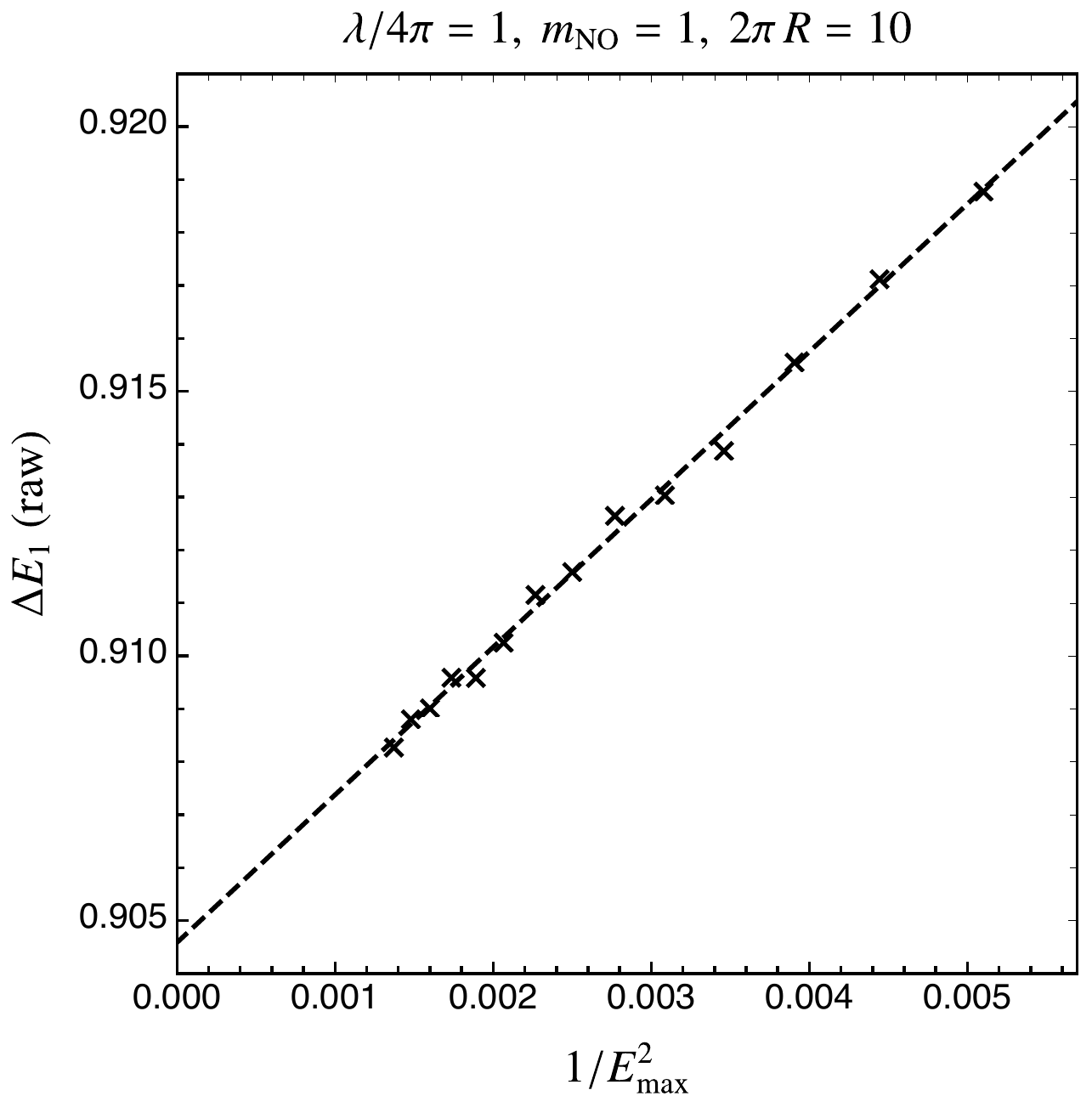}}
    \hfill
    \raisebox{.0cm}{\includegraphics[trim= 0cm	.05cm	0cm	0cm, clip=true, width=0.49\textwidth]{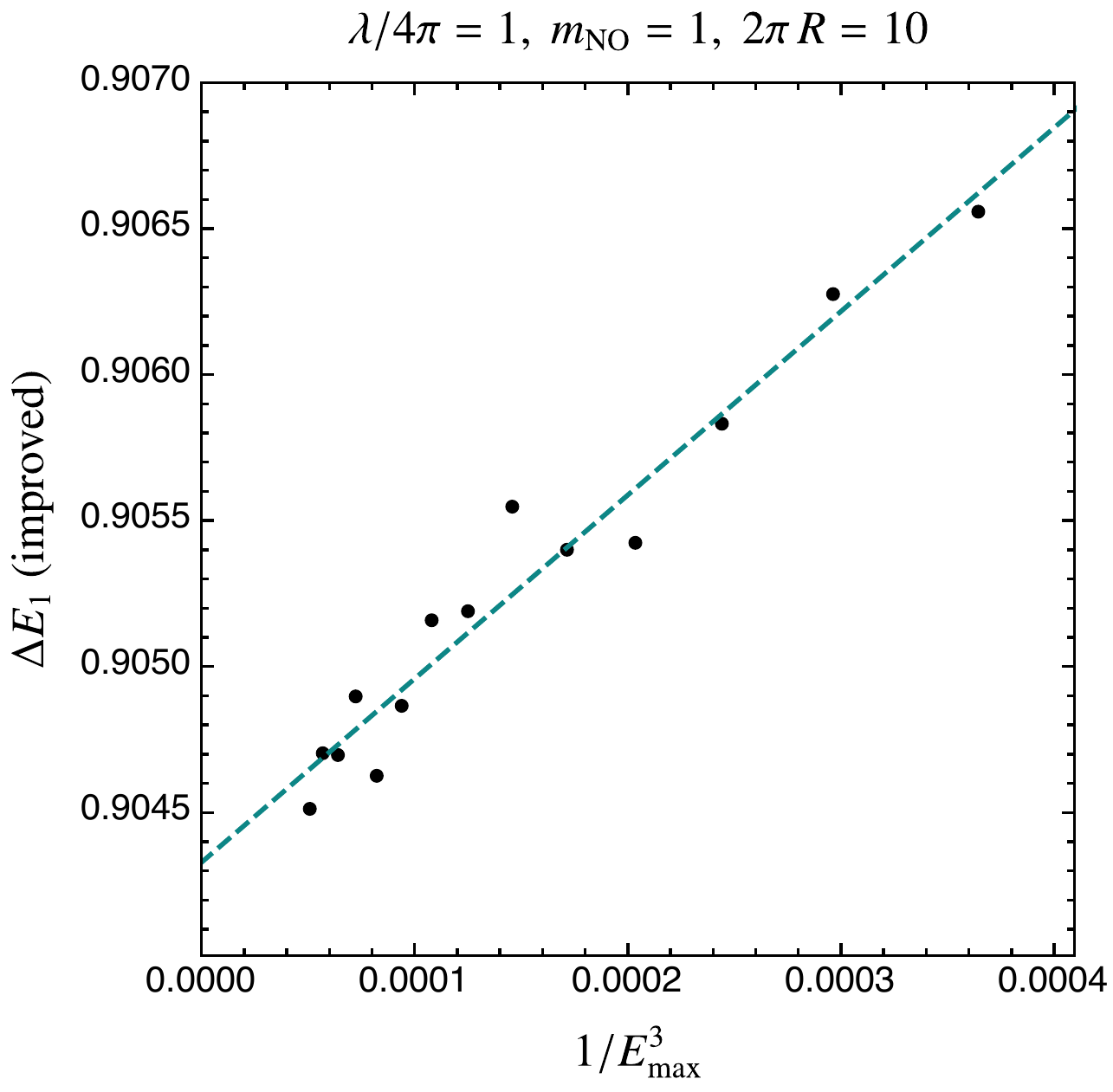}}
    \caption{We highlight the power-law scaling by zooming in on the flat $E_\text{max} \geq 14$ tail and plotting the ground state excitation energy $\Delta E_1$ as a function of $1/E_\text{max}^2$ for the raw truncated theory [left] and as a function of $1/E_\text{max}^3$ for the improved theory [right]. The asymptotic values of $\Delta E_1^\infty$ extracted from the fits are in close agreement: $\Delta E_1^\infty = 0.9046$ (raw) and $\Delta E_1^\infty =  0.9043$ (improved).  
}
    \label{fig:E10_fit_linear}
    \end{minipage}
\end{figure}

We first present results that show the convergence of our method 
as a function of $E_\text{max}$. 
In Figs.~\ref{fig:E10_fit} and~\ref{fig:E10_fit_linear}, we present numerical results for the ground state excitation energies
$\Delta E_n= E_n - E_0$, where $E_{0}$ is the ground state energy, 
and $E_n$ is the energy of the $n^\text{th}$ excited state. When separating states into the $\mathbb{Z}_2$ even and odd bases, we label the energy levels as $\Delta E_n^\pm$. We compare the `raw' theory, obtained by the approximation 
$H_\text{eff} = H + V$, and the improved theory,
which includes the $O(V^2)$ corrections computed in \sec{Matching2DTheory}. 
We fit the results to
\[
\De E_{n}(E_\text{max}) = \De E_n^{(\infty)} + \frac{C_n}{(E_\text{max})^\al},
\]
where $\al = 2$ for the raw calculation, and $\al = 3$ for the improved theory.
As predicted from the general power counting arguments presented above,
the error in the raw result scales as $1/E_\text{max}^2$, whereas the 
improved result has an error that scales as $1/E_\text{max}^3$. 
This is a very important test that shows that our method is working as expected. 

\begin{figure}[!h]
    \centering
    \begin{minipage}{.9\textwidth}
    \centering
    \raisebox{.0cm}{\includegraphics[trim= 0cm	0cm	0cm	0cm, clip=true,  width=0.32\textwidth]{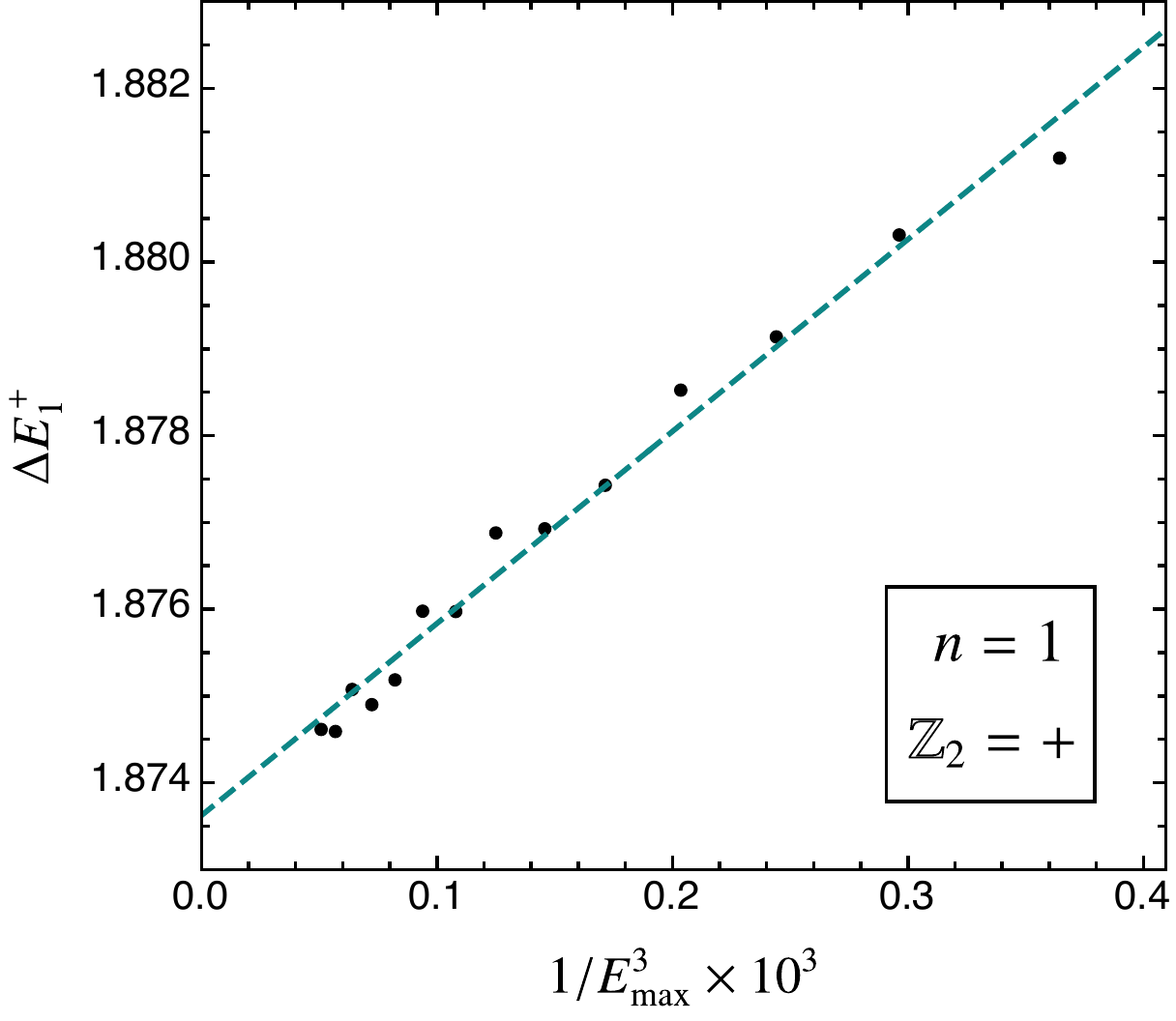}}
    \raisebox{.02cm}{\includegraphics[trim= 0cm	.05cm	0cm	0cm, clip=true, width=0.32\textwidth]{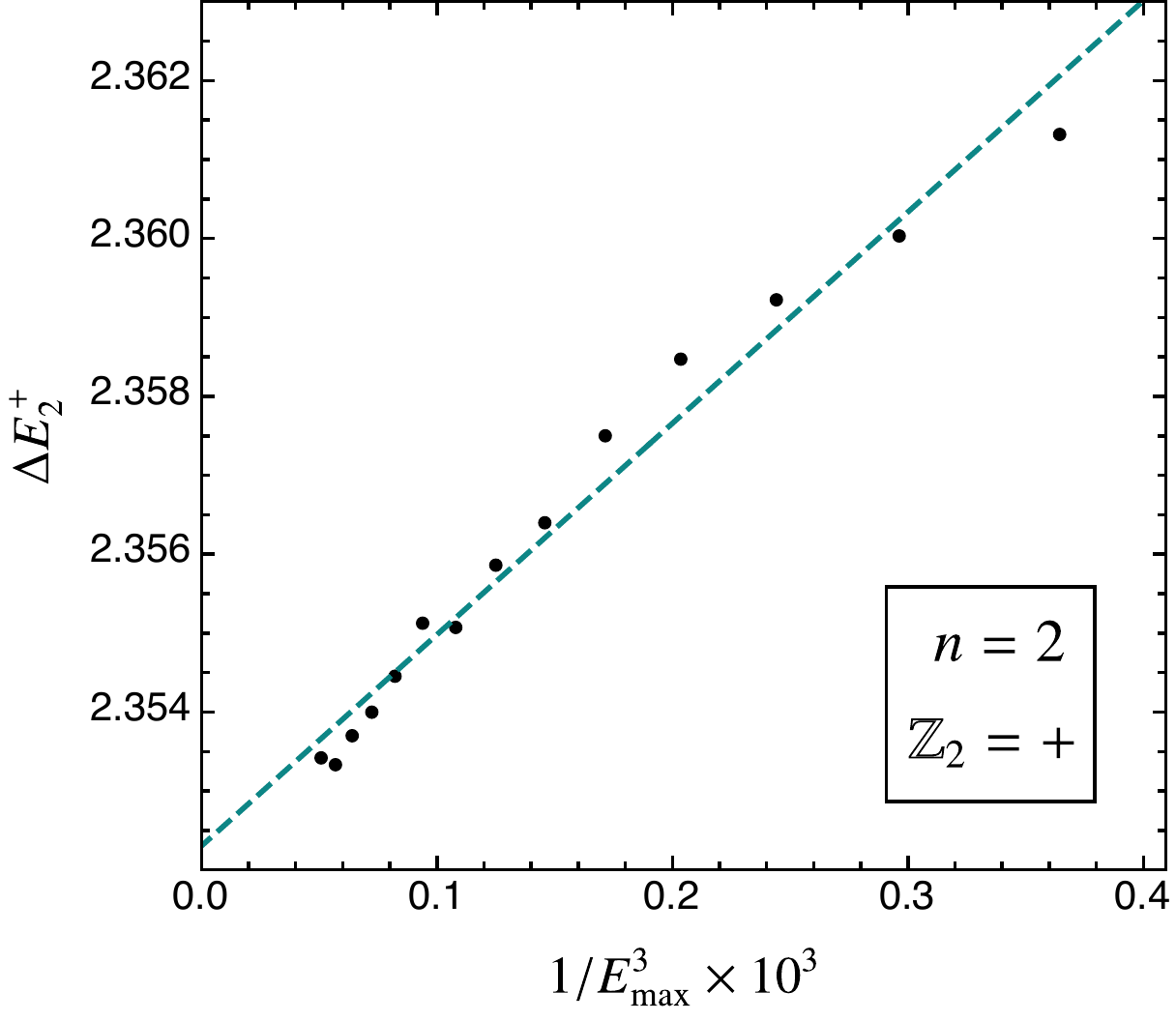}}
    \raisebox{.02cm}{\includegraphics[trim= 0cm	.05cm	0cm	0cm, clip=true, width=0.32\textwidth]{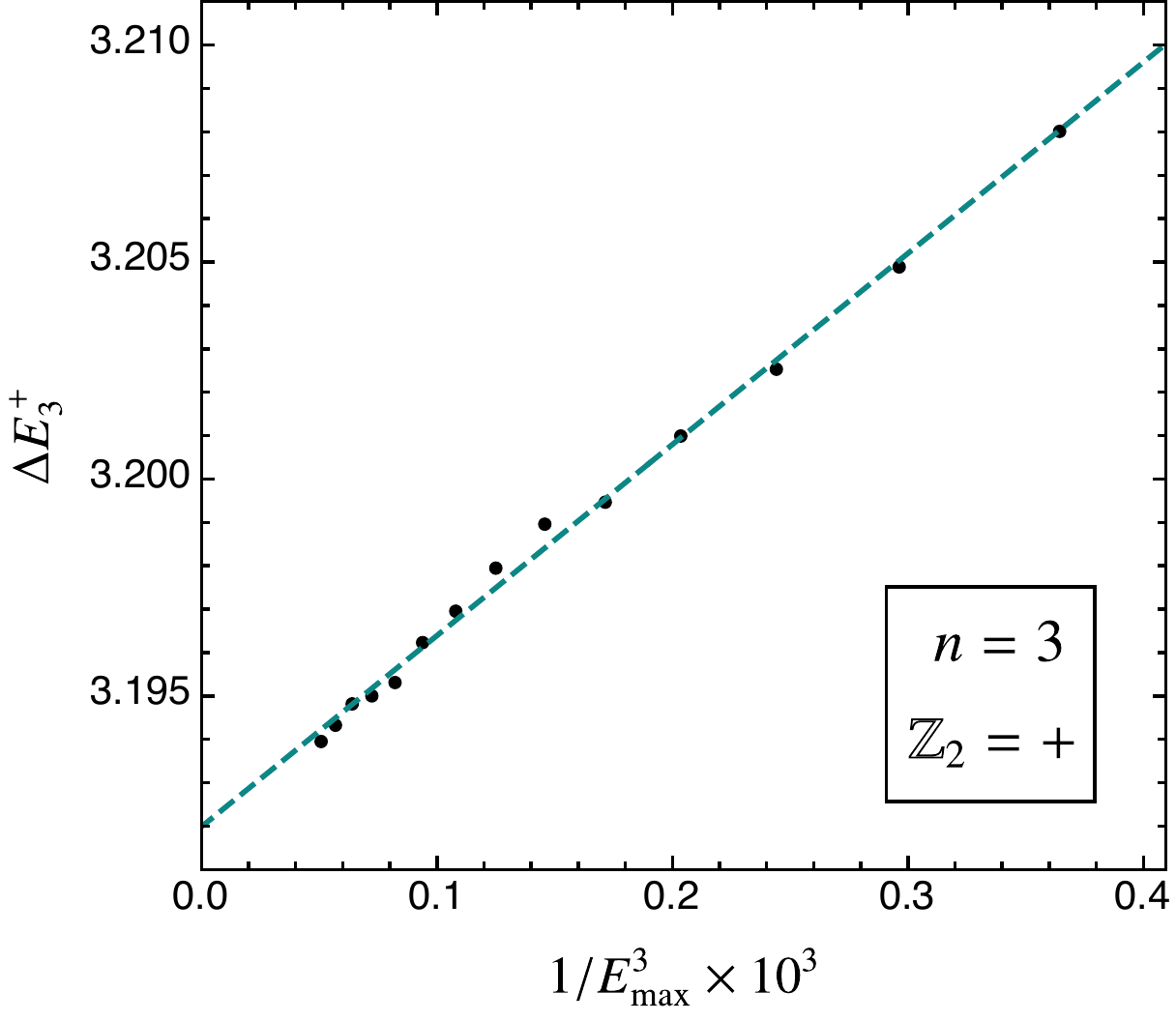}}
    \vspace{12pt}
    \vfill
    \hspace{-.02cm}\raisebox{-.0cm}{\includegraphics[trim= 0cm	.05cm	0cm	0cm, clip=true, width=0.3123\textwidth]{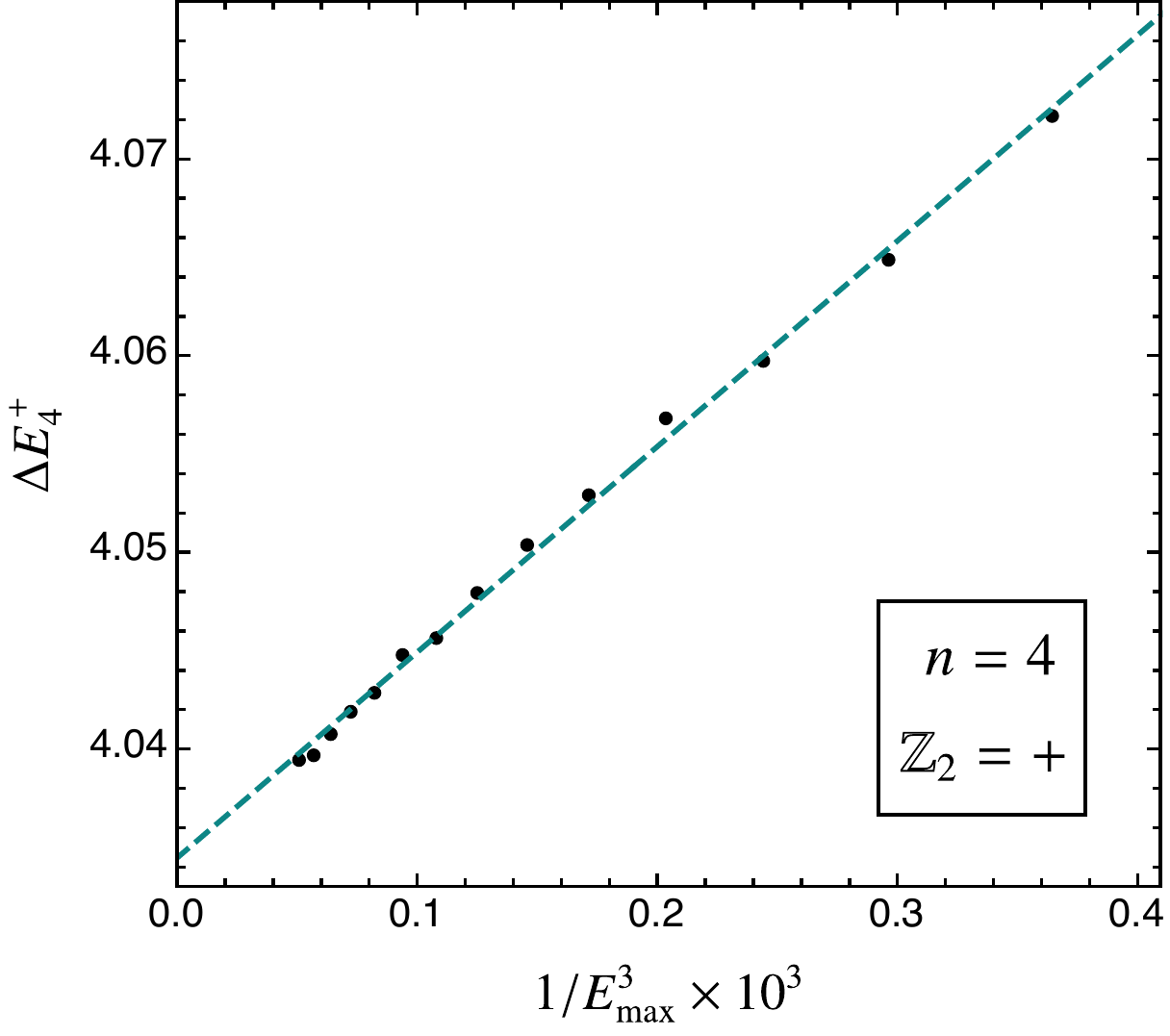}} 
    \hspace{0cm}\raisebox{-.04cm}{\includegraphics[trim= 0cm	0cm	0cm	0cm, clip=true,  width=0.32\textwidth]{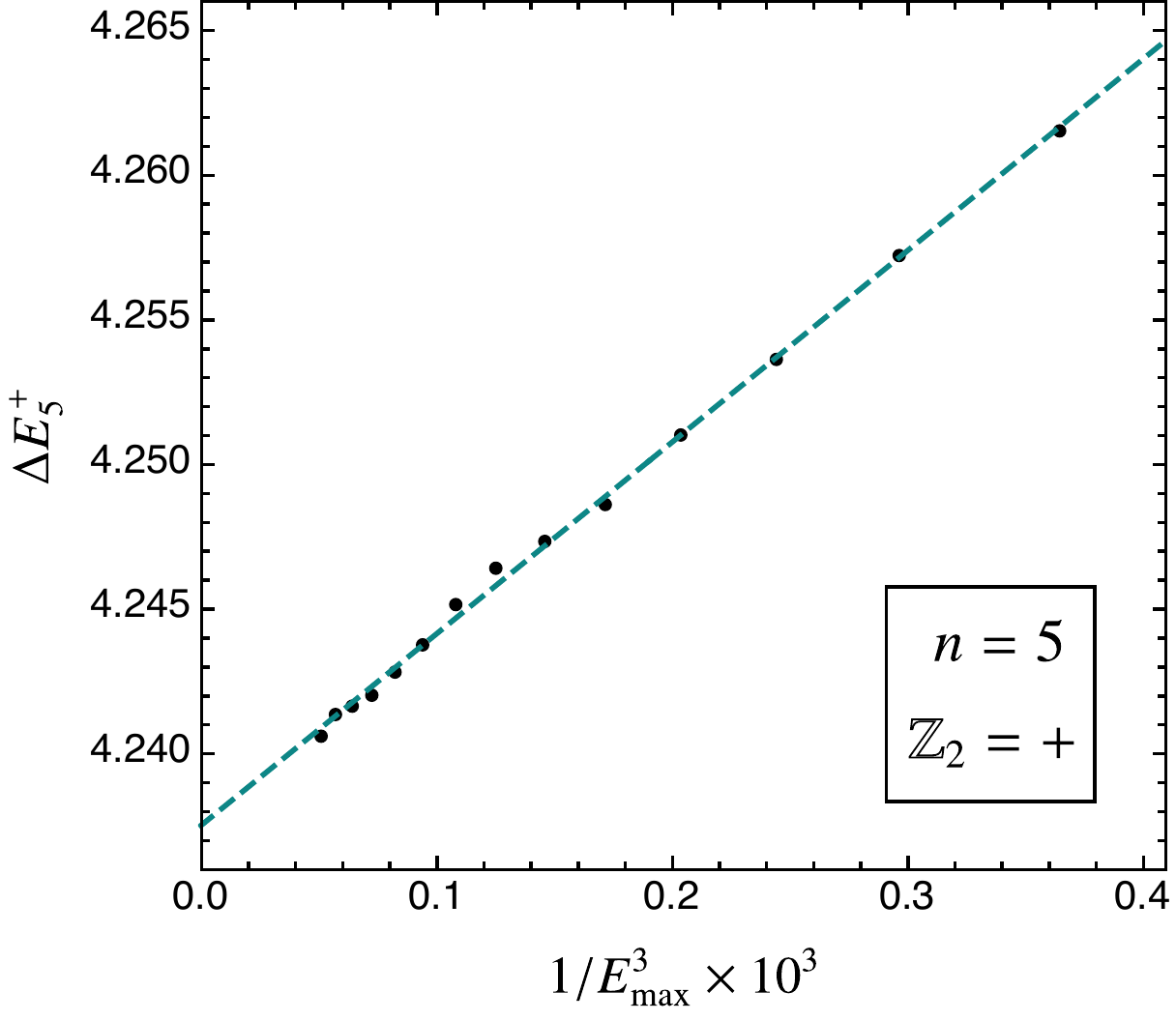}}
    \hspace{.11cm}\raisebox{-.01cm}{\includegraphics[trim= 0cm	.05cm	0cm	0cm, clip=true, width=0.313\textwidth]{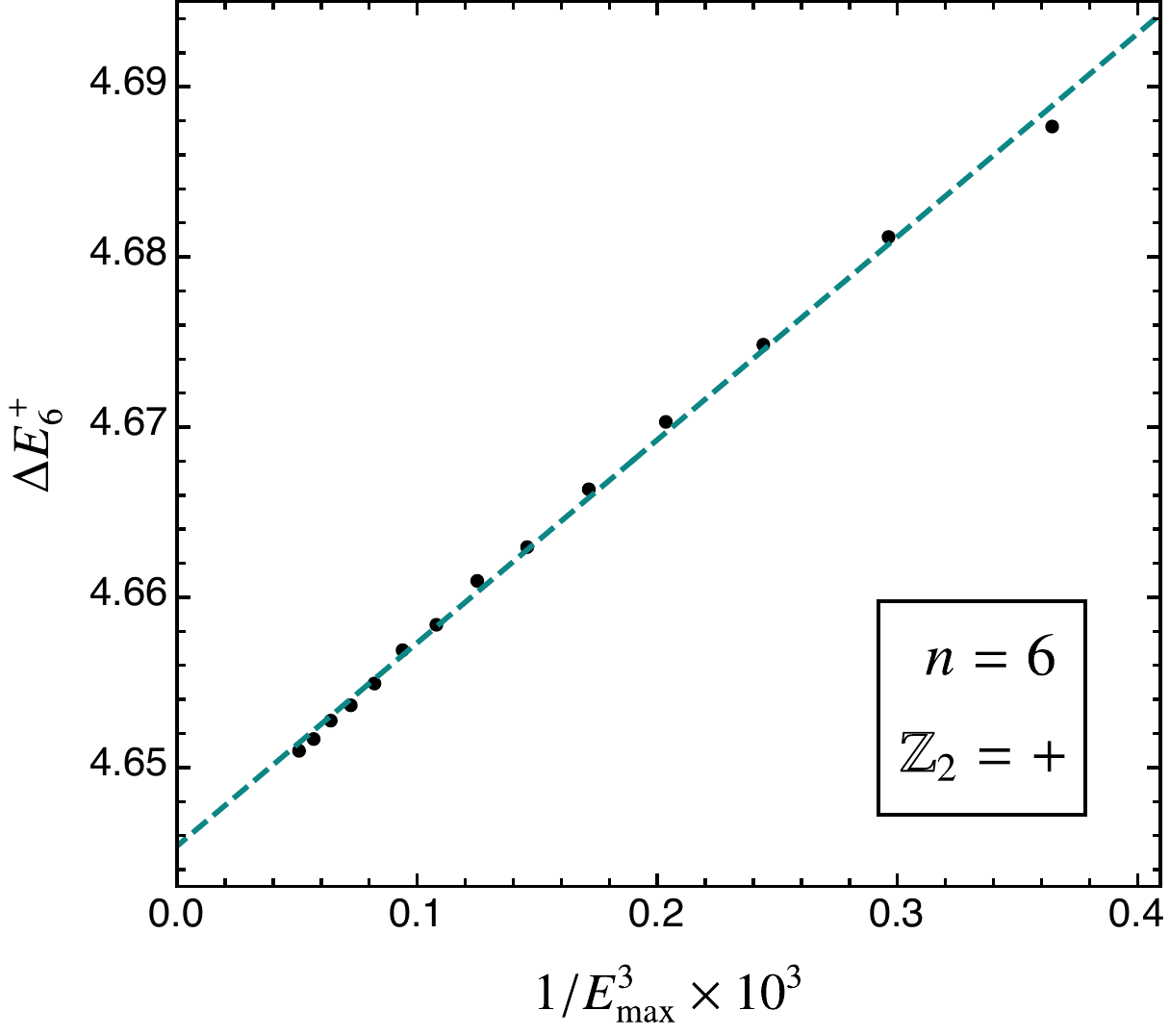}}     \caption{The excitation energy $\Delta E_n$ of the first six $\mathbb{Z}_2$-even excited states in the improved theory as a function of $1/E_\text{max}^3$, including the best fit line. In all panels, $\lambda/4\pi = 1,\ m_\text{NO} = 1,\ 2\pi R = 10$.}
    \label{fig:En-Zp}
    \centering
    \vspace{40pt}
    \hspace{-.125cm}\raisebox{-.0cm}{\includegraphics[trim= 0cm	0cm	0cm	0cm, clip=true,  width=0.3285\textwidth]{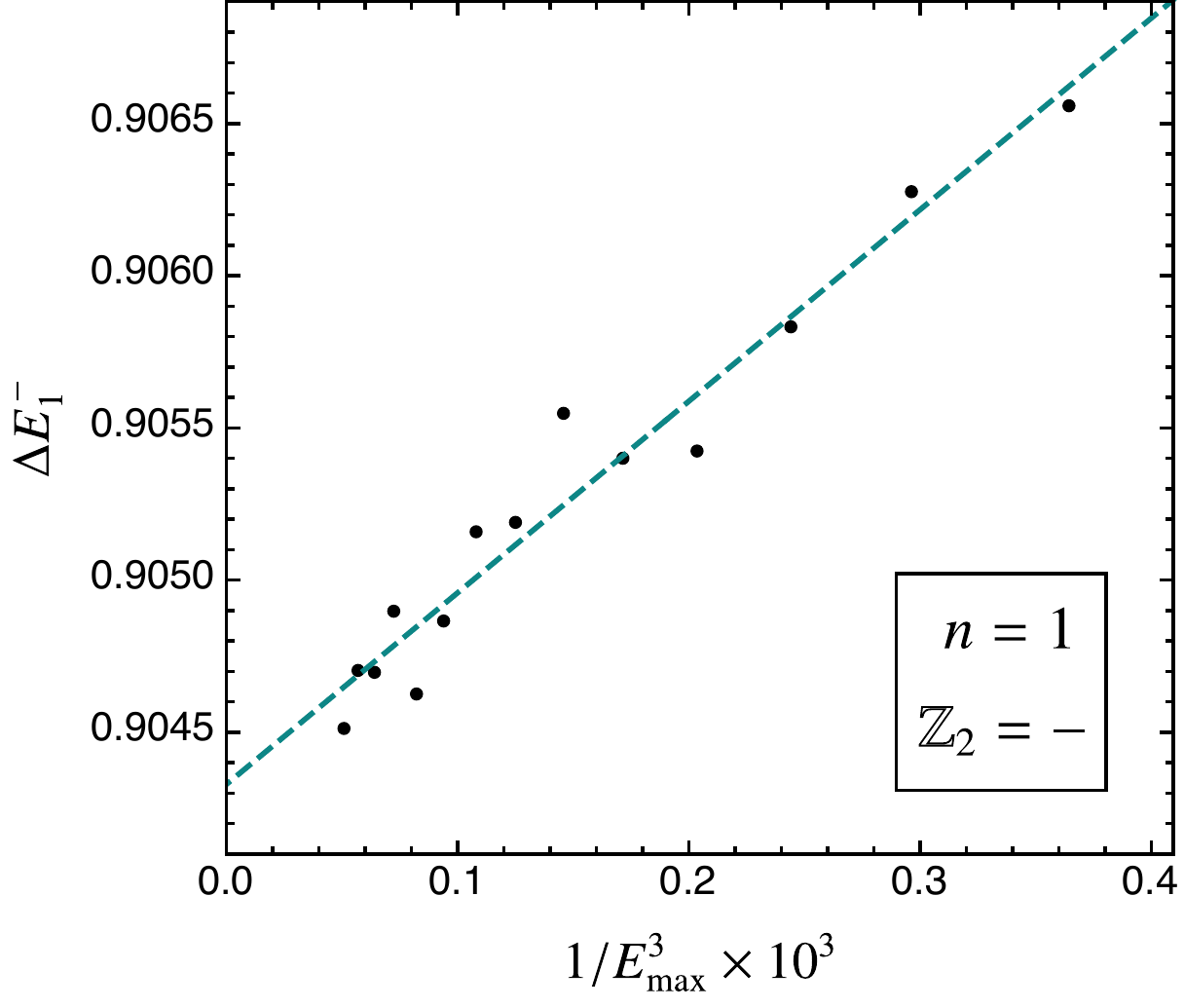}}
    \hspace{-.005cm}\raisebox{.04cm}{\includegraphics[trim= 0cm	.05cm	0cm	0cm, clip=true, width=0.32\textwidth]{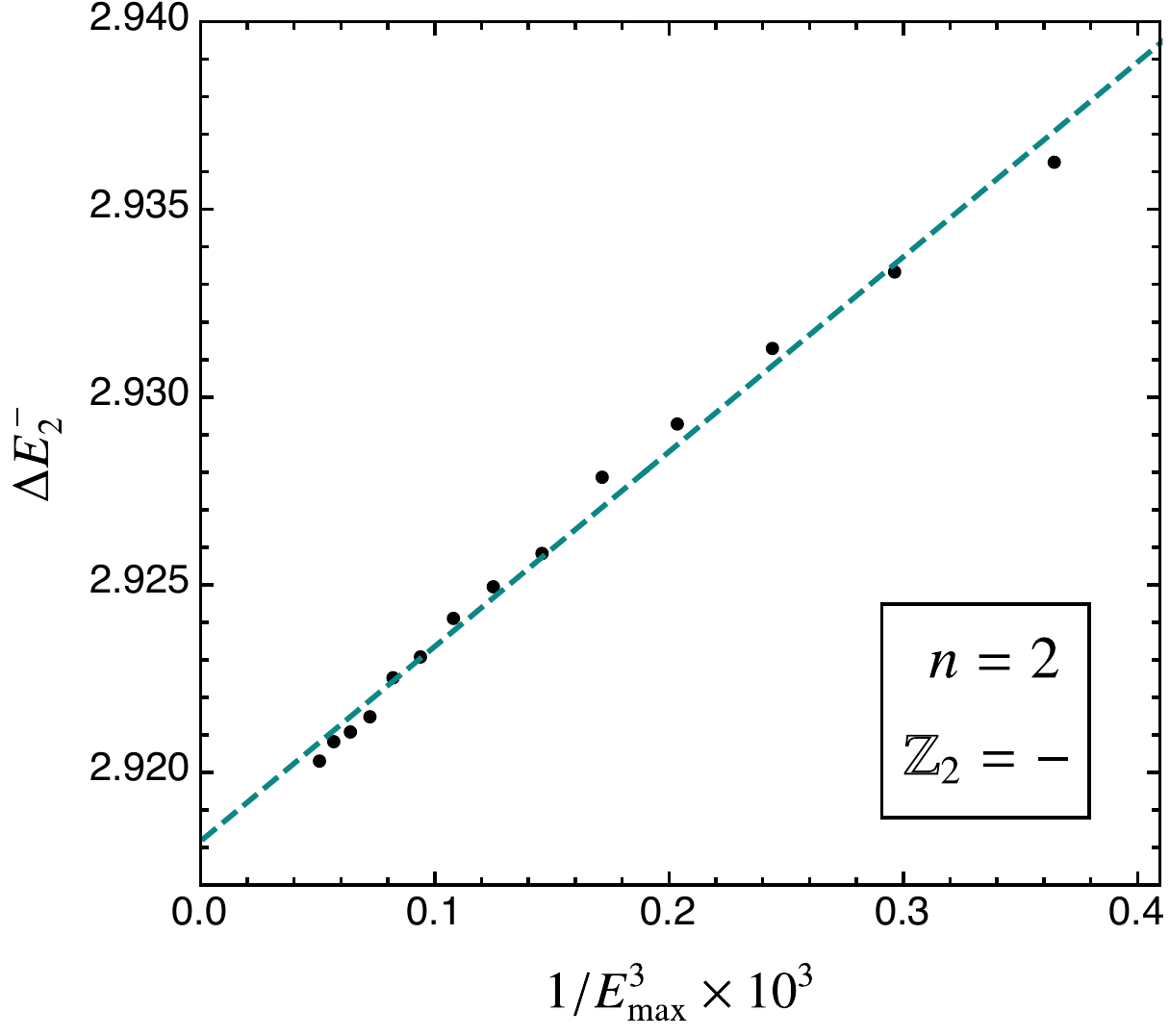}}
    \raisebox{.04cm}{\includegraphics[trim= 0cm	.05cm	0cm	0cm, clip=true, width=0.32\textwidth]{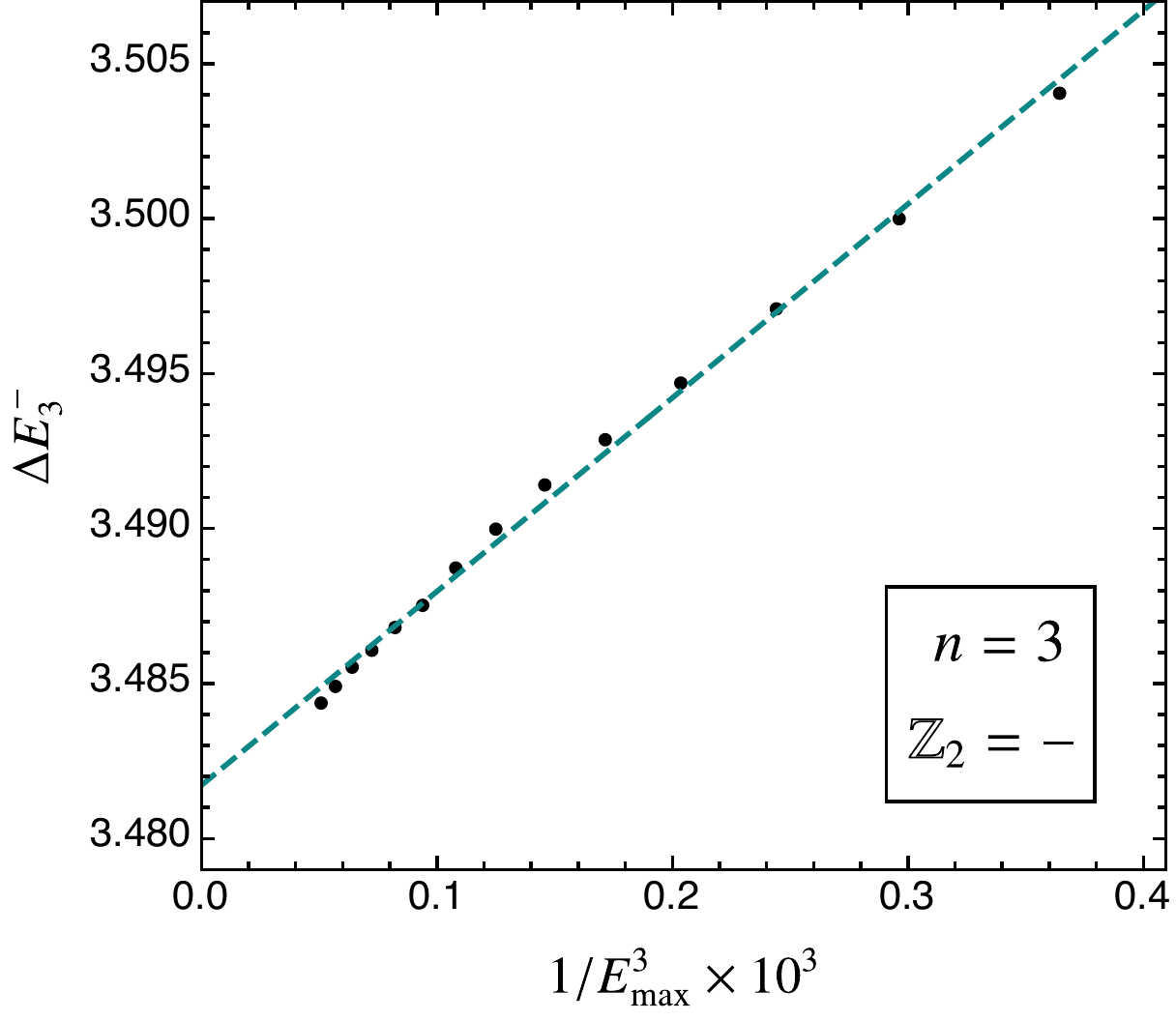}}
    \vspace{12pt}
    \vfill
    \hspace{-.12cm}\raisebox{-.0cm}{\includegraphics[trim= 0cm	.05cm	0cm	0cm, clip=true, width=0.318\textwidth]{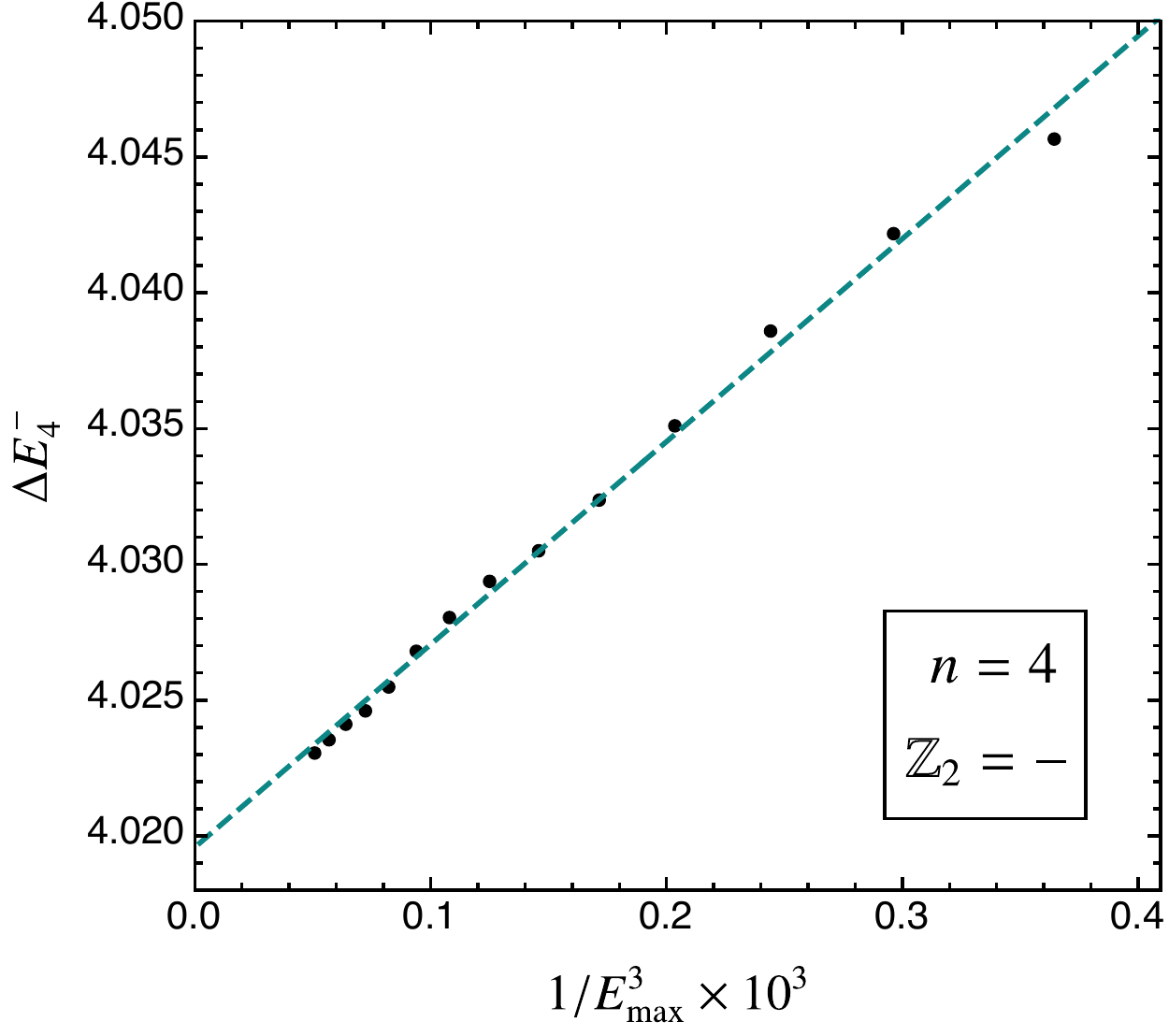}} 
    \hspace{.03cm}\raisebox{-.02cm}{\includegraphics[trim= 0cm	0cm	0cm	0cm, clip=true,  width=0.318\textwidth]{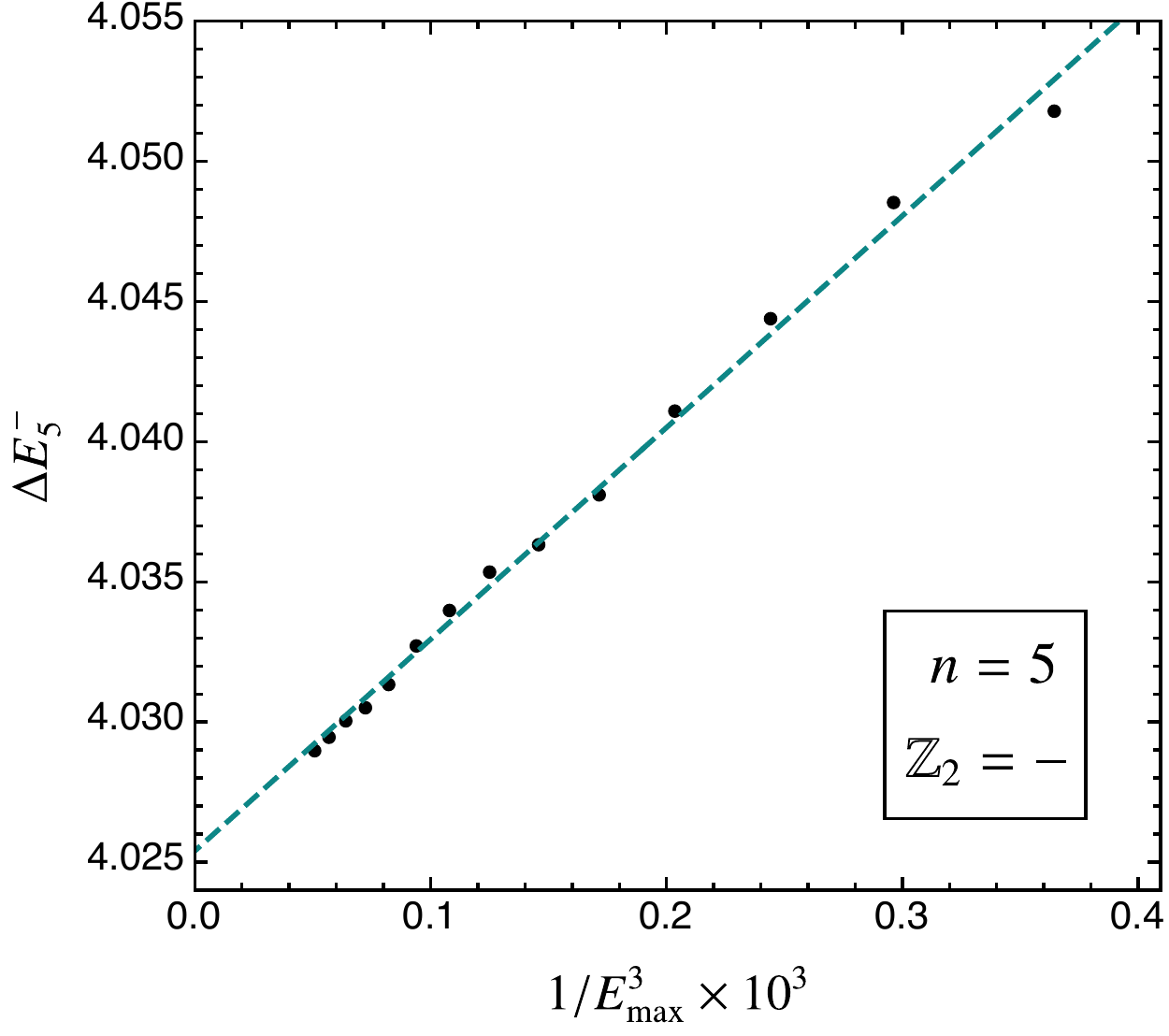}} 
    \hspace{.03cm}\raisebox{-.0cm}{\includegraphics[trim= 0cm	.05cm	0cm	0cm, clip=true, width=0.318\textwidth]{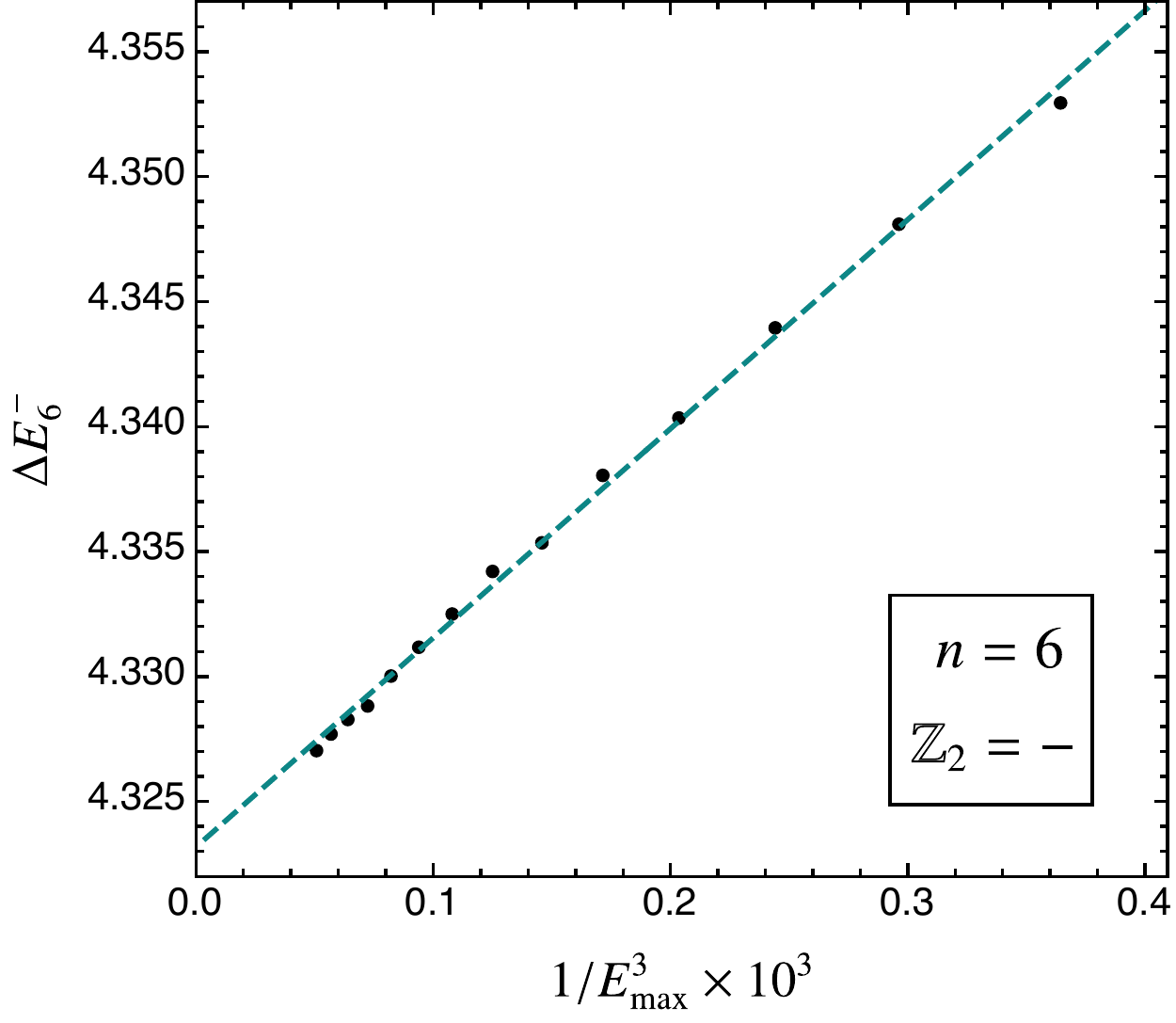}}     \caption{ The excitation energy $\Delta E_n$ of the first six $\mathbb{Z}_2$-odd excited states in the improved theory as a function of $1/E_\text{max}^3$, including the best fit line. In all panels, $\lambda/4\pi = 1,\ m_\text{NO} = 1,\ 2\pi R = 10$.}
    \label{fig:En-Zm}
    \end{minipage}
\end{figure}

Figs.~\ref{fig:En-Zp} and \ref{fig:En-Zm} show that the excitation energy
$\De E_n$ computed in the improved theory
continues to scale as $1/E_\text{max}^3$ for excited states.
The results are shown for states with $Z = \pm 1$, where $Z$ is the $\mathbb{Z}_2$
charge of the states (see \Eq{Zchargedefn}). 
This is encouraging for future work on extracting precision predictions for
physical quantities using our method.

\begin{figure}[t!]
    \centering
    \begin{minipage}{.9\textwidth}
    \centering
    \includegraphics[width=.9\textwidth]{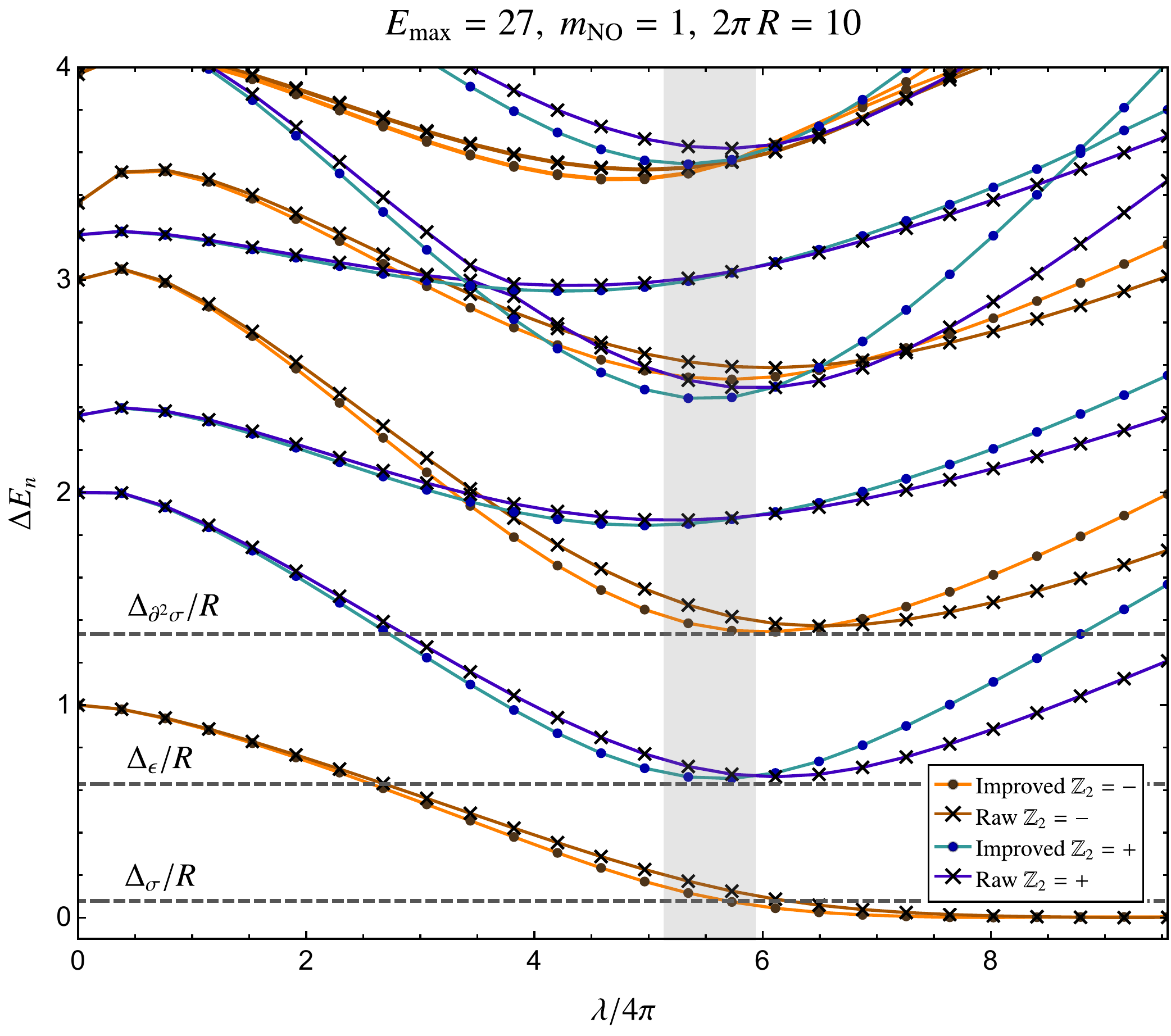}
    \hspace{.4cm}
    \caption{The excitation energy spectra as a function of the $\phi^4$ coupling $\lambda / 4\pi$. The states belonging to the $\mathbb{Z}_2$-even (-odd) basis are shown in blue (orange). The raw truncated theory is plotted in dark blue (dark orange) using cross markers, while the improved theory is plotted light blue (orange) using dot markers. For clarity, lines are drawn to connect the data points within each excited state. The dashed horizontal lines correspond to the known theoretical values for the operator dimensions of the 2D Ising model, and the grey band covers the predicted range of the critical coupling.
    }
    \label{fig:spectrum-vary-la}
    \end{minipage}
\end{figure}

\begin{figure}[t!]
    \centering
    \begin{minipage}{.9\textwidth}
    \centering
    \hspace{.0cm}\raisebox{-.0cm}{\includegraphics[trim= 0cm	0cm	0cm	0cm, clip=true,  width=0.2448\textwidth]{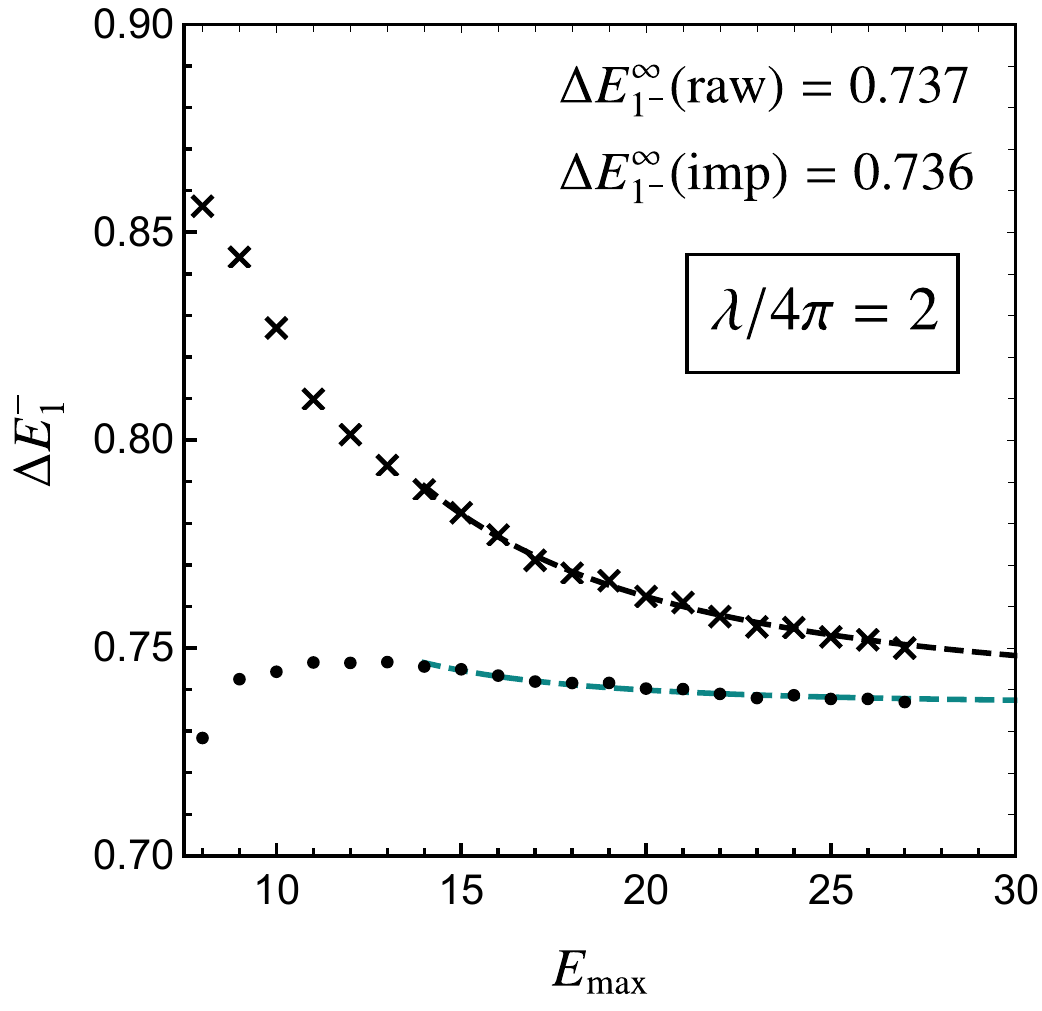}}
    \hspace{.00cm}\raisebox{.005cm}{\includegraphics[trim= 0cm	.05cm	0cm	0cm, clip=true, width=0.2448\textwidth]{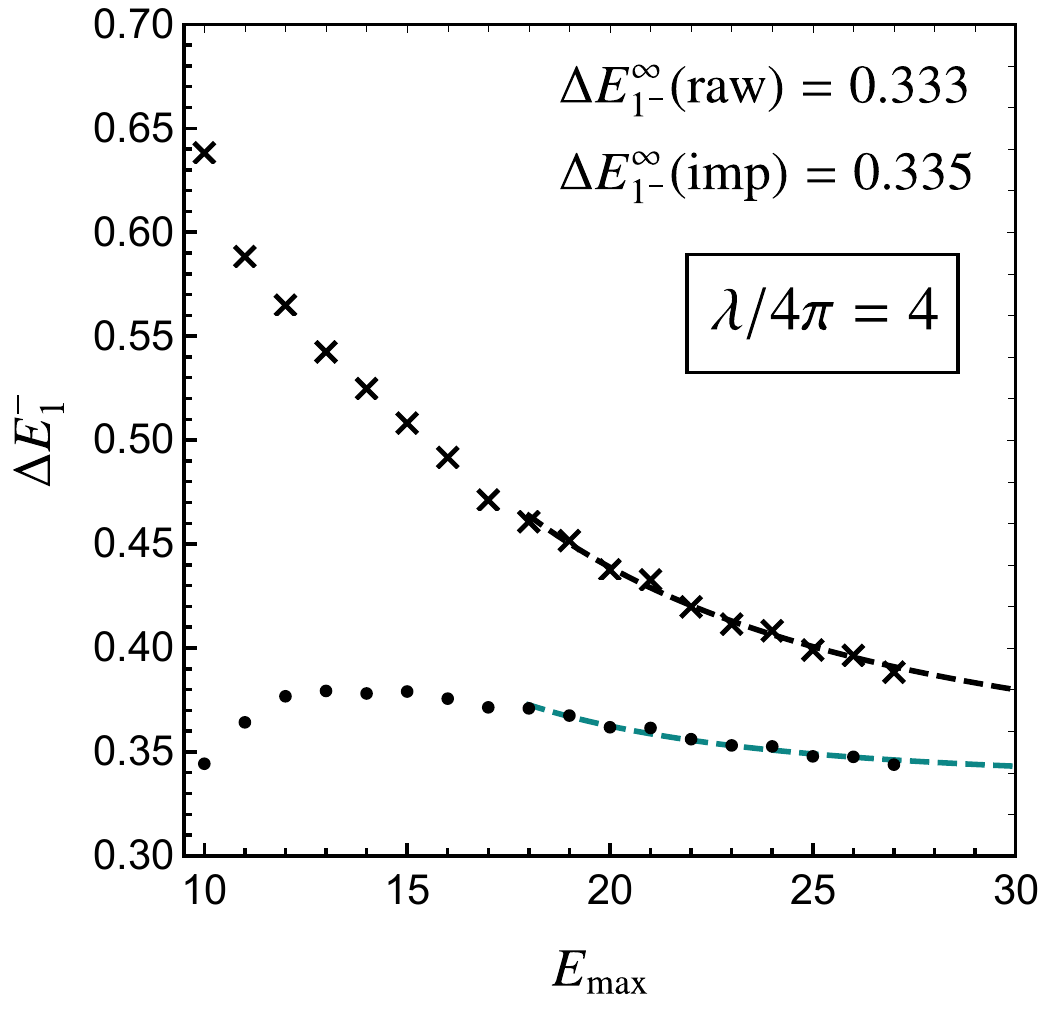}}
    \raisebox{.02cm}{\includegraphics[trim= 0cm	.05cm	0cm	0cm, clip=true, width=0.23868\textwidth]{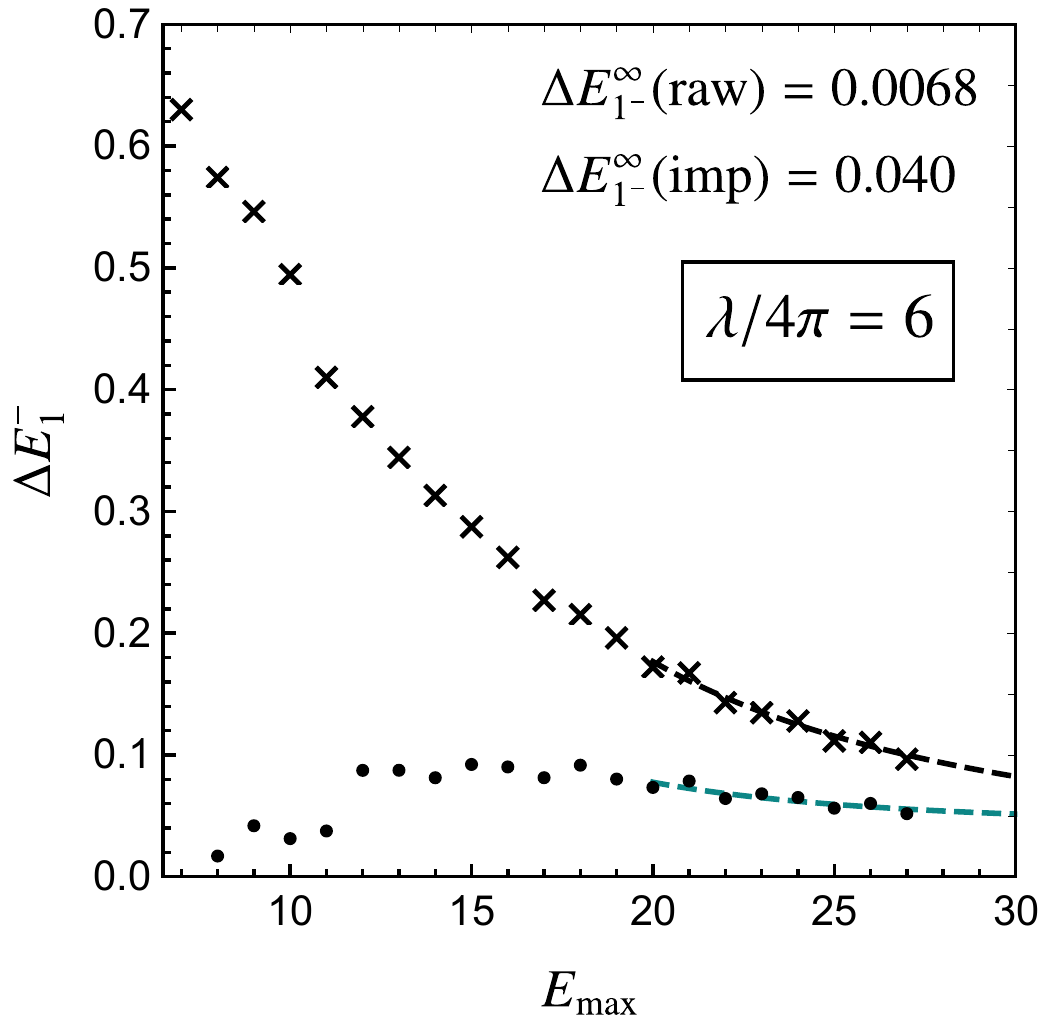}} 
    \hspace{0cm}\raisebox{.005cm}{\includegraphics[trim= 0cm	.05cm	0cm	0cm, clip=true, width=0.2448\textwidth]{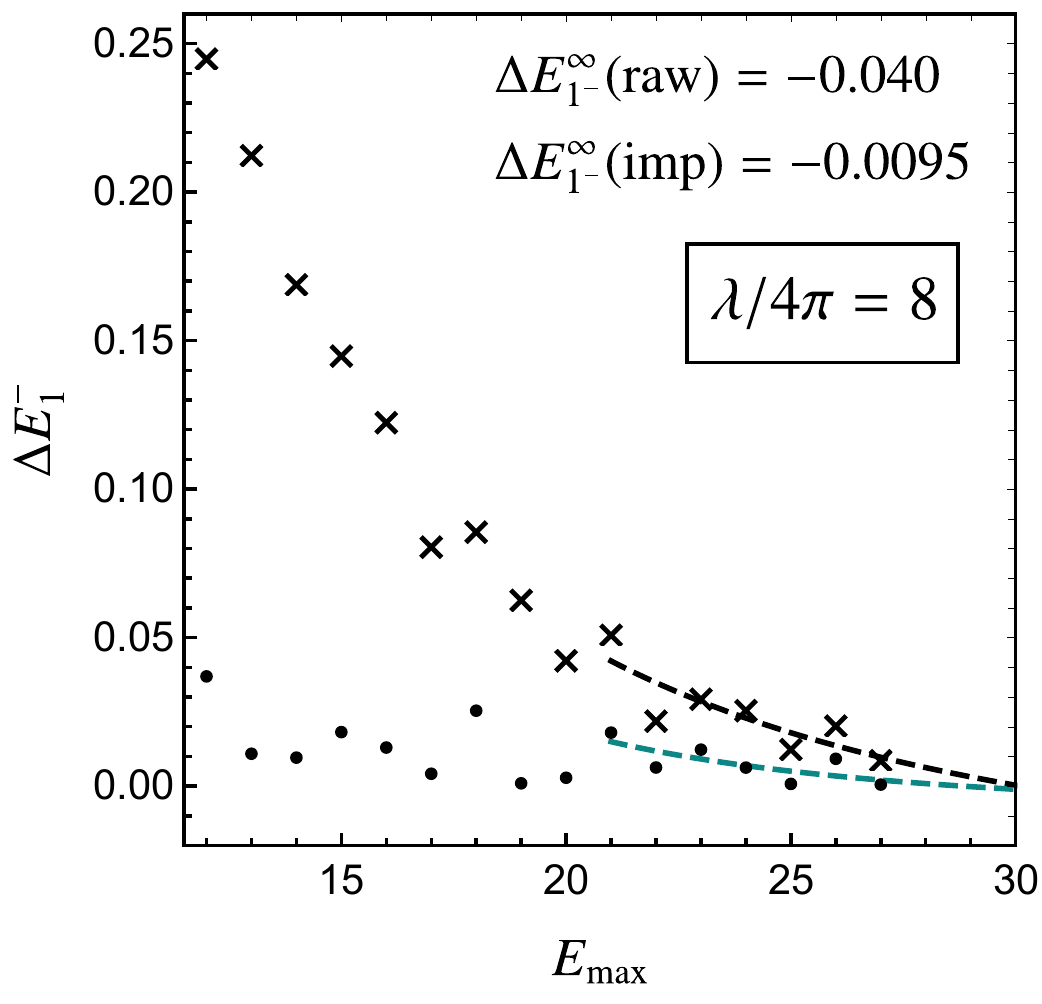}}
 	\vspace{0pt}
 %   \vspace{12pt}
    \vfill
    \hspace*{-.1013cm}\raisebox{-.0cm}{\includegraphics[trim= 0cm	0cm	0cm	0cm, clip=true,  width=0.2455\textwidth]{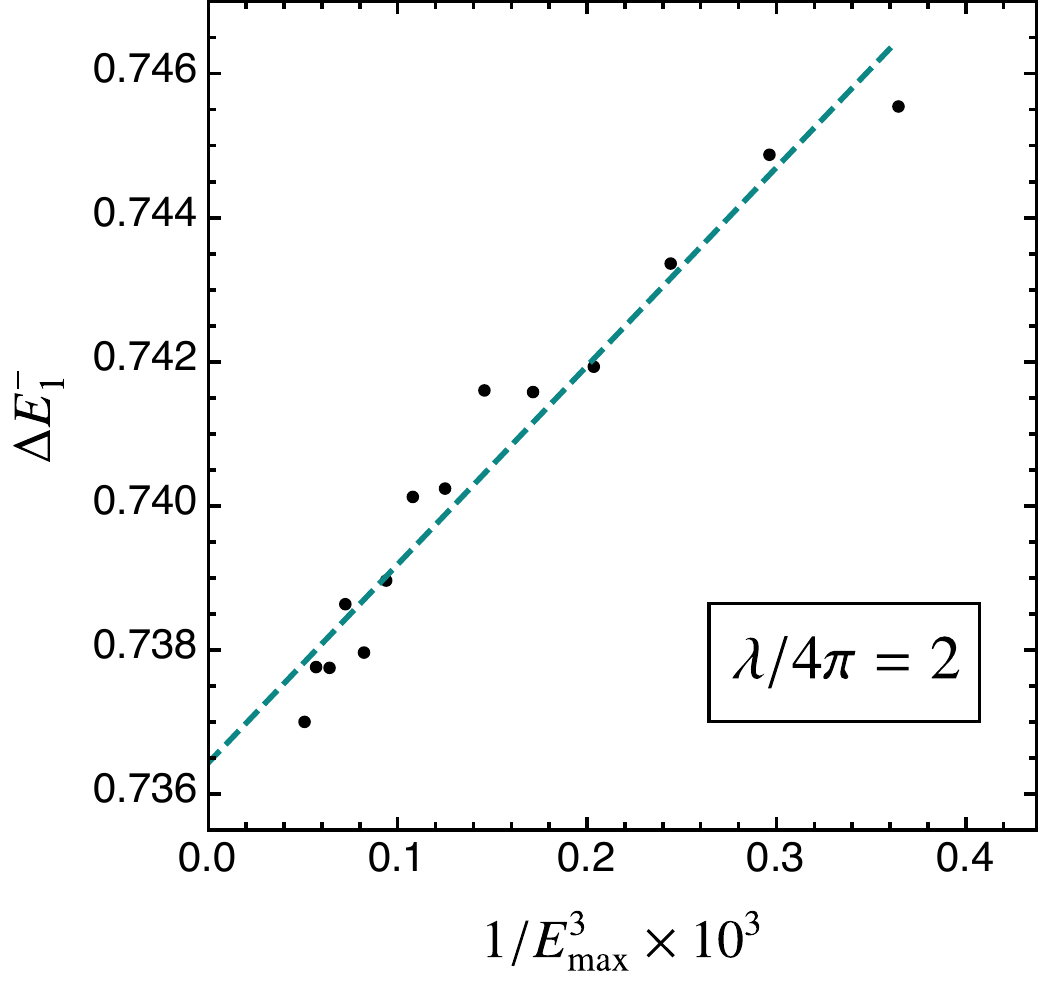}} 
    \hspace*{.094cm}\raisebox{.032cm}{\includegraphics[trim= 0cm	.05cm	0cm	0cm, clip=true, width=0.2387\textwidth]{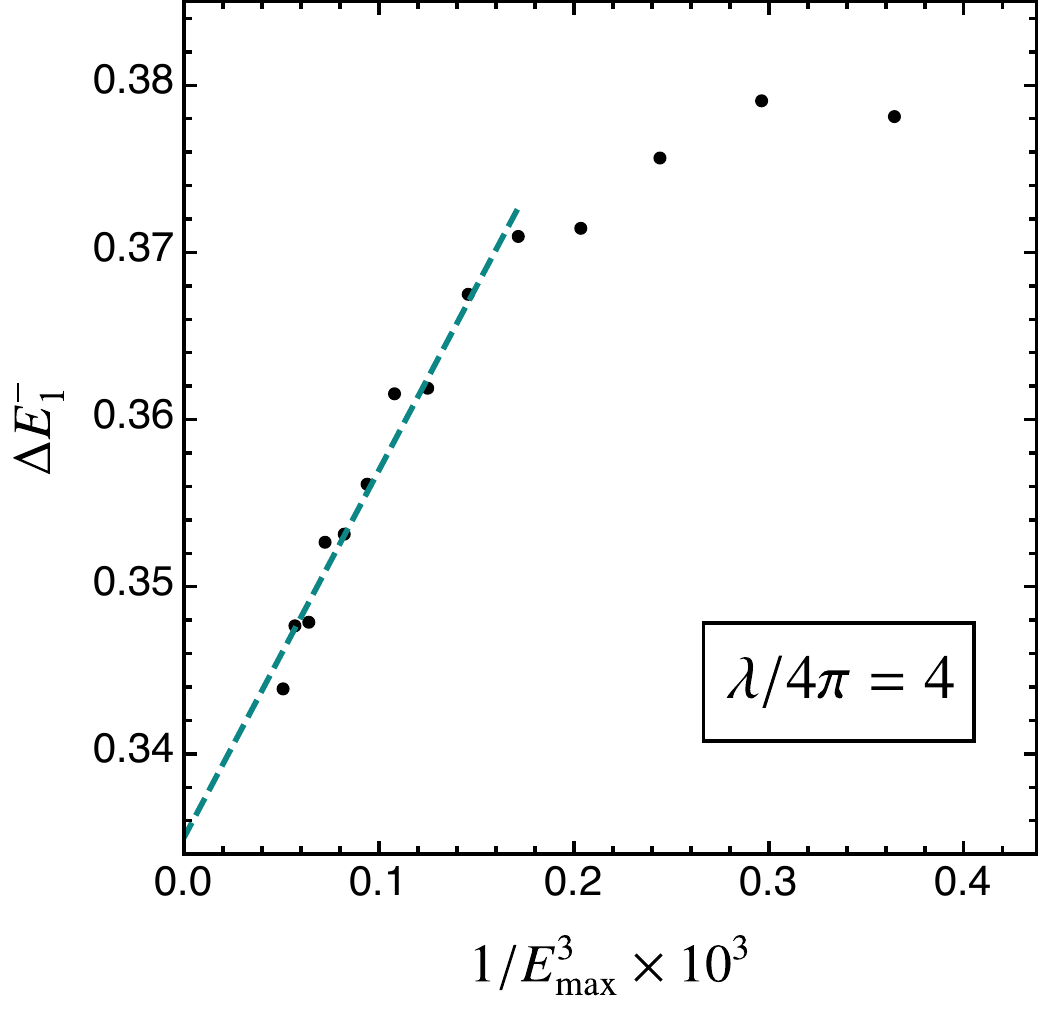}}
    \hspace*{.009cm}\raisebox{.03cm}{\includegraphics[trim= 0cm	.05cm	0cm	0cm, clip=true, width=0.239\textwidth]{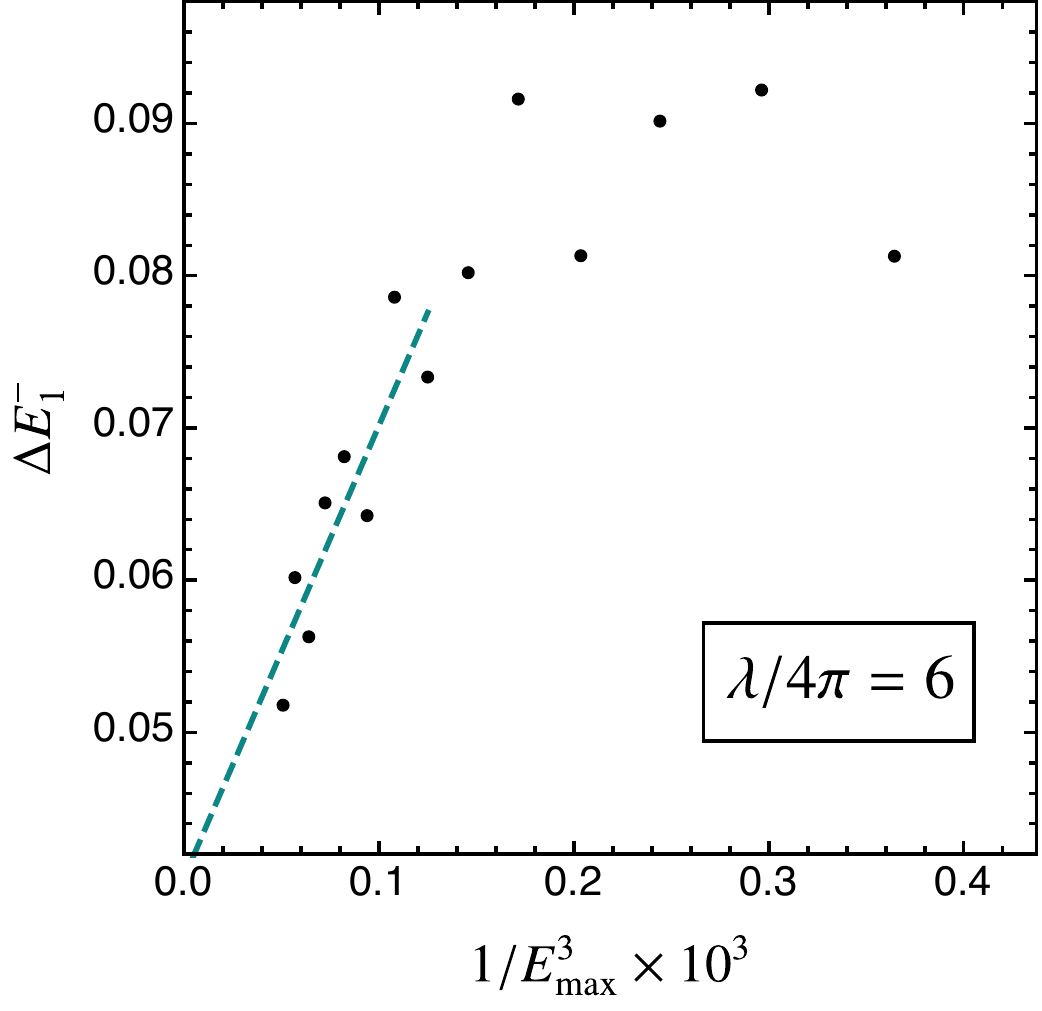}}
    \hspace*{-.021cm}\raisebox{.02cm}{\includegraphics[trim= 0cm	.05cm	0cm	0cm, clip=true, width=0.2455\textwidth]{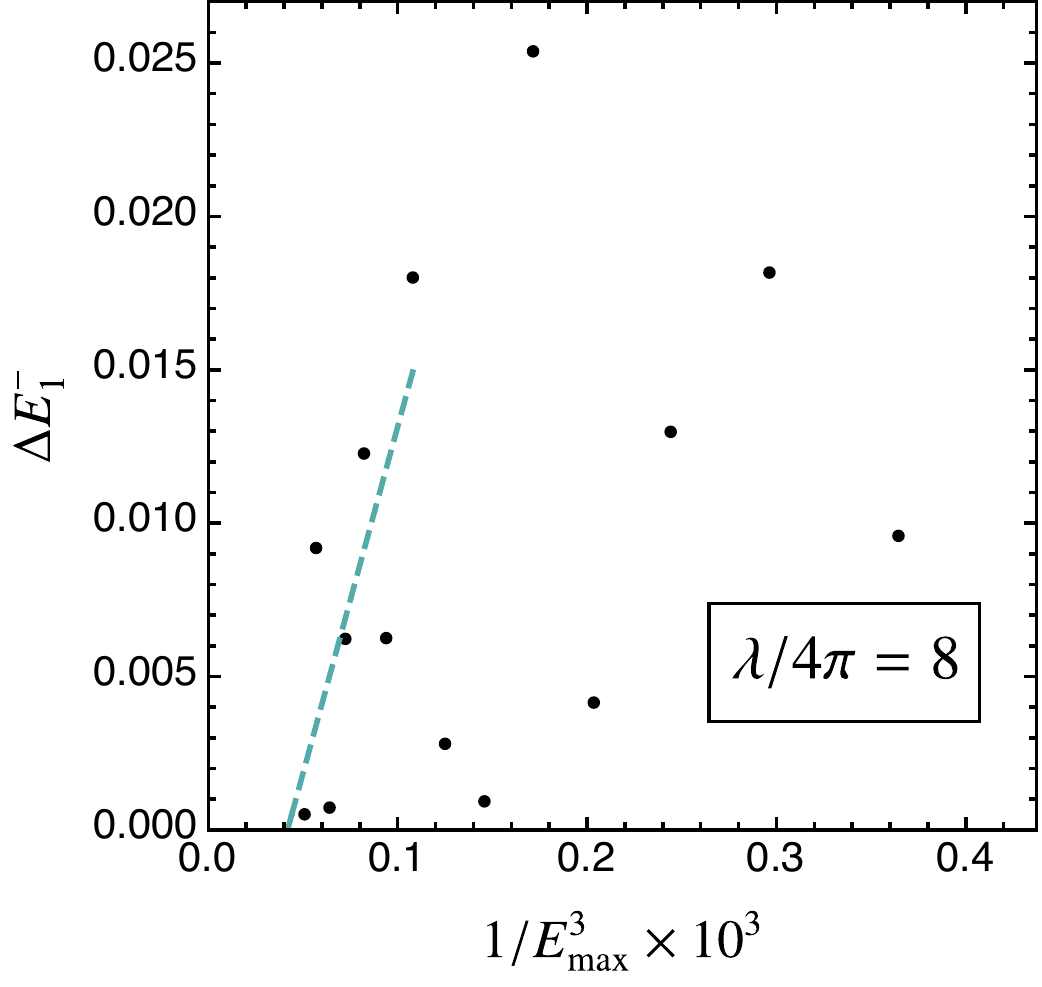}}  
    % - -.0208
%    \hfill
    \vspace{-12pt}
     \caption{The $\mathbb{Z}_2$-odd ground state excitation energy $\Delta E_1^-$ dependence on the energy cutoff $E_\text{max}$ for $\la/4\pi = 2, 4, 6, 8$. In all panels, $m_\text{NO} = 1$ and $2\pi R = 10$. The high $E_\text{max}$ tails of $\Delta E_1$ for the raw and improved theories are fit to $1/E_\text{max}^2$ and $1/E_\text{max}^3$, respectively. 
	The convergence is spoiled by numerical noise for larger
     values of $\la$.
	}
    \label{fig:E1-varyLam}
    \end{minipage}

    \centering
    \begin{minipage}{.9\textwidth}
    \centering
\vspace{45pt}
    \hspace{.0cm}\raisebox{-.0cm}{\includegraphics[trim= 0cm	0cm	0cm	0cm, clip=true,  width=0.2472\textwidth]{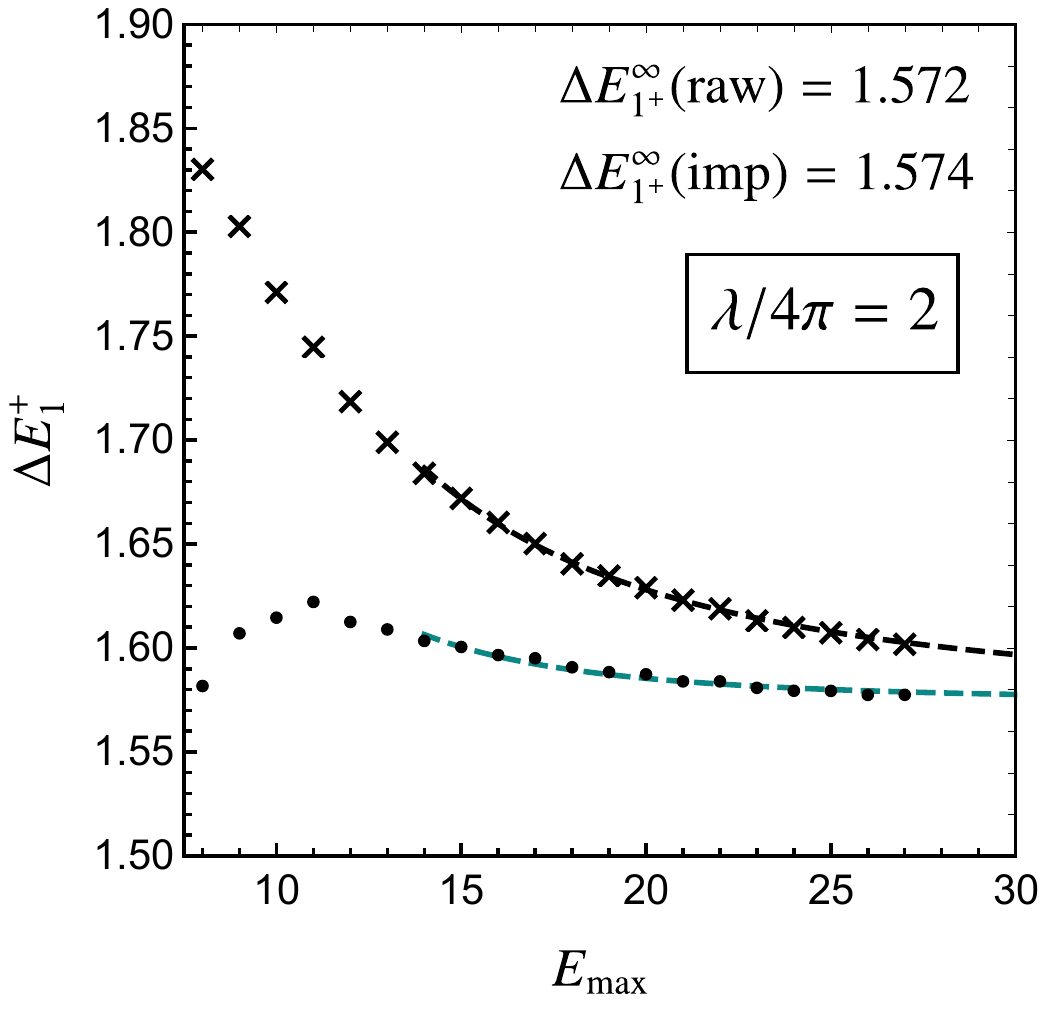}}
    \hspace{.00cm}\raisebox{.03cm}{\includegraphics[trim= 0cm	.05cm	0cm	0cm, clip=true, width=0.241226\textwidth]{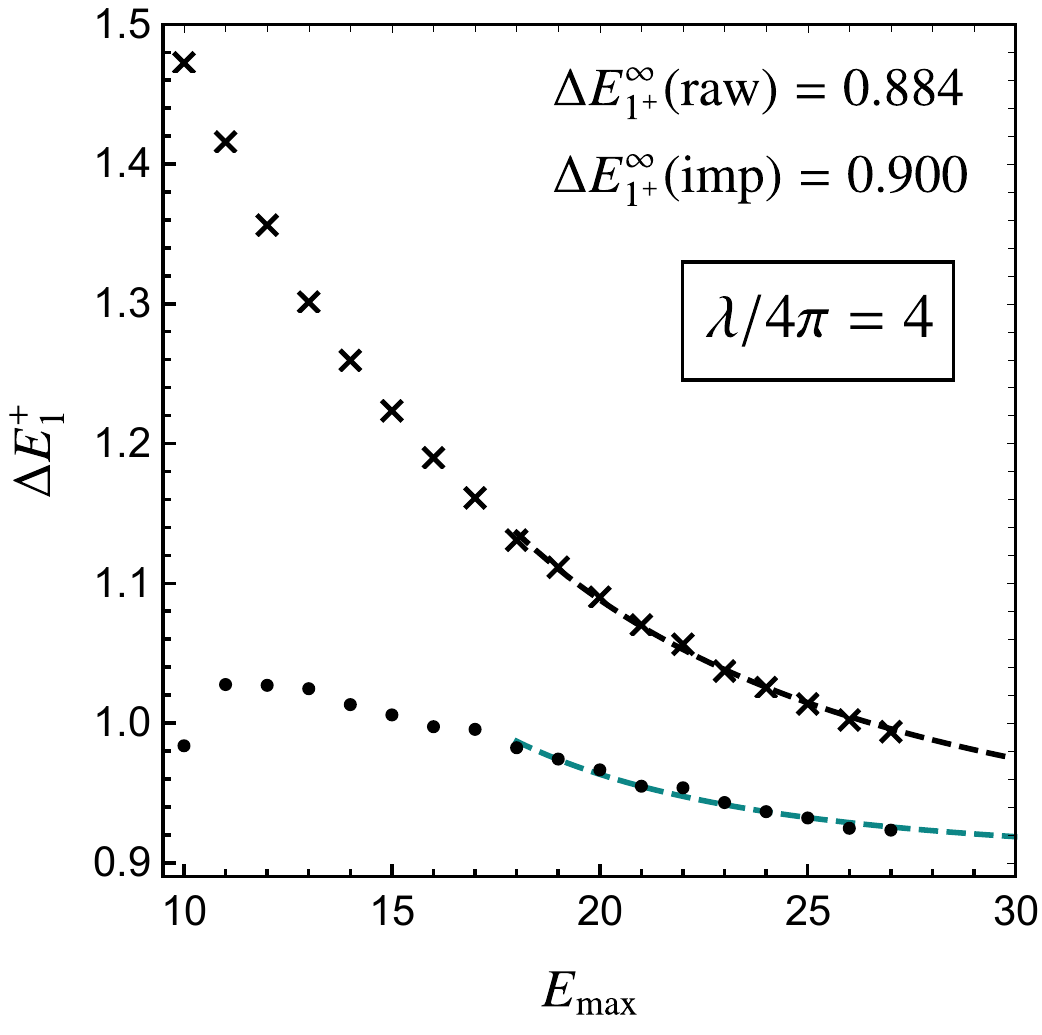}}
    \raisebox{.03cm}{\includegraphics[trim= 0cm	.05cm	0cm	0cm, clip=true, width=0.241329\textwidth]{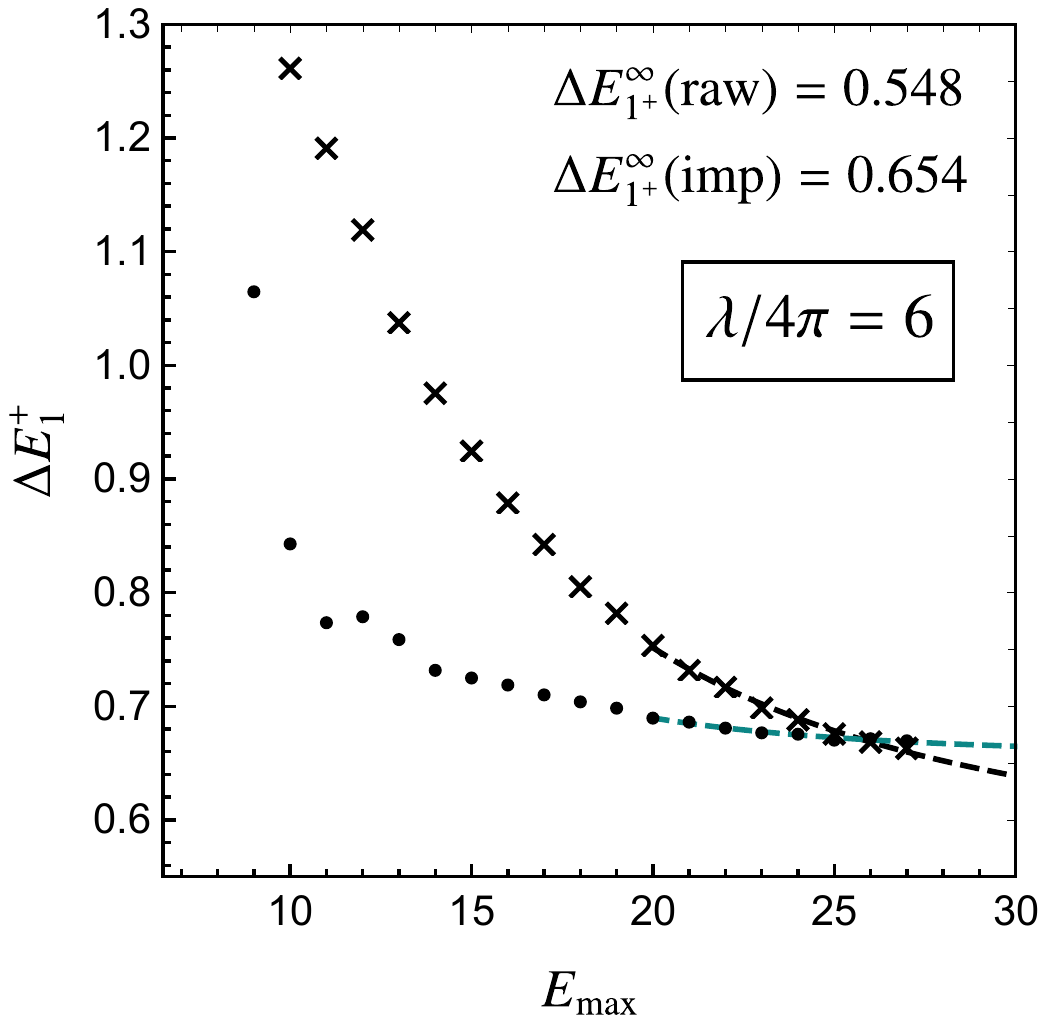}} 
    \hspace{0cm}\raisebox{.03cm}{\includegraphics[trim= 0cm	.05cm	0cm	0cm, clip=true, width=0.241226\textwidth]{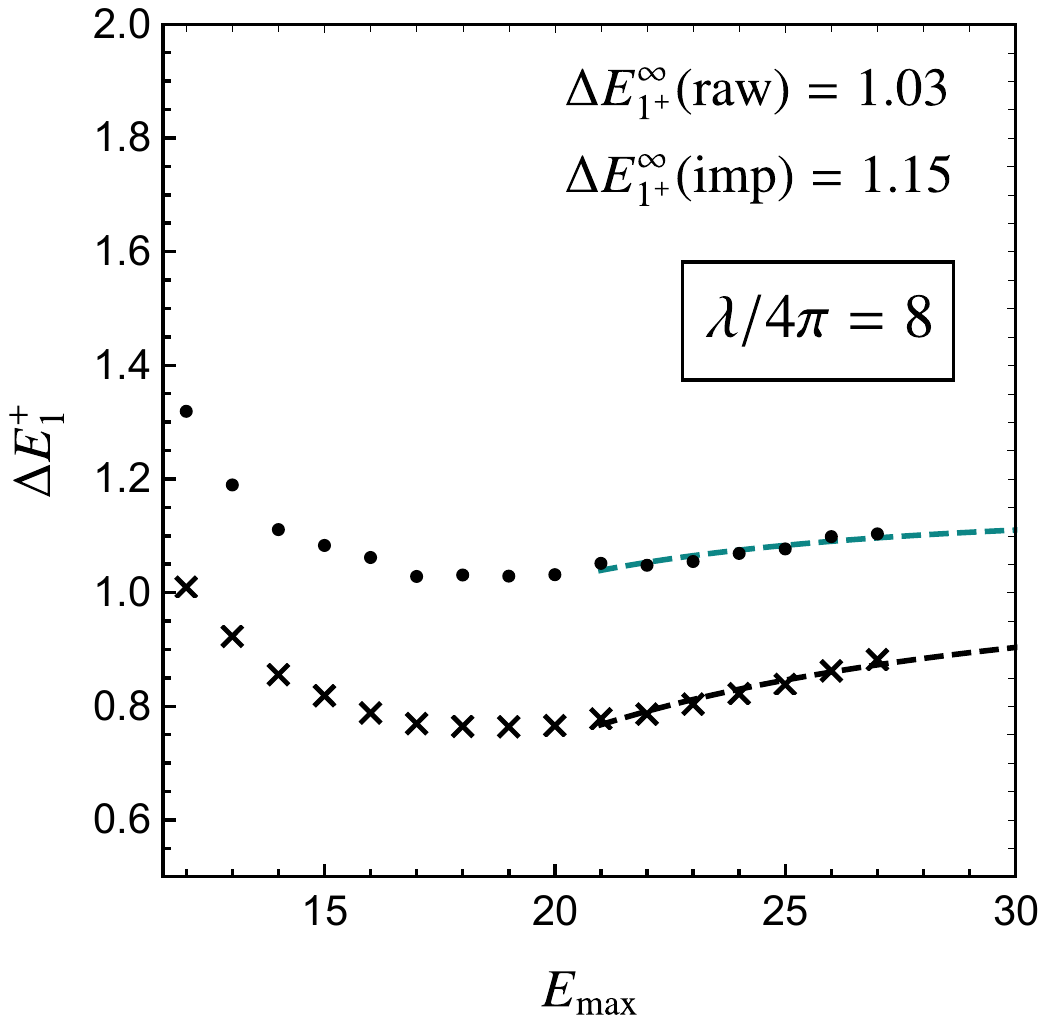}}
    \vfill
    \hspace{-.0cm}\raisebox{-.0cm}{\includegraphics[trim= 0cm	0cm	0cm	0cm, clip=true,  width=0.2412\textwidth]{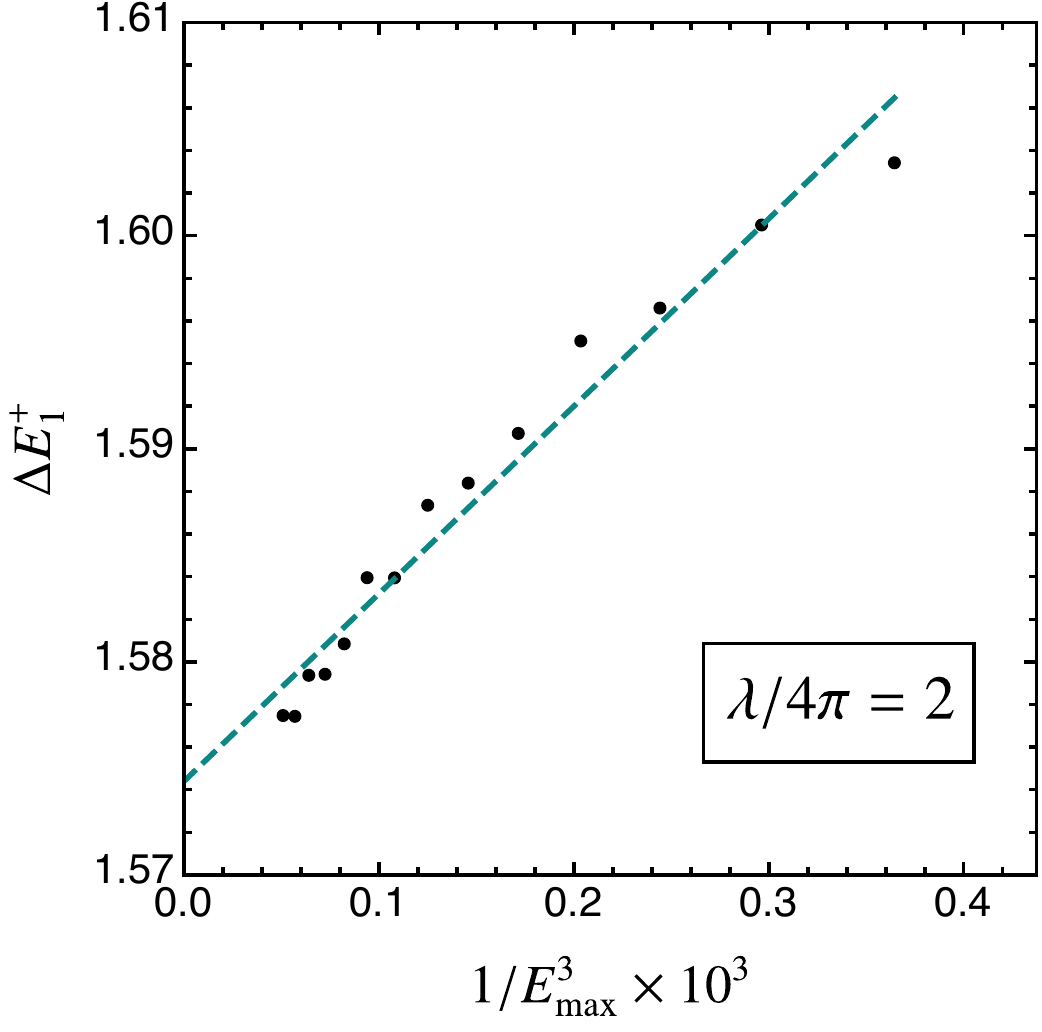}}
    \hspace{0.005cm}\raisebox{.02cm}{\includegraphics[trim= 0cm	.05cm	0cm	0cm, clip=true, width=0.2412\textwidth]{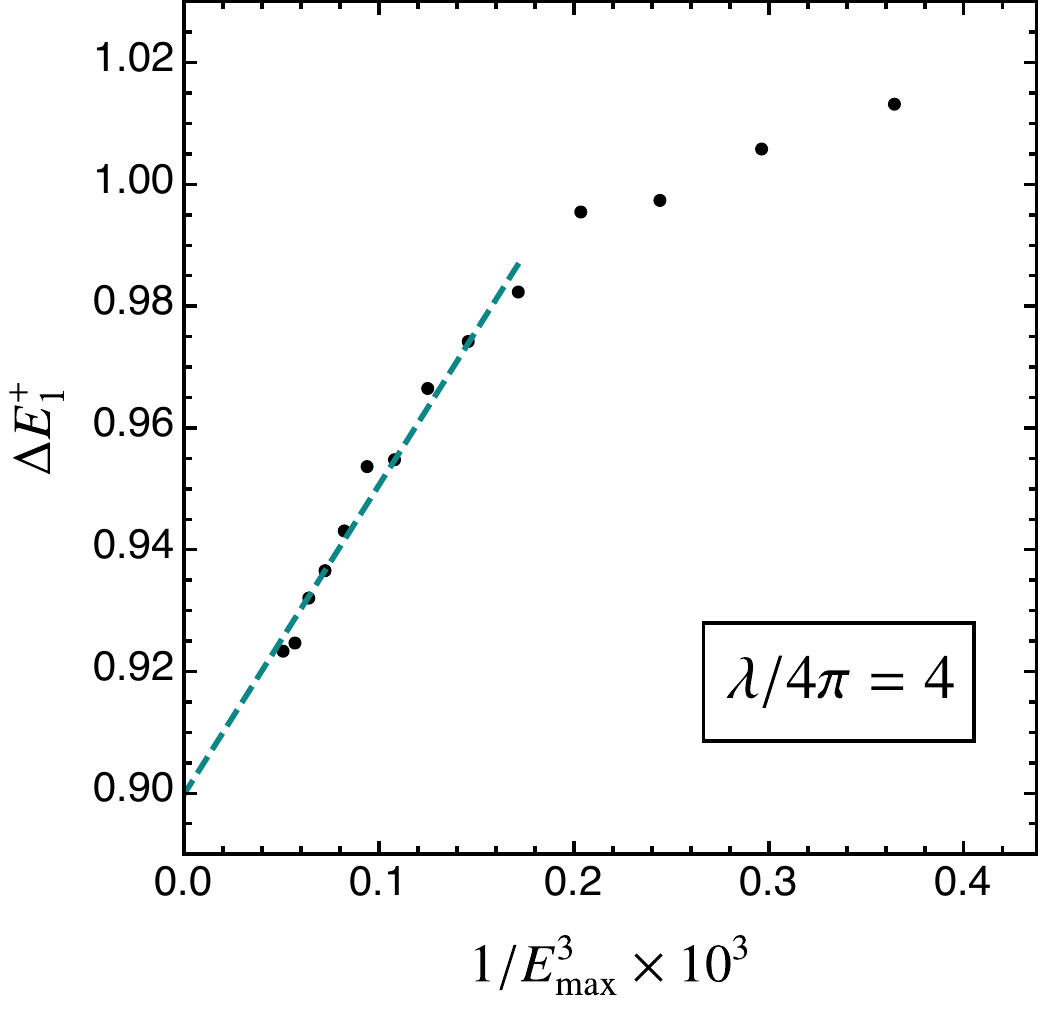}} 
    \hspace{0.0cm}\raisebox{.02cm}{\includegraphics[trim= 0cm	.05cm	0cm	0cm, clip=true, width=0.2415\textwidth]{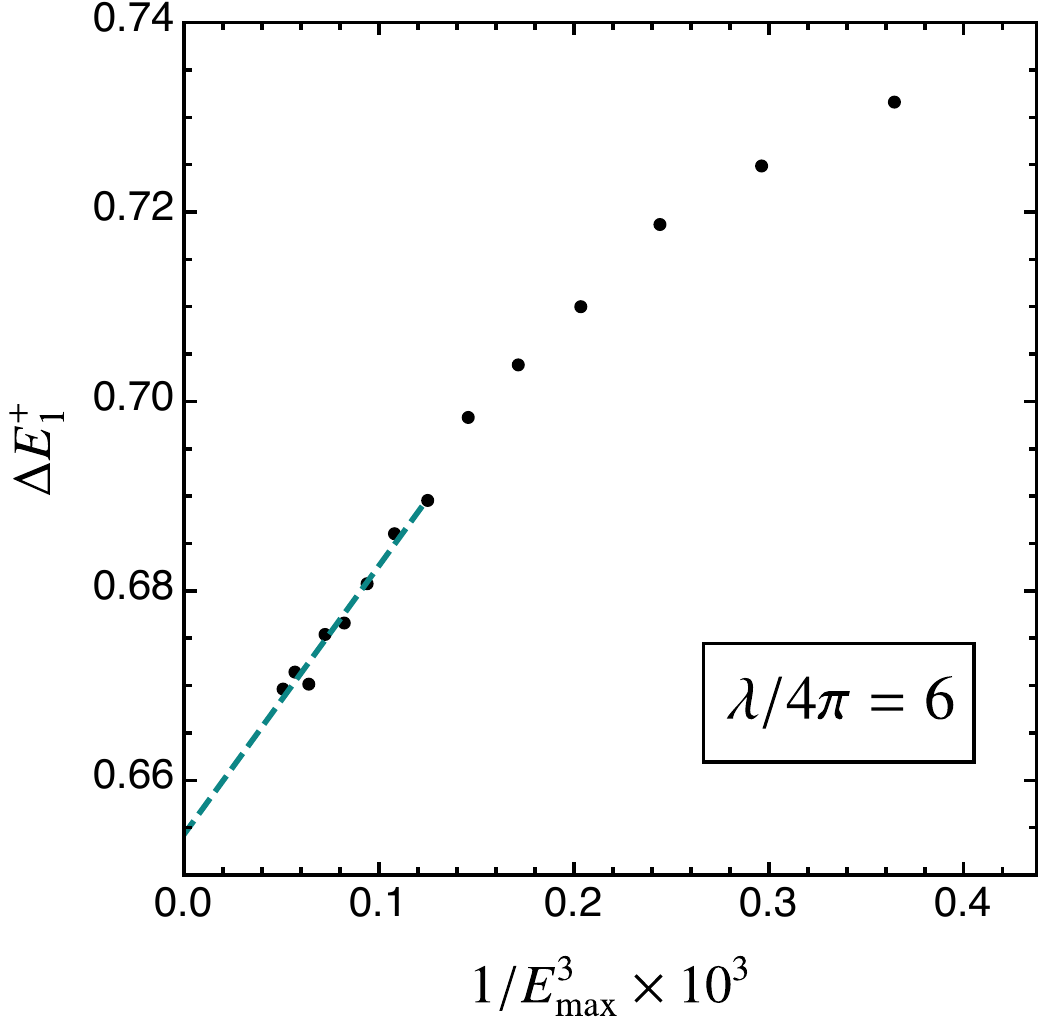}} 
    \hspace{-.00cm}\raisebox{.01cm}{\includegraphics[trim= 0cm	.05cm	0cm	0cm, clip=true, width=0.2412\textwidth]{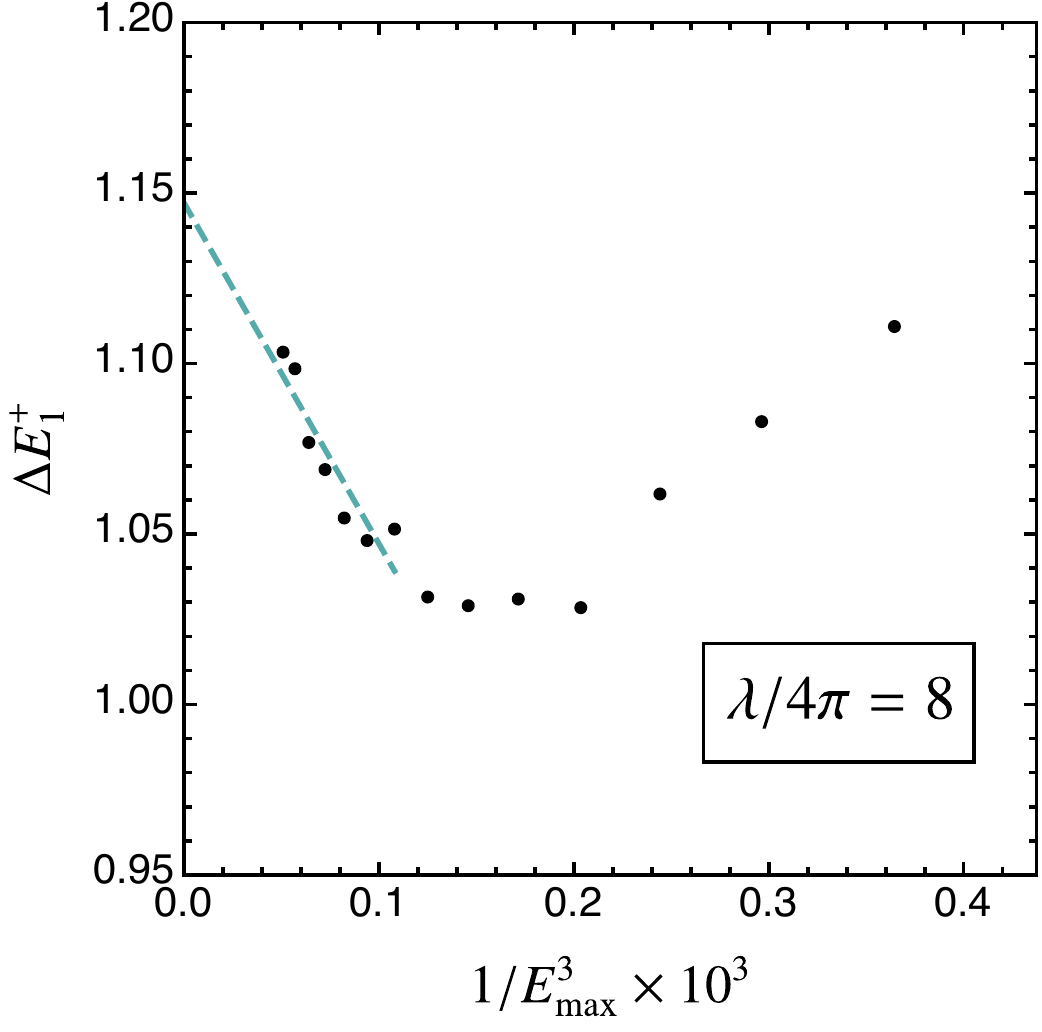}} 
     \caption{The $\mathbb{Z}_2$-even ground state excitation energy $\Delta E_1^+$ dependence on the energy cutoff $E_\text{max}$ for $\la/4\pi = 2, 4, 6, 8$. In all panels, $m_\text{NO} = 1$ and $2\pi R = 10$. The high $E_\text{max}$ tails of $\Delta E_1^+$ for the raw and improved theories are fit to $1/E_\text{max}^2$ and $1/E_\text{max}^3$, respectively. 
    The agreement between the asymptotic values of the raw and improved theories degrades for larger couplings.
     In the bottom panels, we see that the improved result maintains $1/E_\text{max}^3$ scaling to good approximation for strong couplings. 
	}
    \label{fig:E2-varyLam}
    \end{minipage}
\end{figure}

In \Fig{spectrum-vary-la} we present results for the first few excited
states as a function of the coupling $\la$.
The results are computed for $E_\text{max} = 27$, the largest value
used in our computations.
We see that the eigenvalues change significantly over the range
of $\la$ considered, a further indication that we are exploring
strong coupling.  
It is interesting to compare the results in \Fig{spectrum-vary-la}
with theoretical expectations.
It is known that the theory has a second-order phase transition
in the Ising universality class, where
the $\mathbb{Z}_2$ symmetry is spontaneously broken for
$\la$ larger than some critical value.
In finite volume, the signal of $\mathbb{Z}_2$ symmetry breaking is
that the states of the system become degenerate in pairs with
$Z = \pm$.
In \Fig{spectrum-vary-la} we see that this is happening at large $\la$ for the
ground state, but not for the excited states.
On the other hand, the convergence is not good at values of $\la$ far above
the critical point.
We see that the first few eigenvalues of the spectrum line up with values of the operator dimensions of the 2D Ising model
within the expected range of critical coupling \cite{Schaich:2009jk,Milsted:2013rxa,Rychkov:2014eea,Bajnok:2015bgw,Bosetti:2015lsa,Pelissetto:2015yha,Elias-Miro:2017tup,Serone:2018gjo,Heymans:2021rqo}.
It would be interesting to attempt to extract an accurate value for the
critical coupling using our methods taking into account finite volume
corrections, and compare to the results of other methods \cite{Schaich:2009jk,Milsted:2013rxa,Bosetti:2015lsa,Pelissetto:2015yha,Serone:2018gjo,Heymans:2021rqo}.
We leave this for the future.

In Figs. \ref{fig:E1-varyLam} and \ref{fig:E2-varyLam}, we present results on the convergence of $\Delta E_1$
 for $\la/4\pi = 2, 4, 6, 8$ in both the $\mathbb{Z}_2$-odd and -even bases.
The errors generally continue to scale as $1/E_\text{max}^3$
for larger values of $\la$. Numerical noise spoils the convergence of $\Delta E_1^{-}$ for $\la/4\pi=8$ (Fig.~\fig{E1-varyLam}), corresponding to the region where the first $\mathbb{Z}_2$-odd excited state and the vacuum are becoming degenerate.%
\footnote{The \texttt{scipy.sparse.linalg} matrix diagonalization uses the implicitly restarted Arnoldi method~\cite{Lehoucq1997arpack,arnoldi1951principle}, which may not be reliable for systems with nearly degenerate eigenvalues~\cite{baglama1998computation}.}

\begin{figure}[t!]
	\centering
	\begin{minipage}{.9\textwidth}
    \centering
    \includegraphics[trim= 0cm 0cm 0cm 0cm, clip=true, width=.9\textwidth]{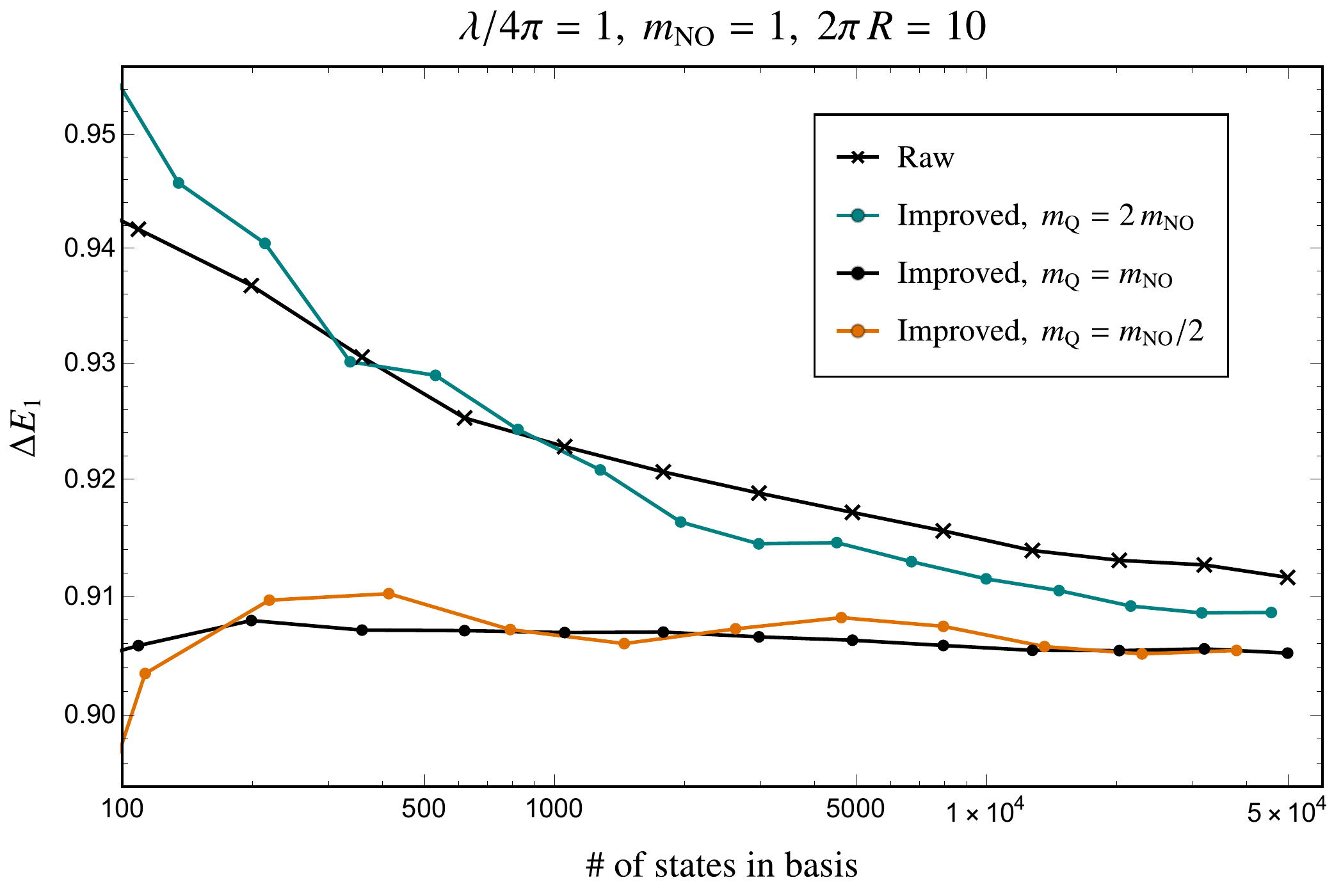}
    \hspace{.5cm}
    \caption{The ground state excitation energy $\Delta E_1$ of the raw theory [crosses] and the improved theory [dots] for $m_\text{Q} = 2 m_\text{NO},\ m_\text{NO},\ m_\text{NO}/2$. Here $\Delta E_1$ is plotted against the size of the basis to better compare the convergence given finite computational resources. The data points are connected by lines to guide the eye. This figure includes data points for theories with up to $5\times 10^4$ states in the basis. With this upper bound, we probe theories with $E_\text{max}$ up to $E_\text{max} = \{16,\ 20,\ 26\}$ for $m_\text{Q} = \{ 1/2,\ 1,\ 2\} \times m_\text{NO}$.
	}
    \label{fig:mQ_basis}
    \end{minipage}
\end{figure}

Finally, in \Fig{mQ_basis}
we show how our results change as a function of the quantization mass $m_\text{Q}$.
In our approach, $m_\text{Q}$ is an arbitrary variational parameter that can be used
to optimize the convergence.
It is defined by taking the particle energies in the Fock space to be
given by $\om_k = \sqrt{(k/R)^2 + m_\text{Q}^2}$ (see \sec{mQ}). 
To provide a `fair' comparison, we vary $m_\text{Q}$ keeping the number of
states in the truncation fixed, since this is what determines the required
computational resources.
Our previous results have been presented with $m_\text{Q} = m_\text{NO}$,
and in \Fig{mQ_basis} we compare these results with 
$m_\text{Q} = 2 m_\text{NO}$ and $m_\text{Q} = 
m_\text{NO}/2$ .
We see that the convergence appears to be best for $m_\text{Q} = m_\text{NO}$. 
Physically, we expect that the optimal value of $m_\text{Q}$ is to take it
equal to the physical mass scale of the theory, since then the Fock states
are expected to give the best variational basis for the interacting states.
It would be interesting to explore the optimal choice of $m_\text{Q}$ in more
complicated situations with more than one scale, for example taking
the volume to be larger to extrapolate to $R \to \infty$.
We also leave this for future work.

\section{Comparison with Previous Work}
\scl{Previous}
In this section, we compare our approach to other approaches to
improving the accuracy of Hamiltonian truncation in quantum field theory.

\subsection{Extending the Basis}
The first attempt to improve the predictions of Hamiltonian truncation
 for quantum field theory appears to be \Ref{Lee:2000gm}.
In this approach, the truncation to the space $\scr{H}_\text{eff}$ spanned
by $H_0$ eigenstates with eigenvalue $E \le E_\text{max}$ is the 
first step in an iterative procedure that aims to improve the convergence.
At each step, one has a finite-dimensional subspace $\scr{H}_\text{eff}^{(n)}$
that is updated by adding and removing vectors to improve the convergence.
Basis vectors are removed if their overlap with the (truncated) eigenvectors
is smaller than some chosen value, and a stochastic algorithm is used to
identify basis states with larger overlap.

This method is entirely numerical, and does not exploit 
the fact that the theory is weakly coupled in the UV.
Also, it may be difficult to extend this method to theories
with UV divergences.
In such theories, the total contribution of modes with high energy
grows with the energy, due to the exponential increase of the
number of modes.
The contribution of individual states above the truncation cutoff 
is exponentially small, but their total contribution is infinite.

\subsection{Renormalization Group Equation}
Another approach to improving Hamiltonian truncation was proposed
in the context of 2D conformal theories perturbed by relevant operators in 
\Refs{Feverati:2006ni,Giokas:2011ix}.
The idea is to define a renormalization
group (RG) equation for the coupling constants of the theory, namely the
coefficients of the relevant operators.
(In the case of 2D $\la \phi^4$ theory, these are the coupling
$\la$ and the mass $m^2$.)
The RG equation is determined by requiring that certain correlation
functions are independent of the truncation cutoff.
Because the perturbations are relevant, the RG equations can be
computed in a perturbative expansion in the couplings.

As in our paper, their approach takes advantage of the weak
coupling of the theory in the UV, and computes a correction to the
truncated Hamiltonian by a perturbative matching calculation.
Our work can be viewed as an extension of this approach that systematically
includes additional terms in the truncated Hamiltonian.

\subsection{Exact Effective Hamiltonian}
\scl{ExactEffH}
This approach was introduced in \Refs{Hogervorst:2014rta,Rychkov:2014eea},
and further developed in \Ref{Elias-Miro:2017tup}.
The starting point is the observation that one can 
rewrite the exact eigenvalue equation \Eq{Heigenvalue}
in terms of an `exact effective Hamiltonian' (our terminology)
acting on the low-energy subspace $\scr{H}_\text{eff}$:
\[
H_\text{exact}(\scr{E}) \ket{\scr{E}}_\text{eff}
= \scr{E} \ket{\scr{E}}_\text{eff},
\eql{Hexact}
\]
where $\ket{\scr{E}}_\text{eff} \in \scr{H}_\text{eff}$.
The exact effective Hamiltonian is given by
\[
H_\text{exact}(\scr{E}) &= H_0 + V + \De H_\text{exact}(\scr{E}),
\]
with
\[
\bra{f} \De H_\text{exact}(\scr{E}) \ket{i}
&= -\bra{f} V P_> (H_0 + P_> V P_> - \scr{E})^{-1} P_> V \ket{i},
\eql{exactHeff}
\]
where $\ket{i}, \ket{f} \in \scr{H}_\text{eff}$
and $P_>$ is the projector onto states with $E > E_\text{max}$.

Note that $\De H_\text{exact}(\scr{E})$ depends on the eigenvalue one
is trying to compute.
In principle, one can vary $\scr{E}$ iteratively to converge to the correct
eigenvalue, but this is numerically expensive.
Instead \Refs{Hogervorst:2014rta,Rychkov:2014eea} diagonalize
$H_\text{exact}(\scr{E}_*)$, where $\scr{E}_*$ is a reference value
chosen to be close to the low-lying eigenvalues one is computing.
Next, these authors evaluate
the \rhs\ of \Eq{exactHeff}
using an approximation that is valid for $E_\al \gg E_\text{max}$.
The resulting leading corrections to the effective Hamiltonian
are local, like the leading $O(V^2)$ corrections we
computed in \sec{Matching2DTheory}.
Numerical diagonalization of this approximation to 
$H_\text{exact}(\scr{E}_*)$ for 2D $\la\phi^4$
theory shows significantly improved convergence, but the errors are larger than the 
$O(1/E_\text{max}^3)$ errors we obtain from our method.

In \Ref{Elias-Miro:2017tup}, an extension of this method was found that
does give $O(1/E_\text{max}^3)$ errors for 2D $\la\phi^4$ theory.
This was accomplished by adding judiciously chosen
additional states above the cutoff (called `tail states' by the authors).
Our approach obtains $O(1/E_\text{max}^3)$ errors without enlarging the basis.

\subsection{Counterterms in Theories with UV Divergences}
\scl{Previous3D}
\Refs{EliasMiro:2020uvk,Anand:2020qnp} were the first to `break the UV divergence barrier' and
successfully apply Hamitonian truncation to
a theory with UV divergences, namely $\la \phi^4$ theory in 3D.
In such theories, the `raw' truncation approximation
$H_{\text{eff}} = H_0 + V$ does not converge as $E_\text{max}$ is 
increased.

\Ref{EliasMiro:2020uvk} analyzed perturbation theory for the energy eigenvalues and identified
UV divergent contributions arising from sums over states with energy 
above $E_\text{max}$.
They then defined a renormalized effective Hamiltonian by subtracting these
contributions, and demonstrated that numerical diagonalization of this
Hamiltonian gives sensible results.

\Ref{Anand:2020qnp} analyzed the theory using a different truncation 
applied to the lightcone Hamiltonian. 
They identified contributions with unphysical non-local cutoff dependence, 
which they interpreted as artifacts of their truncation and therefore subtracted. 
Numerical diagonalization of this
Hamiltonian gives also gives sensible results.

As we discussed in \sec{matchspectrum},
matching the energy eigenvalues is not sufficient to completely
determine the effective Hamiltonian.
Therefore, our work can be viewed as a systematic extension of the
methods of \Refs{EliasMiro:2020uvk,Anand:2020qnp} that allows further 
improvement of the results.

\section{Conclusions}
\scl{Conc}
In this paper, we have formulated Hamiltonian truncation as an effective
field theory (HTET) with a cutoff on the total energy $E_\text{max}$.
This allows us to perform matching calculations to compute corrections to
the effective Hamiltonian as a systematic expansion in powers of $1/E_\text{max}$
using a diagrammatic approach. 
We demonstrated this method by applying it to $\la \phi^4$ theory in 2D.
We computed the leading corrections to the `raw' Hamiltonian truncation in
this model, and carried out numerical tests of the improved truncation.
The calculation involves 2- and 3-loop diagrams with overlapping UV/IR
dominated regions, and we demonstrated that the results are compatible
with the separation of scales, as we expect in effective field theory.
We also performed calculations up to 2 loops in 3D $\la\phi^4$ theory to 
illustrate the application of our methods in theories with 
nontrivial UV divergences.
Our numerical studies of the 2D theory showed that the corrected theory
has an error that scales as $1/E_\text{max}^3$, the theoretically expected
scaling from the power counting of the effective theory.
These results are an indication that our method is working as expected.

There are many directions to explore in future work.
We begin by mentioning several projects that are straightforward
applications of the formalism developed in this paper.
One is to compute the next order corrections to
2D $\la\phi^4$ theory, where the effective Hamiltonian is 
neither local nor Hermitian.
This calculation would give a nontrivial test of the power counting
we proposed for the effective theory.
At this order, the truncation error is expected to scale as $1/E_\text{max}^4$, and
we expect to get predictions accurate to 5 significant digits.
Another project is to finish the calculations  for the leading corrections
to 3D $\la\phi^4$ theory, together with a numerical analysis
to demonstrate the method in a theory in higher dimensions
with nontrivial UV divergences.
A complementary direction is to use Hamiltonian truncation to extract precision
predictions for physical observables, such as the critical coupling,
or physical masses and scattering amplitudes away from criticality.
This would require a careful treatment of finite volume effects and
extrapolation errors.

We also suggest some more speculative directions.
The ultimate goal of Hamiltonian truncation is to improve on existing
numerical methods to study quantum field theories, such as lattice Monte Carlo.
Currently, we are very far from this goal.
For example, there is no general method to treat gauge theories in 
Hamiltonian truncation, and there
has been no successful Hamiltonian truncation calculation in 4 spacetime dimensions.
Hamiltonian truncation suffers from
the fact that the number of states grows exponentially with $E_\text{max}$,
with an exponent that grows with the spacetime dimension.
Nonetheless, we are optimistic.
For example, quantum computers may eventually be able to overcome the exponential
growth in computational complexity in Hamiltonian truncation.
Until then, there are interesting 2D and 3D models that can be studied
using Hamiltonian truncation, including theories that are not accessible
to lattice techniques, such as theories with chiral fermions or supersymmetry.
It is our hope that the conceptual power of understanding Hamiltonian truncation
in the language of effective field theory will help us make 
progress on all these outstanding problems.

\section*{Acknowledgments}

We thank Joan Elias-Mir\'o,
Marat Freytsis,
Brian Henning,
Emanuel Katz,
and
Matthew Walters
for useful discussions,
and Slava Rychkov for comments on the draft. 
The referees contributed a number of useful comments: we thank 
Balt van Rees, Slava Rychkov, and an anonymous referee for their input.
T.~Cohen is supported by the DOE under grant DE-SC-0011640. 
K.~Farnsworth is supported by the Simons Foundation 
under award 658908.  
R.~Houtz is supported by the STFC under grant ST/P001246/1.
M.~A.~Luty is supported by the DOE under grant DE-SC-0009999.

\startappendices
\section*{Appendices}
\section{Renormalization and Matching in 3D $\la \phi^4$ Theory}
\scl{3DTheory}
In this section, we present some renormalization and matching
calculations for 3D $\la\phi^4$ theory.
Although this theory is also super-renormalizable, its UV divergence
structure is more complicated than in 2D $\la\phi^4$ theory.
The calculations in this section illustrate 
that separation of scales continues to work in
theories with UV divergences.

\subsection{Power Counting}
The mass dimensions of the field and couplings in 3D are 
\[
[\phi] = \frac 12,
\qquad
[\la] = 1.
\]
The most general parametric form of the effective Hamiltonian allowed by the power
counting presented in \sec{PowerCounting} is 
\[
H_2 + H_3 + \cdots &\sim \myint \mathrm{d}^2 x \ggap \bigg\{
\la^2 \bigg( E_\text{max} \cdot \id
+ H_0 \ln E_\text{max} + \frac{H_0^2}{E_\text{max}} + \cdots \biggr)
\nn[3pt]
&\qquad\qquad{}
+ \la^2 \phi^2 \biggl( \ln E_\text{max} + \frac{H_0}{E_\text{max}} + \cdots
\biggr)
\nn[3pt]
&\qquad\qquad{}
+ \la^2 \big( \phi^4 + \ii \phi\dot\phi \big)
\biggl( \frac{1}{E_\text{max}}
+ \frac{H_0}{E_\text{max}^2} 
+ \cdots \biggr) 
\nn[3pt]
&\qquad\qquad{}
+ \la^3 \biggl( \ln E_\text{max} \cdot \id
+ \frac{H_0}{E_\text{max}} 
+ \cdots \biggr) 
\nn[3pt]
&\qquad\qquad{}
+ \la^3 \phi^2 \bigg( \frac{1}{E_\text{max}}
+ \frac{H_0}{E_\text{max}^2} + \cdots \bigg)
\cdots \bigg\}.
\]
Note that the power counting suggests that the coefficients of 
$\la^2 H_0$ and $\la^2 \phi^2$
increase logarithmically with $E_\text{max}$.
We therefore expect that the `raw' truncation
$H_\text{eff} \simeq H_0 + H_1$ does not converge
as $E_\text{max}$ is increased as opposed to the 2D theory,
but including the leading $O(\la^2)$ terms that grow with $E_\text{max}$
will give a truncation with errors of order $1/E_\text{max}$.
\Refs{EliasMiro:2020uvk,Anand:2020qnp} successfully carried out
numerical studies of this theory using a different renormalization
method; we compare our approach with theirs in \sec{Previous3D}.

\subsection{Renormalization of the Fundamental Theory}
We write the bare Lagrangian as
\[
\scr{L} = \frac 12 (\d\phi)^2 - \frac 12 m_0^2 - \frac{\la}{4!} \phi^4.
\]
We again regulate the fundamental theory by imposing a momentum
space cutoff $\La$:
\[
\sum_{\vec{k}} \quad\to\quad \sum_{\vec{k}}^{\La} \equiv \sum_{|\vec{k}| \, \le \, \La R} ,
\]
where the sum is over $\vec{k} \in \mathbb{Z}^2$.
This cutoff breaks Lorentz invariance, and therefore we must allow counterterms
that break Lorentz invariance.
However, the cutoff is local, so the counterterms are given by local terms:
\[
\de \scr{L} &\sim 
\la \phi^2 \La
+ \la^2 \phi^2 \ln \La
+ \text{finite},
\]
where we have omitted cosmological constant terms that are independent of $\phi$.
We see that there is no need for wavefunction or coupling constant 
renormalization, as in the 2D theory.
Since we are only interested in differences of energy eigenvalues, we ignore
the vacuum energy divergences and focus on the terms proportional to $\phi^2$.

The Hamiltonian is $H_0 + V$, where
\begin{subequations}
\[
H_0 &= \sum_{\vec{k}}^\La \om\sub{\vec{k}} \ggap a_{\vec{k}}^\dagger a\sub{\vec{k}},
\qquad
\om_{\vec{k}} = \sqrt{|\vec{k}|^2/R^2 + m_\text{Q}^2},
\\[4pt]
V &= \myint \mathrm{d}^2 x \ggap \biggl[ 
\frac 12 (m_V^2 + \de m_V^2) \no{\phi^2}
+ \frac{\la}{4!} \no{\phi^4} \biggr].
\]
\end{subequations}
where
\[
m_V^2 + \de m^2_V = m_0^2 - m_\text{Q}^2
+ \frac{\la}{16\pi^2 R^2} \sum_{\vec{k}}^\La \frac{1}{\om_{\vec{k}}}.
\]
Here $\de m^2_V$ is a counterterm that will be used to absorb the 
$\la^2 \phi^2 \ln\La$ divergence.
The diagrammatic rules for the vertices are then
\begin{subequations}
\[
\eql{3Dvertexrules}
\includegraphics[valign=c,scale=0.75]{Figs/phi4_vert_k_f} 
&= \frac{\la}{(2\pi R)^2} \ggap \de_{\vec{k}_1 + \cdots + \vec{k}_4},
\\[4pt]
\includegraphics[valign=c,scale=0.75]{Figs/phi2_vert_k_f} 
&= m_V^2 
\de_{\vec{k}_1 \vec{k}_2},
\\[4pt]
\includegraphics[valign=c,scale=0.98]{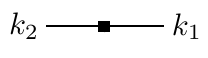}  
&= \de m^2_V \ggap \de_{\vec{k}_1 \vec{k}_2}.
\]
\end{subequations}

As in the 2D theory, we also define a running renormalized mass by subtracting
the contribution of modes with $k > \mu/R$:
\[
m_\text{R}^2(\mu) + \de m^2(\mu)
= m_0^2 + \frac{\la}{16\pi^2 R^2} 
\sum_{ |\vec{k}| \, > \mu R}^\La \frac{1}{\om_{\vec{k}}},
\]
where the $O(\la^2)$ $\de m^2$ is allowed to depend on the
renormalization scale $\mu$.
We will see that separation of scales is manifest in terms of the 
renormalized coupling $m_\text{R}^2$.
Note that if we choose the $O(\la^2)$ counterterms to satisfy
$\de m_V^2 = \de m^2(\mu = 0)$, we have
\[
\eql{mv3d}
m_V^2   = m_\text{R}^2(\mu) - m_\text{Q}^2+ \frac{\la}{16\pi^2 R^2} 
\sum_{|\vec{k}| \, \le \, \mu R} \frac{1}{\om_{\vec{k}}},
\]
just as in the 2D theory.

The $O(\la^2)$
diagrams that we must compute to renormalize the $\phi^2$ term are given
in \Eq{V2diagrams}.
The power counting argument above shows that these contributions are independent
of the external momenta and energies, so we can compute them in the local
approximation.
We obtain%\hspace{-3pt}
\footnote{The results can be read off from the 2D results for the same
diagrams with the replacements $\la \to \la/2\pi R$,
and $\sum_k \to \sum_{\vec{k}}$.}
\[
\eql{sunset3D}
\!\!\!\!
&\includegraphics[valign=c,scale=0.65]{Figs/sunset_f} 
+ \includegraphics[valign=c,scale=0.65]{Figs/sunset2_f}
+ \includegraphics[valign=c,scale=0.65]{Figs/sunset3_f}
+ \includegraphics[valign=c,scale=0.65]{Figs/sunset4_f}
\nn[4pt]
&\qquad\quad{}
\simeq -\frac{\la^2}{768\pi^4 R^4}
\myint \mathrm{d}^2 x\ggap \bra{f} \no{\phi^2} \ket{i}
\times \sum_{\vec{k}_{1,2,3}}^{\La}   \frac{\de_{123,0}}{\om_1 \om_2 \om_3 (\om_1 + \om_2 + \om_3)}.
\]
This diagram contains mixed UV/IR regions where one of the 3 momenta is much
smaller than the others.
In this region, we have
\[
\sum_{\vec{k}_{1,2,3}}^\La  \frac{\de_{123,0}}{\om_1 \om_2 \om_3 (\om_1 + \om_2 + \om_3)}
\simeq 3 \times \sum_{|\vec{k}| \, \gg \, \mu R}
\frac{1}{2 \om_{\vec{k}}^3}
\sum_{|\vec{k}'| \, \ll \, \mu R}
\frac{1}{\om_{\vec{k}'}} + \cdots,
\]
which is finite as $\La \to \infty$.
Therefore, there are no overlapping UV divergences in this diagram.
The remaining diagrams are given by
\[
& \includegraphics[valign=c,scale=0.65]{Figs/snail_f}
+ \includegraphics[valign=c,scale=0.65]{Figs/snail2_f}
+ \includegraphics[valign=c,scale=0.65]{Figs/snail3_f}
+ \reflectbox{\includegraphics[valign=c,scale=0.65]{Figs/snail_f}}
+ \reflectbox{\includegraphics[valign=c,scale=0.65]{Figs/snail2_f}}
+ \reflectbox{\includegraphics[valign=c,scale=0.65]{Figs/snail3_f}}
\nn[4pt]
&\qquad\quad{}
\simeq -\frac{\la \gap m_V^2}{64\pi^2 R^2}
\myint \mathrm{d}^2 x \ggap \bra{f} \no{\phi^2} \ket{i}
\times \sum_{\vec{k}}^\La \frac{1}{\om_{\vec{k}}^3},
\]
which is also finite.

We conclude that the only UV divergent $\phi^2$ term comes from the region where 
all 3 momenta in \Eq{sunset3D} are large.
We therefore define the counterterm $\de m^2(\mu)$ to cancel the contribution
of this region:
\[
\eql{m2ct3D}
\de m^2(\mu) = \frac{\la^2}{768\pi^4 R^4}
\sum_{|\vec{k}_{1,2,3}| > \mu R }^\La  
\frac{\de_{123,0}}{\om_1 \om_2 \om_3 (\om_1 + \om_2 + \om_3)}
\sim \ln(\La/\mu).
\]
Note we use the renormalization scale $\mu$ to define the counterterm. At $O(\la^2)$, we also have contributions to the $\phi^4$ vertex of the form (in the local approximation)
\[
 &\includegraphics[valign=c,scale=0.65]{Figs/V221_f}
+ \includegraphics[valign=c,scale=0.65]{Figs/V222_f}
+ \includegraphics[valign=c,scale=0.85]{Figs/V227_f}
+ \includegraphics[valign=c,scale=0.65]{Figs/V223_f}
+ \includegraphics[valign=c,scale=0.65]{Figs/V224_f}
+ \includegraphics[valign=c,scale=0.65]{Figs/V225_f}
+ \includegraphics[valign=c,scale=0.65]{Figs/V226_f} 
\nn[4pt]
&\qquad\quad{}
\simeq -\frac{\la^2}{256\pi^2 R^2}
\myint \mathrm{d}^2 x \ggap \bra{f} \no{\phi^4} \ket{i}
\times \sum_{\vec{k}}^\La \frac{1}{\om_{\vec{k}}^3} ,
\]
which is finite.

\subsection{Matching and Separation of Scales at $O(V)$}
We now turn to the matching calculation of the effective Hamiltonian,
starting at $O(V)$.
Using the same notation as the 2D theory, we write
\[
H_1 &= \myint \mathrm{d}^2 x \ggap \biggl[
\frac 12 m_1^2 \phi^2
+ \frac{\la_1}{4!} \phi^4
\biggr]
= \myint \mathrm{d}^2 x \ggap \biggl[
\frac 12 m_{V1}^2
\no{\phi^2}
+ \frac{\la_\text{eff}}{4!} \no{\phi^4}
\biggr],
\]
where 
\[
m_{V1}^2 = m_1^2 + \frac{\la_1}{16\pi^2 R^2} \sum_{|\vec{k}| \, \le \, k_\text{max}}
\frac{1}{\om_{\vec{k}}}
\]
and $\om_{\vec{k}_\text{max}} = E_\text{max}$.
Matching the coefficients of the normal-ordered operators then gives
\[
\la_\text{eff} &= \la,
\qquad
m_{V1}^2 = m_V^2.
\]
As in 2D, $m^2_{V1}$ is independent of $\mu$, and
the coefficient of $\phi^2$ in $H_\text{eff}$ at $O(\la)$ is given by
\[
m_\text{Q}^2 + m_1^2 = m_\text{R}^2(\mu = k_\text{max}/R).
\]
We see that separation of scales is manifest in terms of 
$m^2_\text{R}(\mu \sim E_\text{max})$, just as in the 2D theory.

\subsection{Matching and Separation of Scales at $O(V^2)$}
Using the local approximation, the 2-loop diagrams are given by
\[
&\includegraphics[valign=c,scale=0.65]{Figs/sunset_f} + \cdots
- \bigg[ \includegraphics[valign=c,scale=0.65]{Figs/sunset_f} + \cdots
\bigg]_\text{eff}
\nn[4pt]
&\qquad{}
\simeq -\frac{\la^2}{768\pi^4 R^4}
\myint \mathrm{d}^2 x\ggap \bra{f} \no{\phi^2} \ket{i}
\times \sum_{\vec{k}_{1,2,3}}^\La \de_{123} \frac{\Th(\om_1 + \om_2 + \om_3 - E_\text{max})}
{\om_1 \om_2 \om_3 (\om_1 + \om_2 + \om_3)}.
\]
This has a UV divergence that is canceled by the counterterm \Eq{m2ct3D} in the
fundamental theory.
The 1-loop diagram contributions to $\phi^2$ are given by
\[
\!\!\!\!\!\!
&\includegraphics[valign=c,scale=0.65]{Figs/snail_f} + \cdots
- \bigg[ \includegraphics[valign=c,scale=0.65]{Figs/snail_f} + \cdots
\bigg]_\text{eff}
\nn[2pt]
&\qquad{}
\simeq -\frac{\la \gap m_V^2}{64\pi^2 R^2}
\myint \mathrm{d}^2 x \ggap \bra{f} \no{\phi^2} \ket{i}
\sum_{\vec{k}}^\La \frac{\Th(2\om_{\vec{k}} - E_\text{max})}{\om_{\vec{k}}^3}.
\]
Finally the 1-loop contributions to $\phi^4$ are given by
\[
\eql{3dphi4}
 &\includegraphics[valign=c,scale=0.65]{Figs/V221_f}
+ \cdots
- \bigg[ \includegraphics[valign=c,scale=0.65]{Figs/V221_f}+\cdots \bigg]_\text{eff}
\nn[2pt]
&\qquad\quad{}
\simeq -\frac{\la^2}{256\pi^2 R^2}
\myint \mathrm{d}^2 x \ggap \bra{f} \no{\phi^4} \ket{i}
\times \sum_{\vec{k}}^\La \frac{\Th(2\om_{\vec{k}} - E_\text{max})}{\om_{\vec{k}}^3}
\]

The $\phi^2$ term (not normal-ordered) in the effective Hamiltonian is then given by
\[
H_2 = \myint \mathrm{d}^2 x\ggap \frac 12 m^2_2 \phi^2 + \cdots\,,
\]
with 
\[
\nn[-30pt]
m^2_2 &= -\frac{\la^2}{384\pi^4 R^4} 
\sum_{\vec{k}_{1,2,3}}^{\La}  \de_{123,0} 
\frac{\Th(\om_1 + \om_2 + \om_3 - E_\text{max})
- \prod_{i=1}^3 \Th(|\vec{k}_i| - \mu R)}
{\om_1 \om_2 \om_3 (\om_1 + \om_2 + \om_3)}
\nn[4pt]
&\qquad{}
- \frac{\la}{32\pi^2 R^2} m_V^2  \sum_{\vec{k}}^\La \frac{\Th(2\om_{\vec{k}} - E_\text{max})}{\om_{\vec{k}}^3}
\nn[4pt]
&\qquad{} +\frac{3\la^2}{512 \pi^4 R^4} \sum_{\vec{k}}^\La \frac{\Th(2\om_{\vec{k}} - E_\text{max})}{\om_{\vec{k}}^3}\sum_{|\vec{k}'| \, \le \, k_\text{max}} \frac{1}{\om_{\vec{k}'}}.
\]
The second step function in the first sum comes from the counterterm \Eq{m2ct3D} and the third line in the sum is from un-normal-ordering the $\phi^4$ correction \Eq{3dphi4}.
This is UV finite, and can be evaluated numerically. 
We can also see the separation of scales 
if we choose the renormalization scale $\mu \sim E_\text{max}$ in the expression 
of $m_V^2$ \Eq{mv3d}. 
Focusing on the terms with mixed UV/IR behavior, we have 
\[
m^2_2 &\simeq - \frac{\la^2}{128 \pi ^4 R^4}\biggl(\, \underbrace{\frac 12 + \frac 14 - \frac 34}_{{}  = \, 0}\, \biggr)\sum_{|\vec{k}'| \, \ll \, k_\text{max}} \frac{1}{\om_{\vec{k}'}} \sum_{\vec{k}}^\La \frac{\Th(2\om_{\vec{k}} - E_\text{max})}{\om_{\vec{k}}^3} +\cdots
\]
and we see exactly the same cancellation as in
the 2D case \Eq{2dsep}.

\section{A Hermitian Effective Hamiltonian?}
\scl{nonHermitian}

The effective Hamiltonian $H_\text{eff}$ 
defined in this paper
is non-Hermitian.
In this appendix, we 
look for a similarity transformation
\[
H_\text{eff}' = G H_\text{eff} G^{-1}
\]
so that $H_\text{eff}'$ is Hermitian.
We will consider this for the 2D $\la\phi^4$ theory, where
the leading non-Hermitian terms in the $1/E_\text{max}$
expansion involve $\phi^2 H_0$ and $\phi^4 H_0$ (see \Eq{H2E3}).
Writing $\phi^n H_0 = \frac 12 \{ \phi^n, H_0 \}
+ \frac 12 [\phi^n, H_0]$, we have
\[
H_\text{eff} = \text{Hermitian}
+ \frac{\la^2}{E_\text{max}^3} \myint \text{d}x\ggap
\bigg( c_2 [\phi^2, H_0] + c_4 [\phi^4, H_0] \bigg)
+ O(1/E_\text{max}^4).
\eql{HeffnonHerm}
\]
We therefore define $G$ perturbatively
\[
G = 1 + X + O(X^2),
\]
which gives
\[
\eql{Gpert}
H_\text{eff}' = H_\text{eff} + [X, H_\text{eff}]
+ O(X^2).
\]
Comparing \Eqs{HeffnonHerm} and \eq{Gpert},
we see that a natural choice is
\[
X = \ep H_0.
\]
Since
\[
V = \myint \text{d}x \ggap \bigg[ \frac 12 m_V^2 \phi^2
+ \frac{\la}{4!} \phi^4 \bigg]
\]
this gives
\[
[X, H_\text{eff}] = \ep \myint \text{d}x\ggap \bigg(
\frac 12 m_V^2 [H_0, \phi^2]
+ \frac{\la}{4!} [H_0, \phi^4] \bigg)
+ O(\ep\la^2) + O(\ep^2) .
\]
Comparing with \Eq{HeffnonHerm},
we see that we can choose $\ep$ to cancel one of
the two leading non-Hermitian terms, but not both.

\pagebreak

\addcontentsline{toc}{section}{\protect\numberline{}References}%
\bibliographystyle{utphys}
\bibliography{HTET}

\end{document}